\documentclass[a4paper]{JHEP3}

\JHEPspecialurl{http://jhep.sissa.it/JOURNAL/JHEP3.tar.gz}
\usepackage{epsfig,amssymb,amsmath}

\newdimen\mathindent
\renewcommand{\qquad}{\hspace*{25pt}}
\newcommand{\eref}[1]{(\ref{#1})}
\def\uno{1\!\! 1}

\def\rank{{\rm rank\,}}

\newcommand{\eqll}[2]{\begin{equation}{#1}\label{#2}\end{equation}}
\newcommand{\eq}[1]{\begin{equation}{#1}\end{equation}}
\def\integer{{\mathbb{Z}}}
\def\real{{\mathbb{R}}}

\newtheorem{theo}{Theorem}

\title{Symmetry allowed, but unobservable, phases in renormalizable Gauge Field Theory Models}
\author{G. Sartori  and G. Valente \\
 Dipartimento di Fisica,Universit\`a di Padova \\
and INFN, Sezione di Padova\\
 via Marzolo 8, I--35131 Padova, Italy \\
E-mail: \email{gfsartori@padova.infn.it},\email{valente@padova.infn.it}}

\received{February 20, 2004}
\revised{May 1, 2004} \accepted{November 27, 2004}
\preprint{\hepth{}}

\abstract{In Quantum Field Theory models  with spontaneously
broken gauge invariance, renormalizability limits to four the
degree of the Higgs potential, whose minima determine the vacuum
state at tree-level. In many models, this bound has the intriguing
consequence of preventing the observability, at tree-level, of
some phases that would be allowed by symmetry. We show that,
generally, the phenomenon persists also if one-loop radiative
corrections are taken into account. The tree-level unobservability
of some phases is characteristic in two-Higgs-doublet extensions
of the Standard Model with additional discrete symmetries (to
protect against neutral current flavor changing effects, for
instance). We show that an extension of the scalar sector through
suitable singlet fields can resolve the {\em unnatural}
limitations on the observability of all the phases allowed by
symmetry.}


\begin{document}

\section{Introduction}
\label{intro}

In Quantum Field Theory models with spontaneously broken gauge
invariance, at tree level, the true vacuum of the theory is
determined by the location of the absolute minimum of the Higgs
potential, thought of as a function of classical fields $\phi
\in\real^n$. The set $\phi = (\phi_1,\dots ,\phi_n)$ of all the
scalar fields of the model transforms as an $n$-dimensional vector
of the space of a linear representation of the full symmetry group
(gauge group plus, possibly, discrete symmetries) of the
Lagrangian (we shall denote by $G$ the linear group thus defined).
The Higgs potential $V^{(d)}_a(\phi)$ is built as the most general
G-invariant polynomial of given degree $d$ and is characterized
also by a set $a=(a_1,a_2,\dots )$ of {\em independent} real
coefficients determined by external conditions ({\em control
parameters}). Generally the degree of the Higgs potential is
chosen to be four, to guarantee renormalizability of the theory,
and the control parameters are completely free, but for the
constraints on the coefficients of the terms of highest degree in
$\phi$, required to guarantee that $V^{(d)}_a(\phi)$ is bounded
from below.

Owing to $G$-invariance, the absolute minimum of $V^{(d)}_a(\phi)$
is degenerate along a $G$-orbit $\Omega_0$, whose points define
equivalent vacua. The set of subgroups of $G$ that leave invariant
({\em isotropy subgroups of $G$ at}) the points of $\Omega_0$ form
a conjugacy class $[G_0]=\{g\,G_0\,g^{-1}\mid g\in G\}$, that
defines both the {\em orbit type} of $\Omega_0$ and the residual
symmetry of the system after spontaneous symmetry breaking. We
shall think of this symmetry as thoroughly characterizing the {\em
phase} of the system\footnote{We are only interested in the so
called {\em structural} phases and, in this paper, the term phase
will be a synonymous of structural phase.}.

Distinct $G$-orbits can have the same symmetry and orbits with the
same symmetry are said to form a {\em stratum}. Minima of the
Higgs potential located at orbits lying in the same stratum
determine the same phase: {\em there is a one-to-one
correspondence between strata and phases allowed by the
$G$-symmetry}.

An {\em allowed} phase can be dynamically realized as a phase of
the system at tree-level only if the Higgs potential develops an
absolute minimum at an orbit of the corresponding stratum, for at
least a choice of values in the range of the control parameters.
This possibility is strongly conditioned by the degree of the
polynomial $V^{(d)}_a(\phi)$, which has to be chosen $\le 4$, if
one likes to guarantee the renormalizability of the model.

Generally, by varying the values of the control parameters, the
location of the absolute minimum of the Higgs potential can be
moved to different strata. When this happens, structural phase
transitions take place \cite{Landau}:

\begin{enumerate}
\item The transition is said {\em second order} if a continuous
variation of the control parameters
determines a continuous displacement of the location of the
absolute minimum to a contiguous stratum and a consequent
abrupt change of the residual symmetry. The initial and final
symmetries are necessarily linked by a {\em group-subgroup relation}.
\item The transition is said {\em first order} if,  for some
values of the control parameters, the absolute minimum of the potential
coexists with a  faraway local minimum, sitting in a
different (not necessarily contiguous) symmetry stratum. As the
control parameters vary, the local minimum becomes deeper than the
original global minimum, which is first transformed into a
metastable local minimum and, subsequently, may even disappear.
The details of the phase transformation process may be different,
depending on the physical problem one is dealing with. According
to the {\em delay convention} the system state remains in a stable
or metastable equilibrium state until such state disappears.
According to the {\em Maxwell convention}, the
system state always corresponds to the global minimum of the
potential. These two conventions represent extremes in a continuum
of possibilities (see \cite{Gilmore}). Quite recently, the
impact of a delay-like convention ({\em i.e.} the requirement that
the electroweak vacuum is sufficiently long-lived) on the lower
bounds on the Higgs mass has been analyzed \cite{Isidori}.
\end{enumerate}

A phase will be said to be {\em stable} if it is associated to a
{\em non-degenerate} absolute minimum of $V^{(d)}_a(\phi)$ which
is {\em stable in its stratum}, stability being intended in the
sense that small arbitrary
perturbations of the control parameters in their allowed range
cannot push the location of the minimum in a different stratum.
Generally, only stable phases are thought to have non-zero
probability to be observed. Therefore, in this paper, we shall
identify the {\em observable} phases of a model with the allowed
phases which can be stable in the dynamics of the model.

Our attitude in the analysis of the critical points of an Higgs
potential is suggested by Catastrophe Theory, whose aim is to
classify the modifications in the {\em qualitative nature} of the
solutions of equations depending on ({\em control}) parameters, as
these are varied. A particularly interesting
class of equations is formed by gradient systems, {\em i.e.}
autonomous dynamical systems, in which the (generalized) forces
can be derived from the gradient of some potential. In particular,
Elementary Catastrophe Theory studies the way the equilibria of a
potential are modified as the control parameters are varied \cite{book}.
In this framework, a potential is considered as {\em structurally
stable} if its qualitative properties (number and types of
critical points, basin of attractions, etc.) are not changed by a
sufficiently small perturbation of the control parameters. In a
$n$-parameter family of functions, Morse functions\footnote{A
Morse functions is characterized by the fact that its Hessian
matrix is regular at all critical points.} are generic, i.e.\ are
structurally stable. Thus, in a model describing the evolution of
our Universe through a phenomenological potential, consisting in a
$n$-parameter family of functions, it is natural to assume that a
physically realizable phase corresponds to a generic
configuration. Non-Morse potential functions have the role of
organizing the entire qualitative nature of the family of
functions, determining the possible phase transitions.

A model in which all the allowed phases are observable will be
said to be {\em complete}.

It is not difficult to guess that, if in a model the degree of the
Higgs potential is allowed to be sufficiently high, then the model
is complete \cite{AS3}. On the contrary, if the degree of the
potential is limited, for instance to guarantee the
renormalizability of the model, some allowed phases may become
unobservable at tree-level. This fact has been more or less known
since a long time, but has never attracted the due attention,
mainly because, after the paper by S. Coleman and E. Weinberg
(hereafter referred to as CW \cite{CW}) it is widely belived that
the problem can always be removed by radiative corrections, whose
contributions to the ``effective'' Higgs potential consist in
$G$-invariant polynomials in $\phi$, of increasing degrees at
increasing perturbative orders.

One of the main goals of this
paper is to prove that this widespread belief is based on an
unjustified extensive interpretation of the CW results, in the
sense that the inclusion of radiative corrections\footnote{Given the
big difficulties in the calculations of the effective potential at
more than one-loop and in the determination of its absolute
minimum, it is difficult to conceive that it will be possible to
prove or disprove the fact that a complete perturbative solution
of a model is necessarily complete.} is
not in general, sufficient to cure the tree-level
incompleteness of a gauge model. This statement will be proved
to be true in a (SO$_3\times\integer_2$, {\underline 5}) model
studied by CW, in an (SO$_3$, {\underline 5}) variant of the model
and in an (SU$_3$, {\underline 8}) model.

The reason why radiative corrections may result ineffective in
removing a tree-level incompleteness of a model, is due to the
fact that, in the $G$-invariant polynomials in $\phi$, yielding
the contributions of radiative corrections, the coefficients are
well determined functions of the parameters defining the Lagrangian
of the model at tree-level, and cannot, therefore, play the role of arbitrary
independent parameters, like the control parameters\footnote{The
arbitrariness in the choice of the renormalization point is
irrelevant, since a change in this choice leads only to a
reparametrization of the same theory.}.

It is also worth recalling that, in spontaneously broken gauge
symmetries, the exact effective potential is real, while its
perturbative series can be complex (see for instance
\cite{AA,BB,DD,CC}). So, besides the computational difficulties
in the determination of the quantum contributions to the (perturbative)
effective potential, which essentially limit the results to one- or
two-loop effects, particular care has to be taken in the
regions where the effective quantum potential is complex\footnote{Although
fascinating, the interpretation of the possible imaginary
part as a decay rate (\cite{BB}) seems to be somehow ambiguous
(see note nr. 18 in \cite{CC}).}.

We consider quite intriguing the emergence, in the set of allowed phases
of a model, of possible selection rules originating from the constraint posed by
the request of renormalizability.
Our point of view is that renormalizability, which actually has to be
considered as a ``technical'' assumption required to allow a
consistent and significant perturbative solution of the theory,
should not limit the implications of the basic symmetry of the
formalism used to describe the system \cite{Gufan}, {\em not even
at tree-level}. In other words, in our opinion, {\em all the
allowed} phases should be {\em observable} already at tree-level
in a viable model. This attitude, if accepted, may have important
consequences in the study of of Electro-Weak (EW) phase
transitions, in the sense that all the allowed phases have to be
thought, in principle, as possible phases in the evolution of the
Universe \cite{Zar, Zar1}.

In the Standard Model (SM) of EW interactions, although the gauge
boson and fermion structure has been accurately tested,
experimental information about the Higgs sector (HS) is still very
weak (see, for instance, \cite{1,2} and references therein).
Serious motivations are well known for the extension of the scalar
sector; among them we just recall supersymmetry (SUSY) and
baryogenesis at the EW scale, \cite{Rub}. So far, various
extensions of the SM have been devised: the Minimal SUSY SM, the
SM plus an extra Higgs doublet, the MSSM plus a Higgs singlet, the
left--right symmetric model, the SM plus a complex singlet Higgs
(see the introduction to \cite{3} and references therein); quite
recently, even a partly supersymmetric SM has been conceived
(\cite{4}). There is still, therefore, a certain freedom in the
choice of the Higgs sector of the theory. A second goal of this
paper is to show how compatibility between tree-level completeness
and renormalizability can give further inputs in its construction.

In particular we shall show that, while in the basic two-Higgs-doublet
extension of the SM all the allowed
phases are observable, in the most popular models with two Higgs
doublets, if the usual additional discrete symmetries are added to
avoid flavor changing neutral current (FCNC) effects, this is only
true if the Higgs potential is a polynomial of sufficiently high
degree, greater than four.

Nowadays, there is general agreement in considering the Standard
Model (SM) as an effective theory \cite{WW1,WW2,WW3}, since, for
example, higher order (non renormalizable) operators are generally
required to describe non vanishing  neutrino masses. So, the main
attitudes in dealing with phenomenology are either thinking to the
SM as a low energy limit of a (supersymmetric) Grand Unification
Theory (GUT) or to disregard the parent high energy theory and
trying to recover, in a model independent way, some knowledge on
bounds to the GUT unification scale, used to suppress higher order
operator contributions to low energy physics. String theory is a
major (but not the only!) candidate for this high energy theory,
having the capability to include gravitation in a unique
framework.

Despite this and even if it is only a
technical requirement, one may wonder whether
renormalizability can be maintained, without limiting the symmetry
content of the theory. We shall show that, in some tree-level
incomplete two-Higgs-doublet models, symmetry and renormalization
can be reconciled if the Higgs sector
is extended with the addition of one or more scalar singlets, with
convenient transformation properties under the discrete symmetries of
the model.

The paper is organized in the following way. In Section 2, making
systematic use of simple results and techniques of geometric
invariant theory \cite{Mumford}, which strongly simplify the
calculations, we determine all the allowed phases of three simple
models: an (SO$_3\times \integer_2,{\underline 5}$)--model studied
as an example by CW, a simple  (SO$_3,{\underline 5})$ variant of
the same model and, finally,  an (SU$_3$, ${\underline
8}$)--model. We show that all these models are tree-level
incomplete and that the incompleteness is not removed if one-loop
radiative corrections are taken into account. In Section 3 we
justify the formal approach followed in Section 2, recalling the
basic elements of a general approach (orbit space approach) to the
determination of all the allowed phases \cite{AS1, AS2, AS3} of a
gauge model. Section 4 is devoted to the determination of allowed
and observable phases in two Higgs doublet (2HD) extensions of the
Standard Model in different dynamical configurations
(renormalizable and incomplete or non-renormalizable and
complete). In particular, besides the basic 2HD model with gauge
and symmetry group SU$_2\times \mathrm{U}_1$, we shall examine a
2HD model with an additional FCNC protecting  discrete symmetry
and  a model in which the symmetry group is further extended with
the inclusion of a CP-like transformation. In Section 5, we show
that the extensions of these models with the introduction of
convenient additional scalar singlets allows to make them
complete, without giving up renormalizability.

\section{Allowed and observable phases in three simple gauge models}

In this section we shall determine all the allowed phases of three
simple models: an (SO$_3\times \integer_2,{\underline 5}$)--model
studied as an example by CW, a simple (SO$_3,{\underline
5})$ variant  of the same model and an (SU$_3$, ${\underline 8}$)--model. We
show that:

\begin{enumerate}
\item the renormalizable versions of all these models are tree-level incomplete;
\item the incompleteness persists if the tree-level Higgs potential
is replaced with the one-loop effective potential;
\item the incompleteness is completely removed at tree-level if one gives up renormalizability
and allows a sufficiently high degree polynomial Higgs potential.
\end{enumerate}

We shall be highly facilitated in our calculations by a systematic
use of simple techniques and results of geometric invariant theory, that will
be illustrated in a general formulation, in the next section.

\subsection{An (SO$_3$, \underline{5}) gauge model}
The model is a slight modification of an
(SO$_3\times\integer_2,{\underline 5})$ model, studied by CW, that
will be analyzed in the following subsection.

The gauge (and complete symmetry) group of the model is SO$_3$ and
the Higgs fields transform as the components of a vector $\phi$ in
the space of a real five dimensional orthogonal representation of
the group. If the components of $\phi$ are ordered in a $3\times
3$ traceless symmetric matrix $\Phi(\phi)$\footnote{We have slightly
modified the definition by CW, so that the representation
of SO$_3$ turns out to be orthogonal.}:

\begin{equation}
\Phi(\phi) = \frac 1{\sqrt 2}\,\left(\begin{array}{ccc}
\phi_4 + \phi_5/\sqrt 3& \phi_1                  & \phi_2\\
\phi_1                 & -\phi_4 + \phi_5/\sqrt 3& \phi_3\\
\phi_2                 ,& \phi_3                  & -2\phi_5/\sqrt
3
\end{array}\right),\label{matrixphi}
\end{equation}
their transformation properties under a transformation $\gamma\in
\mathrm{SO}_3$ are specified by the following relations:

\begin{equation}\Phi(\phi)\rightarrow \Phi(\phi')=\gamma\,\Phi(\phi)\,
\gamma^{-1},\qquad \phi'_i=\sum_{j=1}^5\,g_{ij}\,\phi_j,
\label{transfprop}\end{equation} which determine the matrices $g$
forming the linear group $G$. This group has only two
basic\footnote{A set $\{p_1,\dots ,p_q\}$ of basic invariant
polynomials is formed by independent invariant polynomials such
that any polynomial invariant in $\phi$ can be written as a
polynomial in $p$.} homogeneous invariant polynomials, that can be
conveniently chosen to be the following:

\begin{equation}p_1(\phi)={\rm Tr}\,\Phi^2(\phi),\qquad p_2(\phi)=
\sqrt 6\,{\rm Tr}\,\Phi^3(\phi).\label{p}\end{equation}
In particular $p_1(\phi)=\sum_{i=1}^5\,\phi_i^2$, assuring that,
as claimed, the group $G$ is a group of orthogonal matrices.

A general fourth degree $G$-invariant polynomial, to be identified
with the Higgs potential of a renormalizable version of the model,
can be conveniently written in the following form, in terms of the
basic polynomial invariants $p=(p_1,p_2)$:

\begin{equation}V^{(4)}_a(\phi)=\widehat V^{(4)}_a(p(\phi)),\label{V4}
\end{equation}
where

\begin{equation}\widehat V^{(4)}_a(p)=a_1\,p_1 + a_2\,p_2 + a_3\,
p_1^2,\qquad a=(a_1,a_2,a_3) \label{Vp}\end{equation}
and $a_3$ has to be positive to guarantee that the potential is
bounded from below for arbitrary $a_1$ and $a_2$.

A determination of the stationary points of $V^{(4)}_a(\phi)$ with
standard analytic methods is not easy, even in this simple case,
so it is convenient to tackle the problem in a cleverer way.
Since, as stressed in the Introduction, one is essentially
interested only in  the location of the $G$-orbit at which
$V^{(4)}_a(\phi)$ takes on its minimum, and a $G$-invariant
function is a constant along a $G$-orbit, it will be advantageous
to express $V^{(4)}_a$ as a function of the $G$-orbits. The
approach followed by CW points in this direction. In fact, they
exploit the fact that the matrix $\Phi(\phi)$ can be diagonalized
by an SO$_3$ transformation (see (\ref{transfprop})). This means
that in every $G$-orbit there is at least a point $\phi$
represented by a diagonal matrix $\Phi(\phi)$, so the minimization
problem can be tackled with the additional conditions $\phi_1 =
\phi_2 = \phi_3 = 0$, an expedient that make it easily solvable.
Even if effective in simple cases, like the one we are
considering, this approach has two shortcomings:

\begin{enumerate}
\item The diagonalization of $\Phi(\phi)$ does not lead to a unique result.
From a geometrical point of view, the different results correspond
to the distinct intersections of the $G$-orbit through $\phi$ with
a convenient orthogonal hyperplane and, in the case of compact
groups, these intersections are always multiple. In the present
case, for fixed $(x,y)\in\real^2$ and $\phi_1 = \phi_2 = \phi_3 =
0$, the six distinct points $(\phi_4,\phi_5)=(\pm x,y),\ (\pm (x -
\sqrt 3 y)/2,\ -(\sqrt 3 x + y)/2),\ ((\pm x + \sqrt 3
y)/2,(-\sqrt 3 x + y)/2)$ lie on the same orbit. Thus, the
coordinates $(\phi_4,\phi_5)$ do not yield a one-to-one
parametrization of the orbits of $G$.
\item The approach cannot be generalized to an arbitrary compact linear group $G$.
\end{enumerate}

\FIGURE{\epsfig{file=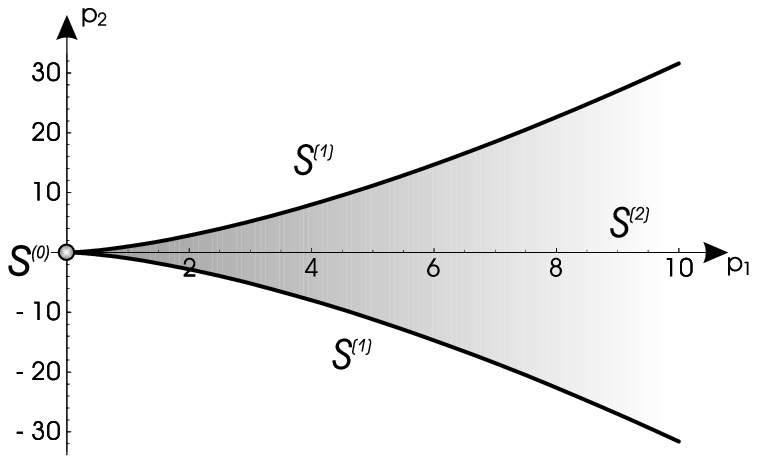,width=8cm}\caption{Orbit space and
symmetry strata of (SO$_3,\underline{5})$ gauge
model.}\label{Nuo1}}

Both these difficulties can be overcome using a fundamental
property of the basic invariants of any compact linear group: at
distinct orbits, $p$ takes on distinct values (see next section).

Therefore, $(p_1,p_2)$ can be used as coordinates of the orbits of
$G$ and the minimization of $V^{(4)}_a(\phi)=\widehat
V^{(4)}_a(p(\phi))$ can be reduced to the minimization of
$\widehat V^{(4)}_a$, thought of as a function of the independent
variables $(p_1,p_2)$.

 This choice yields, as additional
significant bonus, a sensible reduction of the degree of the
polynomial to minimize: $\widehat V^{(4)}_a(p)$ is only second
degree in $p_1$ and linear in $p_2$. The sole price to pay for
these advantages is that the range of $p$ does not coincide with
the real $p$-plane and the minimization problem for $\widehat
V^{(4)}_a(p)$ has to be dealt with as a constrained minimization
problem. But this is a solvable problem. The range of $p$ can,
in fact, be easily determined in the following way. The
rectangular matrix formed by the gradients of the basic
invariants, multiplied by its transpose, defines a positive
semi-definite matrix

\eq{P_{\alpha\beta}(\phi)=\sum_{i=1}^5\,\frac{\partial
p_\alpha(\phi)}{\partial \phi_i}\, \frac{\partial
p_\beta(\phi)}{\partial \phi_i},\qquad \phi\in\real^5,\ \alpha,\beta=1,2,}
whose elements are $G$-invariant polynomials in $\phi$, since the
group $G$ is a group of orthogonal matrices.
\FIGURE{\epsfig{file=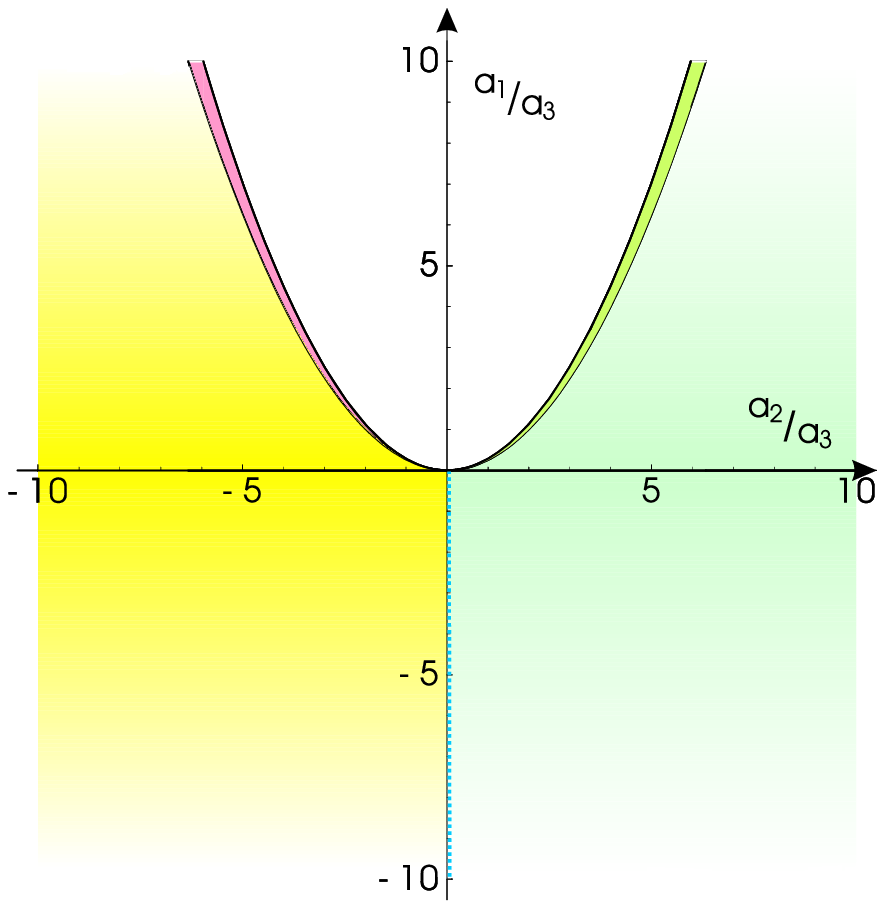,width=7cm}\caption{Representation
 of the solution of the
minimization problem for the (SO$_3,{\underline 5}$ )--model in
the space of the control parameters.}\label{Nuo2}}

In fact, taking into
account also the homogeneity properties of $p_1$ and $p_2$, one
easily finds

\eqll{P(\phi)=\widehat P(p(\phi)),\quad \widehat
P(p)=\left(\begin{array}{rr} 4\,p_1& 6\,p_2 \\  6\,p_2& 9\,p_1^2
\end{array}\right). }{matP}

 The
following conditions, as\-su\-ring the se\-mi-po\-si\-ti\-vi\-ty of the matrix
$\widehat P(p)$, define the range $p(\real^5)$ in the $p$-space
(see Fig.~\ref{Nuo1}):

\begin{equation}p_1^3 - p_2^2 \ge 0.\label{rangep}\end{equation}

As stated in the Introduction, the points $p$ in the range
$p(\real^5)$ of $p(\phi)$ are in a one-to-one correspondence with
the $G$-orbits. Thus, the algebraic set $p(\real^5)$ can be
identified with the {\em orbit space $\real^5/G$ of $G$}.

The validity and meaning of  condition (\ref{rangep}), and,
particularly, of the limiting cases, can be easily understood if it is
written in terms of $\phi$, for $\phi_1 = \phi_2 = \phi_3 = 0$:

\begin{equation}p_1^3(\phi) - p_2^2(\phi)\mid_{\phi_1 = \phi_2 = \phi_3 = 0} =
\phi_4^2\left(\phi_4^2 - 3\phi_5^2\right)^2 \ge 0.\label{rangep1}
\end{equation}

A remarkable fact, which is characteristic of the orbit spaces of
all compact groups, is the following. Being the orbit space a
connected semi-algebraic set, it presents a natural geometric
stratification (disjoint partition) in connected manifolds ({\em
primary strata}), each primary stratum being open in its
topological closure and contained, but for the highest dimensional
one which is unique ({\em principal stratum}), in the boundary of
a higher dimensional primary stratum. In the present case the
primary strata (shown in Fig.~\ref{Nuo1}) correspond to the
following algebraic manifolds $W^{(i)}_{j}$ (the apex $i$
indicates the dimension and $j$ is an order index):

\eq{\begin{array}{lcl} W^{(0)}:\ p_1=0=p_2;&\hspace{1em}& W^{(1)}_1:\ p_1 > 0 = p_2 - p_1^{3/2};\\
W^{(1)}_2:\ p_1 > 0 = p_2 + p_1^{3/2}; &\hspace{1em}& W^{(2)}:\ p_2^2 <
p_1^3.\end{array}} {\em The symmetry strata are formed by one or
more primary strata with the same dimensions.} This property
reduces the determination of all the symmetry strata, {\em i.e.\/}
of all the allowed phases, to the determination of the solutions
of the equation

\eq{g\,\Phi(\phi)\, g^{-1} =  \Phi(\phi),\qquad g\in \mathrm{
SO}_3,\label{g}} only for a few configurations of $\Phi(\phi)$,
one for each primary stratum, and, for each primary stratum,
$\Phi(\phi)$ can be chosen in such a way that the solution turns
out to be particularly simple.

Good choices of $\phi$ for an easy determination of the solutions
of (\ref{g}) are, for instance, $\phi=(0,0,0,0,-1)$,
$\phi=(0,0,0,0,+1)$ and $\phi=(0,0,0,1,0)$  for $W^{(1)}_1$,
$W^{(1)}_2$ and $W^{(2)}$, respectively.
Using also (\ref{transfprop}) one easily concludes that

\begin{itemize}
\item[i)] The $G$-orbit corresponding to the point $\phi = 0$ is
represented by the tip of the orbit space and forms a stratum
$S^{(0)}$, formed by a unique orbit with symmetry $[SO_3]$,
corresponding to the phase ${\mathcal F}^{(0)}$ with unbroken
symmetry.
\item[ii)] The other SO$_3$-orbits lying on the boundary of the orbit space,
characterized by the conditions $p_1^3 = p_2^2 > 0$ $(\phi_1 =
\phi_2 = \phi_3 = \phi_4^2\left(\phi_4^2 - 3\phi_5^2\right)^2 = 0
\ne \phi_4^2 + \phi_5^2$), share the same symmetry [SO$_2$]. They
form, therefore, a unique stratum $S^{(1)}$, corresponding to a
phase ${\mathcal F}^{(1)}$: $S^{(1)} = W^{(1)}_1 \cup W^{(1)}_2$.
\item[iii)] {\em Generic} SO$_3$-orbits, corresponding to interior points of
the orbit space, characterized by $p_1^3 > p_2^2$ ($\phi_1 =
\phi_2 = \phi_3 = 0 < \phi_4^2\left(\phi_4^2 -
3\phi_5^2\right)^2$) have trivial symmetry (isotropy subgroup
$\{\uno\}$). They form, therefore a unique stratum $S^{(2)}$,
corresponding to a phase ${\mathcal F}^{(2)}$ with completely
broken symmetry: $S^{(2)} = W^{(2)}$.
\end{itemize}

 Let us now examine the tree-level observability of the
three {\em allowed} phases just found, in the assumption that the
dynamics in the Higgs sector is determined by the potential
(\ref{Vp}). For this purpose, we have only to check whether, for
each stratum $S^{(i)}$, there is an exclusive three dimensional region
$R^{(i)}$ in the space of the control parameters $a=(a_1, a_2,a_3)$,
such that, for $a\in R^{(i)}$ the function $\widehat V^{(4)}_a(p),\
p\in \real^n/G$ has a stable absolute minimum located in $S^{(i)}$.

For general values of $a_2$, $\widehat V^{(4)}_a(p)$ is linear
in $p_2$. So, one immediately realizes that, for any fixed value of
$p_1$, it takes on its absolute minimum when $p_2$ is
maximum (for $a_2 < 0$) or minimum (for $a_2 > 0$), that
is on the boundary of the orbit space, formed by the union of the
strata $S^{(0)}$ and $S^{(1)}$. Only for $a_2=0$ and $a_1< 0$ the
absolute minimum is located in the principal stratum $S^{(2)}$,
but it is degenerate along a line $p_1 = -a_1/(2a_3)$, which also
crosses the stratum $S^{(1)}$. Any perturbation $\delta a_2\ne 0$
would move it to $S^{(1)}$, so it is unstable.

A complete analytic solution of the minimization problem leads to
the following results (see Fig.~\ref{Nuo2}):

\begin{enumerate}
\item For $a_1<a_2^2/(4a_3)\ne 0$ (open region below the lower
parabola in Fig.~\ref{Nuo2}), the absolute minimum is stable in
the stratum $S^{(1)}$.
\item For $a_1>a_2^2/(4a_3)$  the absolute minimum is stable in
the stratum $S^{(0)}$.
\item For $9a_2^2/(32a_3) > a_1 > a_2^2/(4a_3)$ (open region between
the two parabolas in Fig.~\ref{Nuo2}) the absolute minimum in
$S^{(0)}$ is stable and coexists with a higher stable local
minimum in $S^{(1)}$.
\item For $a_1<0=a_2$, the absolute minimum is degenerate and, therefore, \
unstable, across the strata $S^{(1)}$ and $S^{(2)}$.
\item The parabola  of equation $a_1=a_2^2/(4a_3)$ is a critical line,
formed by first order phase transition points.
\end{enumerate}

If the evolution of the Universe were described by
this model, it would be represented by a continuous line in the
space of the control parameters $(a_1,a_2,a_3)$. The only
observable phase transitions would be first order transitions
between the phases ${\mathcal F}^{(0)}$ and ${\mathcal F}^{(1)}$.
These results can be
easily understood graphically, noting that the level curves
(equipotential lines) in the $p$-plane are parabolas (see
Fig.~\ref{Nuo3}). Therefore, the critical lines in the plane
$(a_1/a_3,a_2/a_3)$ can be easily determined (see Fig.~\ref{Nuo2}).

\FIGURE[ht]{\epsfig{file=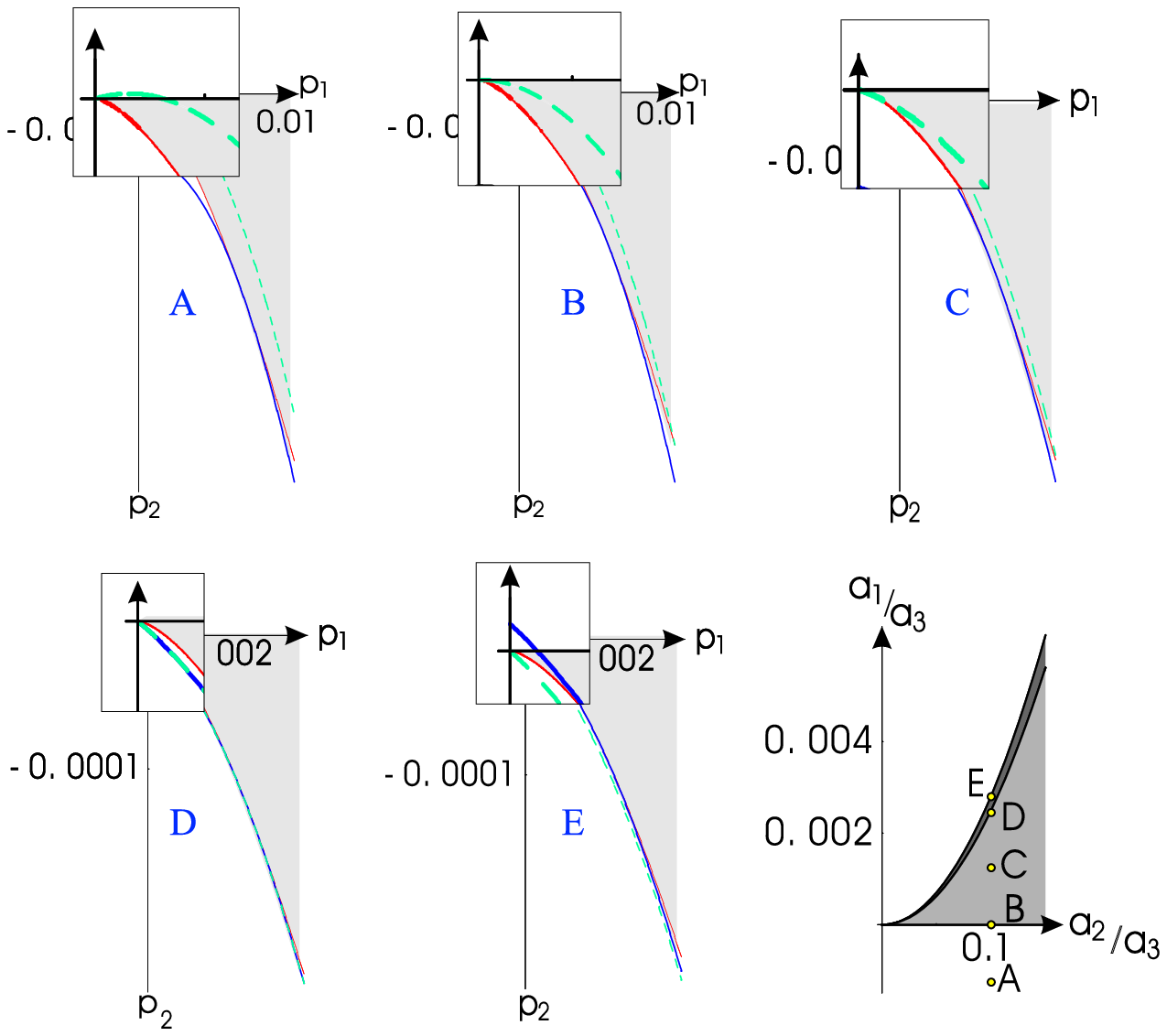,height=4.7in}\caption{Level
curves corresponding to critical points located in the primary
strata $W^{(1)}_2$ (blue, continuous line) and $W^{(0)}$ (dashed
green line) for the (SO$_3,{\underline 5})$ gauge model
(renormalizable version). The border of the orbit space in drawn
in red. Cases A ... E refer to the values of the  control
parameter shown in the last figure (for $a_3=1$ and
$a_2/a_3=0.1$). The region near the origin of the plane $(p_1,
p_2)$ is magnified, to show that the level curve for $W^{(0)}$ is
tangent to the $p_1$ axis, in figure B. In figure D a typical
coexistence of two degenerate minima is shown.}\label{Nuo3}}

\subsubsection{A non-renormalizable version of the model}

Let us now show that if the Higgs potential is chosen as a general
$G$-invariant polynomial of degree six, so that it contains also a
term proportional to $p_2^2(\phi)$, then all the allowed phases
turn out to be observable at tree-level. Like in the previous
subsection, let us define $V^{(6)}_a(\phi)=\widehat
V^{(6)}_a(p(\phi))$, through the relation

\begin{equation}\widehat V^{(6)}_a(p)=a_1\,p_1 + a_2\,p_2 + a_3 p_1^2 + a_4\,p_1\,p_2 +
a_5\,p_1^3 + a_6\,p_2^2.\label{Vp1}\end{equation}

 For arbitrary values of $a=(a_1,\dots ,a_6)$, the
restriction of the function $\widehat V^{(6)}_a(p)$ to the orbit
space is bounded from below if $a_5 > 0$ and $a_6 > -a_5$.

The fact that all the allowed phases are observable can be proved
through explicit standard calculations, but we prefer a more
intuitive approach, that can be generalized to the case of an
arbitrary linear compact group $G$, with a free basis of basic
polynomial invariants.

\FIGURE[ht]{\epsfig{file=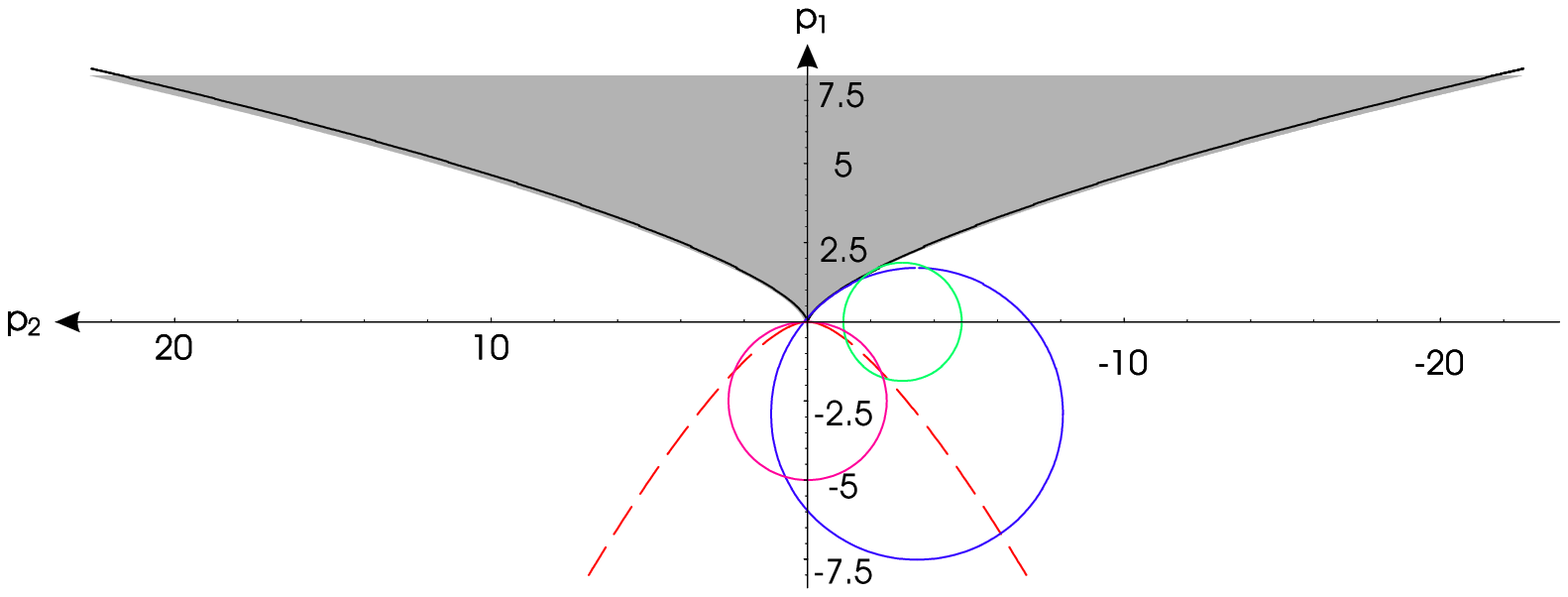,width=14cm}\caption{Level curves
corresponding to critical points located in the primary strata
$W^{(1)}_2$ (green circle) and $W^{(0)}$ (magenta circle) for the
complete non renormalizable version of the (SO$_3,{\underline 5})$
gauge model, for $a=a^{(0)}=(a_1,a_2,1,0,0,1)$. The dashed red
curve represents the Maxwell catastrophe projection, which is the
evolute of the curve $S^{(1)}$, that is the locus of the points in
which degenerate absolute minima coexist in $S^{(1)}$ and $S^{0}$.
The values of the control parameter can be recovered through the
identification $a_1=- 2 \eta_1$ and $a_2=- 2 \eta_2$. The equation
of the critical curve is $6561 {p_2}^4 + 288 (1 + 54 p_1) \,
{p_2}^2 + 512 p_1 (1 + 6 p_1)^2=0$. The region external to the
orbit space can be divided into three connected parts: when $\eta$
is in the first region, extending towards $p_1 \rightarrow
-\infty$ the minimum is in $S^{(0)}$, while in the other two
regions, the minimum lies in the singular stratum $S^{(1)}$ (on
$W^{(1)}_1$ and $W^{(1)}_2$, respectively, for $\eta_1>0$ and
$\eta_1<0$.)}\label{Nuo4}}

Let us denote by $a^{(0)}$ the following particular choice of
values for the control parameters $a=(a_1,\dots ,a_6)$:

\eq{a^{(0)} = (a_1,a_2,1,0,0,1),} where the
coefficients $a_1$ and $a_2$ of the linear terms in $\widehat V^{(6)}_a(p)$, are
still considered as arbitrary parameters. Then, $\widehat
V^{(6)}_{a^{(0)}}(p)$ can be rewritten in  the following
suggestive form:

\eq{\widehat V^{(6)}_{a^{(0)}}(p)=\left(p_1 - \eta_1\right)^2 +
\left(p_2 - \eta_2\right)^2 + C, \qquad p\in p(\real^5),} where,

\eq{\eta_1 = -a_1/2,\qquad \eta_2= -a_2/2,\qquad C = -{\eta_1}^2
-{\eta_2}^2.} In the $p$-space $\real^2$, the polynomial $\widehat
V^{(6)}_{a^{(0)}}(p)-C$ represents the squared distance between
the point $\eta$ and the point $p$ of the orbit space.
It is therefore clear that

\begin{enumerate}
\item For $\eta$ interior to the orbit space, $V^{(6)}_{a^{(0)}}(p)$ has a stable
(against small perturbations of its coefficients $\eta_1$ and
$\eta_2$) absolute minimum at $p=\eta\in S^{(2)}$.
\item For $\eta$ exterior to the orbit space, but not too far from, and $\eta_1 > 0$,
$\widehat V^{(6)}_{a^{(0)}}(p)$ has a stable absolute minimum at
the closest point of $S^{(1)}$ to $\eta$.
\item For $\eta$ exterior to the orbit space and $\eta_1 < 0$,
$\widehat V^{(6)}_{a^{(0)}}(p)$ has a stable absolute minimum in
$S^{(0)}$.
\item For $\eta_1=0$, $\widehat V^{(6)}_{a^{(0)}}(p)$ has an unstable absolute
minimum at $p=0$.
\item For $\eta\in S^{(1)}$, $\widehat V^{(6)}_{a^{(0)}}(p)$
has an unstable absolute minimum at $p=\eta$.
\end{enumerate}

Let us now show that, the conclusions listed in the first three items
above, about the existence of stable absolute minima in all the strata,
continue to hold for $\widehat V^{(6)}_a(p)$, if $a$ ranges in a convenient
domain, close to $a^{(0)}$.

To this end, in the range of the vector control parameters $a$
($a_5>0$), for each arbitrarily chosen couple of non critical
values of $(a_1,a_2)$ ($a_1\,a_2\ne 0$), let us consider points
$a$ which are sufficiently close to $a^{(0)}$. Continuity reasons
assure that, for every choice of $a$ in this region, the
polynomial $\widehat V^{(6)}_a(p)$ has a stable {\em local}
minimum close to the location $p^{(0)}$ of the minimum of
$\widehat V^{(6)}_{a^{(0)}}(p)$ and sitting on the same stratum.
In fact, the existence of a stationary point of $\widehat
V^{(6)}_{a}(p)$ close to $p^{(0)}$ is guaranteed by the inverse
functions theorem and this stationary point is surely an absolute
minimum, since $\widehat V^{(6)}_a(p),\ p\in p(\real^5)$ is a
convex function\footnote{The matrix of second order derivatives of
$\widehat V^{(6)}_a(p)$ with respect to $p$ is positive definite
in a neighborhood of $a^{(0)}$, for positive values of $a_6$ and
$p_1$.}.

Let us denote by $v_a$ the minimum value of $\widehat
V^{(6)}_a(p),\ p\in p(\real^5)$. Then the analytic justification
just given is easy to understand from an analysis of the family of
equipotential lines $\widehat V^{(6)}_a(p)=v_a$, for $a$ near to
$a^{(0)}$ (see Fig.~\ref{Nuo4}).

\FIGURE[ht]{\epsfig{file=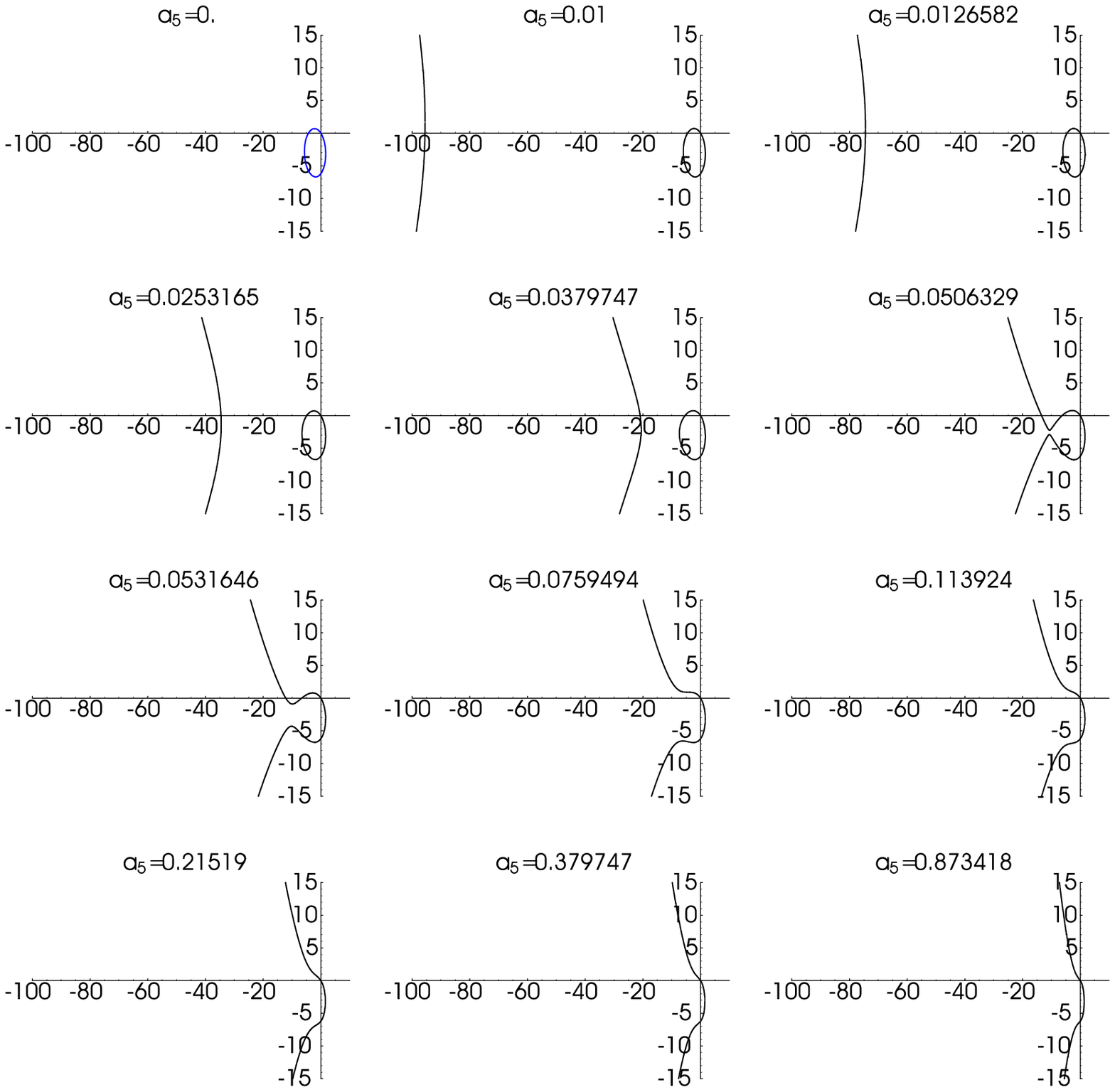,width=5in}\caption{Level curve
deformation for the non-renormalizable version of the
(SO$_3,{\underline 5}$) gauge model. The following values for the
control parameters have been chosen: $a_1 \approx -4.3$, $a_2
\approx 6.18918$, $a_3= a_4 = 0.1$, $a_6 \approx 0.998196$, in
such a way that the centre of the ellipsis corresponding to
$a_5=0$ is in point $(-2,-3)$ in the orbit space plane
$(p_1,p_2)$. For this  configuration the ellipsis turns out to be
tangent to the orbit space primary stratum $W^{(1)}_2$ in the
point of abscissa $p_1 \approx 1.07322$. The sequence of plots
shows how the aspect of the level curve changes according to
different values of $a_5$ parameter.}\label{Nuo5}}

\begin{enumerate}
\item For $a=a^{(0)}$ these lines are obviously circles centered at $\eta$.
They are tangent to the orbit space for $\eta$ exterior to the
orbit space (at a point of $S^{(1)}$, for $\eta_1 > 0$ and at
$S^{(0)}$ for $\eta_1 < 0$), while they reduce to the point $\eta$,
for $\eta$ interior to the orbit space.
\item When $a_3$, $a_4$ and $a_6$ are slightly perturbed, the circle is slightly deformed
to an ellipsis centered in a point near to $\eta$, but, for the rest, the
situation does not essentially change.
\item When also $a_5$ is raised to a small positive value, the geometry of the equipotential
lines abruptly changes, but the conclusions remain essentially the
same indicated above: the ellipsis is
further slightly deformed, but remains a closed curve around $\eta$,
and a new open branch of the algebraic curve is generated. The
new branch, however, is confined to the far negative
$p_1$-half-plane, if $a_5$ is sufficiently small. So it does not intersect the orbit space and
cannot, consequently, host a  minimum of the Higgs
potential.
\end{enumerate}

Before analyzing, in the next subsection, the effects of the
contributions of one-loop radiative corrections to the effective
potential, let us remark the following interesting aspect of the
orbit space approach to the minimization problem we have followed.
It has been proved in \cite{Tal} that all the linear compact
groups, whose base of invariant polynomials reduces to two
elements with the same degrees $(d_1,d_2)$, have isomorphic orbit
spaces. This means that the basic polynomial invariants can be
chosen so that the $\widehat P$-matrix has a universal form,
which, for $(d_1,d_2)=(2,3)$ is specified in (\ref{matP}). Since
the general form of a given degree Higgs potential only depends on
the number and degrees of the basic polynomial invariants, for all
the groups under consideration, the minimization problem of the
Higgs potential at tree-level reduces to the same geometrical
problem. Only the symmetry of the possible phases depend on the
particular symmetry group. The third sample model studied below,
will yield an example of this phenomenon.

\subsubsection{One-loop radiative corrections}
In this subsection, we shall prove that the one-loop radiative
corrections are not sufficient to make observable the phase
${\mathcal F}^{(2)}$, which is tree-level unobservable.

As stressed in CW, the vector bosons are responsible for the
dominant radiative contributions $V_{\rm g}$ to the one-loop
effective potential. In order to calculate $V_{\rm g}$, which is
an SO$_3$-invariant function, in terms of the ``coordinates'' $p$,
let us denote by $\langle\cdot ,\cdot\rangle$ the euclidian scalar
product in $\real^5$, by $T^a$, $a=1,2,3$ the generators of the
Lie algebra of the matrix group $G$ and by $\epsilon^a$ the usual
generators of the Lie algebra of the group SO$_3$:

\eq{\epsilon^a_{ij} = -\epsilon_{aij},\qquad a,i,j=1,2,3.} Then,

\eq{T^a\cdot \Phi = \epsilon^a\,\Phi - \Phi\,\epsilon^a.}

The explicit form of $V_{\rm g}(\phi)$ is the following \cite{CW}:

\eqll{V_{\rm g}(\phi) = \frac 3{64\pi^2}\, {\rm
Tr}\left[M^4(\phi)\,\ln M^2(\phi)\right], }{mass} where the matrix
elements of $M^2(\phi)$ are defined by

\eq{M_{ab}^2(\phi) = g^2\,\langle T_a\phi,T_b\phi\rangle.}

The eigenvalues of the matrix $M^2(\phi)$ are SO$_3$-invariant
algebraic functions of $\phi$ and the sum of all the products of
$k$ distinct eigenvalues can be easily calculated as the sum
$\sigma_k$ of the principal minors of order $k$, $k=1,2,3$. They
are $G$-invariant polynomials in $\phi$ and can, therefore, be
written as polynomial functions of $p$. A direct calculation gives
\DOUBLEFIGURE{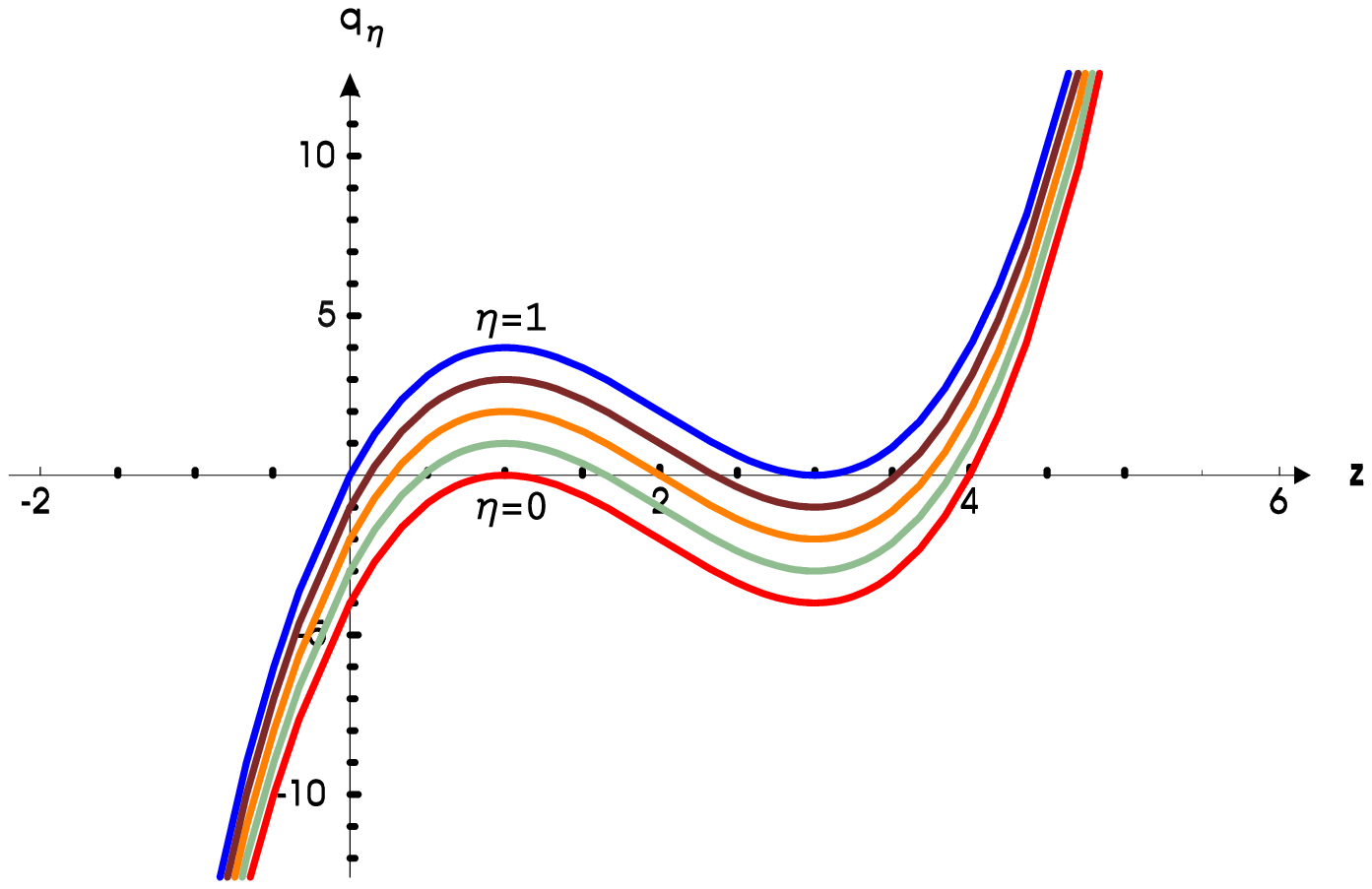,width=6cm}{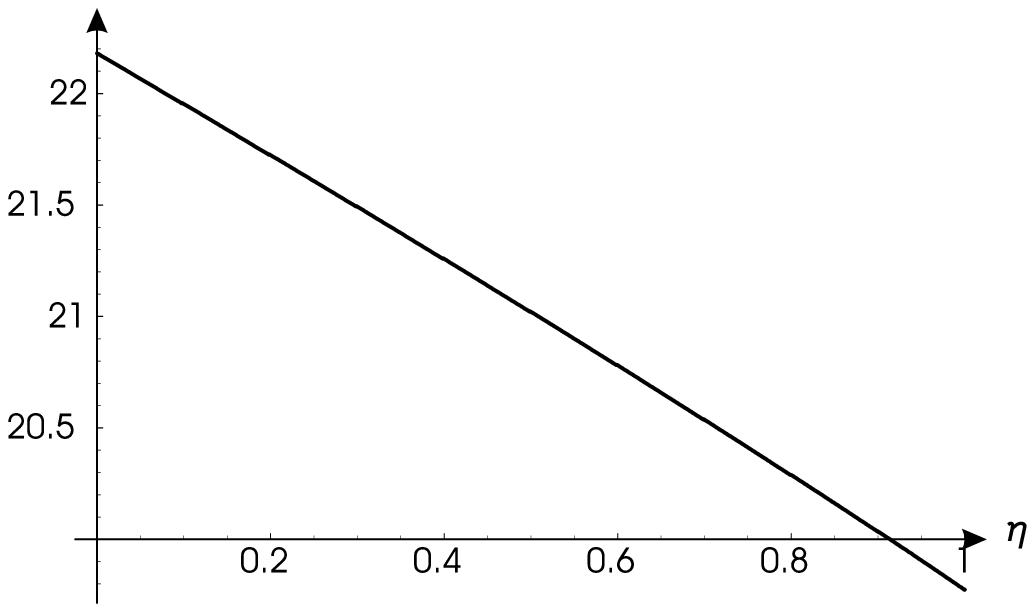,width=6cm}{\label{Nuo6}
Family of functions $q_\eta(z)$, in the interval $[0,1]$ for
$\eta=0,\; 1/4,\; 1/2,\; 3/4,\; 1$. }{\label{Nuo7} Graph of the
function $f(\eta)=\sum_{i=1}^3 z_i^{2}(\eta) \mathrm{ln}
\left(z_i(\eta)\right)$}

\eq{\begin{array}{l}
\sigma_1 = 6\,p_1,\qquad \sigma_2 = 9\,p_1^2,\\
\\
 \sigma_3 ={\displaystyle
4 p_1^3\left(1 - \frac { p_2^2}{p_1^3}\right)}.
\end{array}}

This means that the eigenvalues of $M^2(\phi)$ can be written in
the form $p_1z_i(\eta)$, where the $z_i$ are the roots of the
following polynomial in $z$:

\eq{q_\eta(z) = z^3 - 6\,  z^2 + 9 \, z - 4 \left(1 -
\eta\right)\,,\quad \eta = p_2^2/p_1^3 . \label{e22}}

Like $V(\phi)$, also $V_{\rm g}(\phi)$ can be, more economically,
thought of as a function $\widehat V_{\rm g}(p)$ of $\eta
= p_2^2/p_1^3$ in the orbit
space of $G$:

\eq{\widehat V_{\rm g}(p) = \frac 3{64\pi^2}\,
\sum_{i=1}^3\,p_1^2\,z_i^2(\eta)\,\ln (p_1\,z_i(\eta)).
\label{e23}} The  function $f(\eta)=\sum_{i=1}^3\,z_i^2(\eta)\,\ln
(z_i(\eta))$ is plotted in Fig.~\ref{Nuo7}, which confirms that,
for every fixed value of $p_1>0$, $\widehat V_{\rm g}(p)$ has an
absolute minimum for $p_2^2=p_1^3$. The minimum is degenerate:
$p_2=\pm p_1^{3/2}$, but both locations sit on the stratum
$S^{(1)}$. Thus, the substitution of the tree-level renormalizable
Higgs potential with the one-loop effective potential, cannot but
enforce the tree-level choice of the stratum $S^{(1)}$ as location
of the absolute minimum.

\subsubsection{The Coleman-Weinberg  (SO$_3\times \integer_2$, \underline{5}) gauge model}
If in the model just studied the symmetry group is extended to
SO$_3\times \integer_2$, where $\integer_2$ is the discrete group
generated by the transformation $\phi\rightarrow -\phi$, one gets
one of the models discussed in CW as examples. We shall continue
to use the notation introduced for the (SO$_3\,,\;{\underline
5}$)--model.

The extension of the symmetry leads to the following modifications
of the results obtained for the original (SO$_3$,\; ${\underline
5}$)--model.

There are still two basic homogeneous invariant polynomials, but
${\rm Tr}\,\Phi^3(\phi)$ is not invariant under reflections of
$\phi$, so the choice of the basic polynomial invariants has to be
modified. A possibility is the following:

\begin{equation}p_1(\phi)={\rm Tr}\,\Phi^2(\phi),\qquad p_2(\phi)=
6\,\left({\rm Tr}\,\Phi^3(\phi)\right)^2.\label{p1}\end{equation}
The matrix $\widehat P(p)$, built from the gradients of the basic
invariants $p_1(\phi)$ and $p_2(\phi)$ just defined, has the
following form:

\eqll{\widehat P(p)=\left(\begin{array}{cc} 4\,p_1& 12\,p_2 \\
12\,{p}_2& 36\,{p}_1^2\,{p}_2 \end{array}\right) }{matP1}
and the conditions that guarantee its semi-positivity and define the
range $p(\real^5)$ in $\real^2$ are the  following (see
Fig.~\ref{Nuo8}):

\begin{equation} {p}_1^3 \ge p_2 \ge 0.\end{equation}

Also in this case it is easy to identify four primary strata
$W^{(i)}$, defined by the following relations:

\eq{\begin{array}{l}
W^{(0)}:\ p_1=0=p_2;\qquad W^{(1)}_1:\ {p}_1 > 0 = {p}_2 - {p}_1^3;\\
W^{(1)}_2:\ {p}_1 > 0 = {p}_2 ; \qquad W^{(2)}:\ 0 < {p}_2 <
{p}_1^3.\end{array}}

\FIGURE[ht]{\epsfig{file=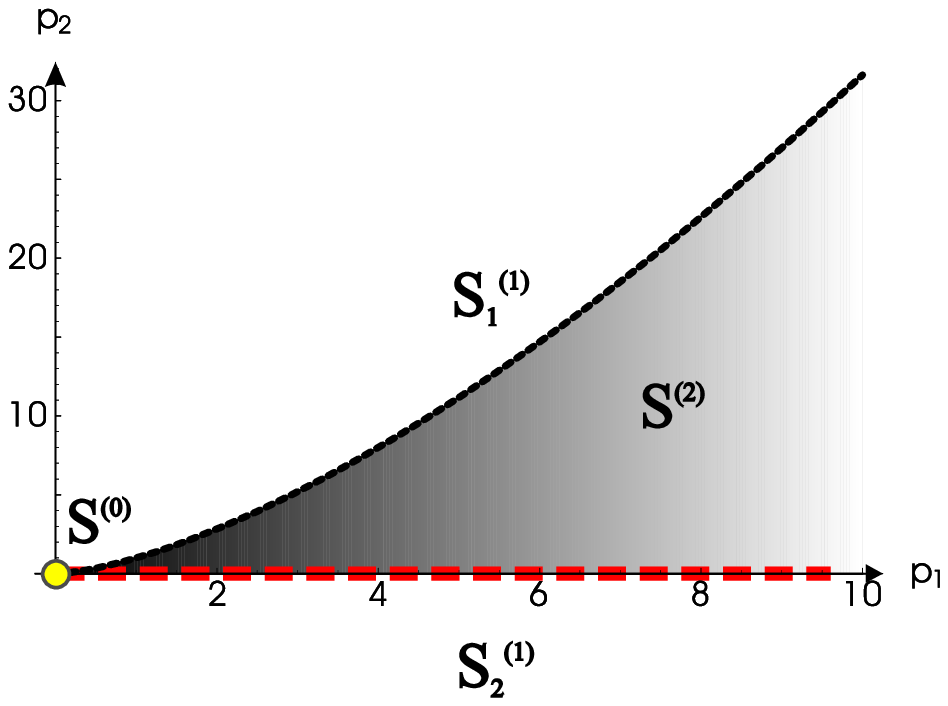,width=6.9cm}\caption{Orbit space
and symmetry strata of the (SO$_3\times \integer_2$, ${\underline
5})$ gauge model. }\label{Nuo8}}

 By selecting a convenient point ($\phi_1 = \phi_2 = \phi_3 = 0$)
in an arbitrarily chosen orbit in each of the primary strata, and
determining the corresponding isotropy subgroup, one easily finds
that the primary strata coincide with symmetry strata.
 Their residual symmetries are: [SO$_3\times\integer_2$],
[SO$_2$], $[\integer _2]$ and $[\uno]$, respectively for
$S^{(0)}$, $S^{(1)}_1$, $S^{(1)}_2$ and $S^{(2)}$. Let us describe the details of the calculation
only in the case of the new phase associated to the stratum
$S^{(1)}_2$.

For $\phi_1 = \phi_2 = \phi_3 = 0$, the condition $p_2(\phi)=0$
reduces to $\phi_5(3\phi_4^2 - \phi_5^2)=0$. It is sufficient to
take into consideration only one of the solutions of this
condition, since, for fixed values of $p_1(\phi) =\phi^2_4 +
\phi^2_5$, the other solutions lie on the same orbit and
variations of $p_1$ lead to orbits of the same stratum. The
solution $\phi_5=0$ corresponds to a matrix

\begin{equation}
\Phi_0 = \frac 1{\sqrt 2}\,\left(\begin{array}{ccc}
\phi_4 & 0       & 0\\
0      & -\phi_4 & 0\\
0      & 0       & 0
\end{array}\right),\label{matrixphi1}
\end{equation}
 which is invariant only under the transformations of the
``parity'' subgroup $\integer'_2$ of SO$_3\times \integer_2$
generated by the transformation resulting from a parity
transformation $\phi\rightarrow -\phi$, followed by an SO$_3$
transformation, generated by the matrix

\begin{equation}
\gamma = \left(\begin{array}{ccc}
0 &  -1 & 0\\
-1&  0  & 0\\
0 &  0  & -1
\end{array}\right)\,,
\end{equation}
that exchanges the two non trivial elements of $\Phi_0$.

In a renormalizable version of the model, the Higgs potential can
be written in the form of (\ref{V4}), (\ref{Vp}), with $a_2=0$ (a
choice that, in this case, is rigorously required by the symmetry
of the model), and $a_3>0$, to guarantee that the potential is
bounded from below.

In this model, the potential $\widehat V(p)$ is independent of
$p_2$. As a consequence, in the $p$-space, equipotential lines
reduce to straight-lines $p_1=c=$const. It is, therefore, clear
that, for $a_1 > 0$, $\widehat V(p)$ has a stable minimum at $p_1
= 0 = p_2$ ($c=0$), while for $a_1 < 0$ its minimum is degenerate
along the straight-line $p_1=-a_1/a_3$, which crosses all the
strata, but for $S^{(0)}$. As a consequence, the minimum is unstable and
 only the phase ${\mathcal F}^{(0)}$ is observable at tree-level.

If one gives up renormalizability and allows Higgs potentials of
arbitrarily high degree, then it would be easy to show, arguing as
in the preceding subsection, that at degree six also the phases
${\mathcal F}^{(1)}_1$ and ${\mathcal F}^{(1)}_2$ become tree-level
observable, while degree twelve (the degree of ${p}_2^2(\phi)$)
has to be reached in order to make sure that all the allowed
phases are tree-level observable.

The dominant one-loop radiative corrections $V_{\rm g}$ are the
same calculated in the case of the SO$_3$ model. As emphasized in
CW and recalled in the preceding subsection, for $a_1<0$ they
constrain the minimum of the effective potential in the stratum
$S^{(1)}_1$ (${p}_2={p}_1^3$), with symmetry [SO$_2$], and the
minimum is stable. Thus, at one-loop, the phase ${\mathcal
F}^{(1)}_2$ is observable, but it is clear that ${\mathcal F}^{(1)}_2$
and ${\mathcal F}^{(2)}$ remain unobservable.

The results we have found are in complete agreement with the
results in CW\footnote{The existence of the allowed phase with
symmetry $[\integer_2]$ had not been noticed by CW,
but a complete classification of all the phases allowed by the
symmetry was not relevant for their aims.}, but a comparison of
the SO$_3$ and the SO$_3\times \integer_2$ models makes it
difficult for us to share the enthusiasm manifested by CW for the
fact that in the SO$_3\times\integer_2$ model ``there is nothing
in the symmetry properties of this theory that guarantees that the
minimum of $V$ will obey Eq.(6,18)\footnote{In CW, Equation (6.18)
corresponds to the conditions which determine a residual symmetry
[SO$_2$].}. Thus, ..., if a massless photon emerges, it will be as
a consequence of detailed dynamics, not just of trivial group
theory.'' In fact, in the (SO$_3$, ${\underline 5}$) model, which
has the same gauge group, the emergence of a massless photon can
be stated already at tree-level: it is a consequence of trivial
group theory. In the SO$_3\times\integer_2$ model, the exceeding
degeneracy of the absolute minimum of the Higgs potential, which
prevents the choice of the true vacuum at tree-level, is an
artifact due to the combined effects of the additional discrete
symmetry $\integer_2$ and the limit imposed by renormalizability
on the degree of the Higgs potential. The introduction of the
additional discrete reflection symmetry, justified in CW ``to
simplify the problem'', does, indeed, strongly modify the symmetry
of the model and its allowed and observable phases: a new allowed
phase is generated and the allowed phase with symmetry [SO$_2$] is
made unobservable at tree-level. Discrete symmetries play
important roles, not only from the phenomenological point of view
and in the characterization of the allowed phases, but also in the
selection of the observable ones.

\subsection{An (SU$_3$, \underline{8}) gauge model}
Let us consider a model with gauge group SU$_3$ and an
octet $\phi$ of real Higgs fields, transforming as a vector in the
space of the adjoint representation of the group. Like in the
models studied above, the linear group $G=$(SU$_3$, ${\underline
8}$) admits only two basic homogeneous invariant polynomials, that
can be conveniently chosen to be the following:

\begin{equation} \label{YT7}
p_1(\phi)=\sum_{i=1}^8\,\phi_i^2,\qquad p_2(\phi)= \sqrt
3\,\sum_{i,j,k=1}^8\,d_{ijk}\phi_i\,\phi_j\,\phi_k,
\end{equation}

\noindent where the $d_{ijk}$ is the usual completely symmetric
Gell-Mann \cite{GM} tensor.
The Higgs potential can be written, in terms of the basic
invariants $p=(p_1,p_2)$ defined in (\ref{YT7}) as in (\ref{V4}),
(\ref{Vp}). The $\widehat P(p)$-matrix and, therefore, the orbit
space, turns out to be isomorphic to the orbit space of the linear
group (SO$_3$, ${\underline 5}$).

Therefore, the geometric aspects of the minimization problem, both
in the re\-nor\-ma\-li\-za\-ble version of the model and in a
non\--re\-nor\-ma\-li\-za\-ble one, with a Higgs potential of
degree six, are exactly the same solved in the case of the
(SO$_3$, ${\underline 5})$ model.

Only the symmetries of the four primary strata have to
be recalculated in the model we are discussing. As well known they
are [SU$_3$] (stratum $S^{(0)}=W^{(0)}$), [SU$_2 \times \mathrm{U}_1$]
(stratum $S^{(1)}=W^{(1)}_1\cup W^{(1)}_2$), [$\mathrm{U}_1\times\mathrm{U}_1$]
(stratum $S^{(2)}=W^{(2)}$), respectively. Also in this case there are
three allowed phases and only ${\mathcal F}^{(0)}$ and ${\mathcal
F}^{(1)}$ turn out to be tree-level observable in a renormalizable
version of the model, while all the allowed phases are tree level
observable if a non-renormalizable Higgs potential of degree six
is allowed.

\DOUBLEFIGURE{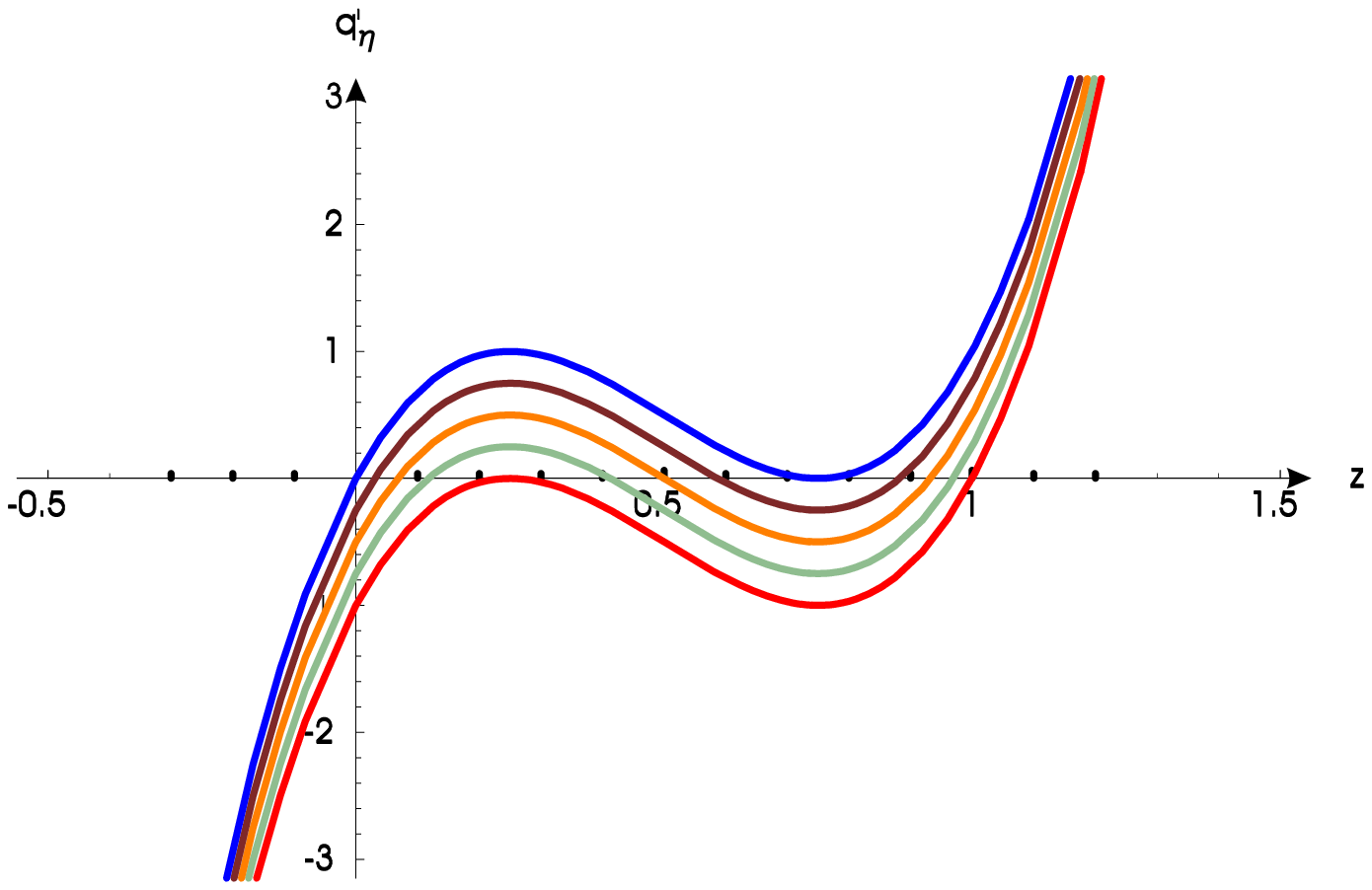,width=6cm}{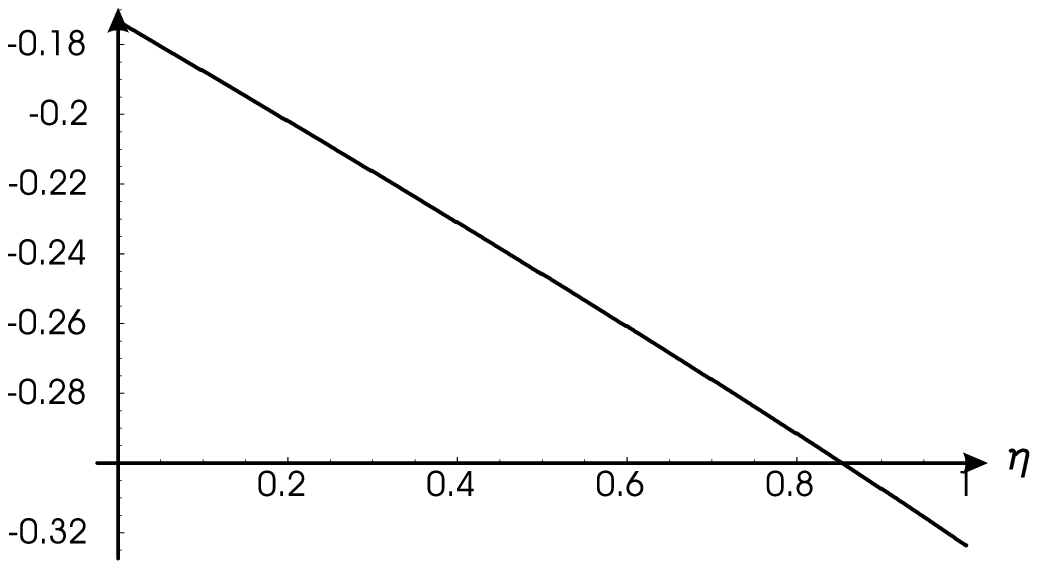,width=6cm}{\label{Nuo9}
Family of functions $q'_\eta(z)$, in the interval $[0,1]$, for
 $\eta=0,\; 1/4,\; 1/2,\;
3/4,\; 1$. }{\label{Nuo10} Graph of the function
$f(\eta)=\sum_{i=1}^3 z_i^{2}(\eta) \mathrm{ln}
\left(z_i(\eta)\right)$, where $z_i(\eta)$ are the roots of the
polynomial $q'_\eta(z)$ .}

 As for the one-loop radiative corrections due to the vector
bosons, the explicit form of $V_{\rm g}$ is given in (\ref{mass}),
where

\eq{M_{ab}^2(\phi) = g^2\,\langle F_a\phi,F_b\phi\rangle,\qquad
F_{a,ij}=f_{aij},\qquad a,i,j=1,\dots ,8} and the $f_{aij}$ are
the usual SU$_3$ completely antisymmetric structure constants.

The
squared mass matrix $M^2$ has three non zero distinct eigenvalues
that, also in this case, are algebraic $G$-invariant functions of
$\phi$.

The same procedure followed in the case of the SO$_3$ model, allows
to express them in the form $p_1\,z_i$, where the $z_i$'s are the roots
of the following polynomial in $z$:

\eq{q'_\eta(z) = (-1 + z)(-1 + 4\,z)^2 + \eta\,, \quad \eta =  \dfrac{p_2^2}{p_1^3}. \label{e32}}

The range of the adimensional variable $\eta$ is the interval
$[0,1]$. The polynomial functions $q'_\eta(z)$ are plotted, for
different values of $\eta$, in Fig.~\ref{Nuo9} and $V_{\rm g}$ as
a function of $\eta$ is plotted in Fig.~\ref{Nuo10}, for fixed
values of $p_1$. The contribution $V_{\rm g}$ to the one-loop
radiative corrections has evidently an absolute minimum for $\eta
= 1$. We conclude, therefore, that, like in the case of the SO$_3$
model, one-loop radiative corrections are not sufficient to make
observable the phase ${\mathcal F}^{(2)}$ (not observable at
tree-level).

\section{The geometrical invariant theory approach to spontaneous symmetry breaking}

The approach followed in the previous section to determine all the
allowed phases in a gauge model can be formulated on an absolutely
general and rigorous ground \cite{AS1, AS2, AS3}. Let us briefly recall the
basic elements.

Let $\phi$ denote the set of real scalar fields of the model to be
thought of as a vector $\phi\in \real^n$ ({\em vector order parameter\/}),
transforming according to a real orthogonal
representation\footnote{This is not a restrictive assumption,
since the internal symmetry group is a compact group.} of the
gauge group: $\phi\rightarrow g\cdot\phi$. We shall denote by $G$ the
group of real orthogonal $n\times n$ matrices $g$.

The Higgs potential $V^{(d)}_a(\phi)$ is a $G$-invariant real
polynomial function of $\phi$ with real coefficients $a_i$ ({\em
control parameters}) and degree $d$. The observable phases of the
system are determined by the location of the points of stable
global minimum of $V^{(d)}_a(\phi)$. Owing to $G$-invariance, the
Higgs potential is a constant along each $G$-orbit, so, each of
its stationary points is degenerate along a whole $G$-orbit.
Minima lying on the same $G$-orbit define equivalent vacua. Since
the isotropy subgroups $G_\phi$ of $G$ at points $\phi$ of the
same $G$-orbit are conjugate in $G$ ($G_{g\phi}=g\,G_\phi
\,g^{-1}$), only the conjugacy class $[G_\phi]=\{g\,G_\phi\,
g^{-1}\mid\,g\in G\}$ formed by the isotropy subgroups of $G$ at
the points of the orbit of minima, {\em i.e.}\ the {\em symmetry}
or {\em orbit-type} of the orbit hosting the absolute minimum, is
physically relevant, and defines the {\em symmetry} of the
associated stable phase.

The set of all $G$-orbits, endowed with the quotient
topology\footnote{ $G$-orbits are compact manifolds and the
distance between two orbits is defined as the distance between the
underlying manifolds.} and differentiable structure, forms the
{\em orbit space}, $\real^n/G$, of $G$. The subset of all the
points lying in $G$-orbits of the same orbit-type forms a {\em
symmetry stratum} of $\real^n$ and the image in the orbit space of
a {\em symmetry stratum} of $\real^n$ forms a {\em stratum} of
$\real^n/G$. Phase transitions take place when, by varying the
values of the control parameters, the absolute minimum of
$V^{(d)}_a(\phi)$ is shifted to an orbit lying on a different
stratum.

If $V^{(d)}_a(\phi)$ is a polynomial in $\phi$ of sufficiently
high degree, by varying the control parameters, its absolute
minimum can be shifted to any stratum of $\real^n/G$. So, {\em the
strata are in a one-to-one correspondence with the allowed
phases}. On the contrary, extra restrictions on the form of the
Higgs potential, not coming from G-symmetry requirements (e.g.,
the assumption that it is a polynomial of degree $\le 4$ in the
Higgs fields), can prevent
its global minimum from sitting, as a stable (against
perturbations of the control parameters) minimum, in particular
strata and make, consequently, the corresponding allowed phases
dynamically unattainable at tree-level.

Being constant along each $G$-orbit, the Higgs potential can be
conveniently thought of as a function defined in the orbit space
$\real^n/G$ of $G$. This fact can be formalized using some basic
results of invariant theory. In fact, every $G$-invariant
polynomial function $F(\phi)$ can be built as a real polynomial
function $\widehat F(p)$ of a {\em finite} set, $\{p_1(\phi),
\dots ,p_q(\phi)\}$, of basic {\em homogeneous} polynomial
invariants ({\em minimal integrity basis of the ring $\real\left[\real^n\right]^{G}$ of
$G$-invariant polynomials}, hereafter abbreviated in MIB) \cite{Hilbert}:

\eqll{F(\phi)=\widehat F(p(\phi)),\quad \phi\in\real^n}{F} and the
range $p(\real^n)$ of the {\em orbit map}, $\phi \mapsto p(\phi) =
(p_1(\phi), \dots ,p_q(\phi))$ yields a diffeomorphic
realization of the orbit space of $G$, as a connected
semi-algebraic set in $\real^q$, {\em i.e.}, as a subset of
$\real^q$, determined by algebraic equations and inequalities.
Thus, the elements of an integrity basis can be conveniently used
to parametrize the points of $p(\real^n)$ that, hereafter, will be
identified with the orbit space $\real^n/G$.

The elements of a minimal integrity basis need not, for general
compact groups, be algebraically independent. If they are not so,
the linear group $G$ is said to be {\em non-coregular} and the
algebraic relations among the elements of its MIB's are called
{\em syzygies}. The number $q_0$ of algebraically independent
elements in a MIB  is $n - {\rm Dim} (\Omega_{\rm p})$, where
${\rm Dim}(\Omega_{\rm p})$ is the dimension shared by all the
generic ({\em principal}) orbits\footnote{The dimension of an
orbit equals the dimension of $G$ minus the common dimension of
the isotropy groups at points of the orbit.} $\Omega_{\rm p}$ of
$G$. The linear groups studied in the preceding section are all
coregular. Examples of non coregular groups will be met in the
second part of the paper. \label{defnoncor}

The orbit space of $G$ presents a natural geometric {\em
stratification}, like all semi-algebraic sets. It can, in fact, be
considered as the disjoint union of a {\em finite number} of
connected semi-algebraic subsets of decreasing dimensions ({\em
primary strata}), each primary stratum being a connected manifold
open in its topological closure and lying in the boundary of a
higher dimensional one (but for the highest dimensional stratum,
which is unique and called {\em principal stratum}). The primary
strata are the connected components of the {\em symmetry strata}.
All the connected components of a symmetry stratum have the same
dimension and the symmetries of two bordering symmetry strata are
related by a group--subgroup relation, the orbit-type of the lower
dimensional stratum being larger: {\em more peripheric strata have
larger symmetries}.

If the only $G$-invariant point of $\real^n$ is the origin, there
are no linear invariants and in $\real^n/G$ there is only one
0-dimensional stratum, corresponding to the origin of $\real^q$.
All the other strata have at least dimension 1, since the isotropy
subgroups of $G$ at the points $\phi\in\real^n$ and $\lambda\,
\phi$, $\lambda\in\real$, are equal and, therefore, the points
$(p_1,\dots ,p_q)$ and $(\lambda^{d_1}p_1,\dots
,\lambda^{d_q}p_q)$ ($d_i$ the degree of the basic invariant
$p_i(\phi)$) sit on the same stratum. This fact, added to the
homogeneity of the basic invariants and of the relations defining
the strata, shows also that a complete information on the
structure of the orbit space and its stratification can be
obtained from its intersection with the image in $\real^n/G$ of
the unit sphere of $\real^n$.

The semialgebraic set $p(\real^n)$, yielding an image of the orbit
space of $G$ in the $p$-space, and its stratification has been
shown to be determined by the points $p\in \real^q$, satisfying
the following conditions \cite{AS1, AS2, AS3}:

\begin{theo} Let $G$ be a compact linear group acting in
$\real^n$, $\{p_1(\phi),\dots ,p_q(\phi)\}$ a MIB of $\real \left[\real^n\right]^{G}$
and $Z\subseteq \real^q$ the algebraic variety of the relations
among the $p_i(\phi)$'s.  Then, $p(\real^n)$ is the unique connected
semi-algebraic subset of $Z$ where the matrix $\widehat P(p)$,
defined by the  following relations, is positive semi-definite:

\begin{equation}
\widehat P_{ab}(p(\phi)) = \sum_{j=1}^n\partial_j
p_a(\phi)\,\partial_j p_b(\phi),\qquad \phi\in \real^n\,,\quad a=b=1,\,\ldots,\,q.
\end{equation}
The $k$-dimensional primary strata of $p(\real^n)$ are the
connected components of the set $\widehat{W}^{(k)}=\{p\in\real^q \mid
\widehat P(p)\ge 0,\ \rank\widehat P(p)=k\}$; they are the images of the
connected components of the $k$-dimensional isotropy type strata
of $\real^n/G$.  In particular, the set of the interior points of
$p(\real^n)$ is the image of the principal stratum.
\end{theo}

In the following, the primary and symmetry strata will always be
denoted by $W^{(i)}_j$ and $S^{(i)}_k$ respectively, where the
apex $i$ gives the dimension of the stratum and $j$ or $k$ are
order numbers, which will be omitted if not necessary.

There is always at least a $G$-invariant polynomial of degree two,
that, in this section, we shall denote by $p_1$:

\eq{p_1(\phi)=\sum_{i=1}^n\,\phi_i^2. \label{quadr}} With this
convention, $\partial p_1(\phi)=2\phi_i$, so that the first row
and column of the matrix $\widehat P(p)$ are completely determined
to be $P_{1i}(p)=P_{i1}(p)=2d_i\,p_i$ by Euler equation, owing to
the homogeneity of the polynomials $p_i(\phi)$. Moreover, the
image in orbit space of a sphere of $\real^n$, centered in the
origin, is the intersection of the orbit space with the linear
variety of equation $p_1=$ const. This intersection is necessarily
a compact subset, so any continuous function of $p$ certainly has
an absolute minimum for a fixed value of $p_1$.

By defining, according to \eref{F},

\begin{equation}
\widehat V^{(d)}_a(p(\phi))= V^{(d)}_a(\phi),\qquad \phi\in \real
^n,\label{1}
\end{equation}
the range of $V^{(d)}_a(\phi)$ coincides with the range of the
restriction of $\widehat V^{(d)}_a(p)$ to the the orbit space
$p(\real^n)$ and the local minima of $V^{(d)}_a(\phi)$ can be
computed as the local minima of the function $\widehat
V^{(d)}_a(p)$ with domain $p(\real^n)$.

In detail, denoting by $f_\alpha(p)=0$, $\alpha=1,\dots k$ a
complete set of independent equations of the stratum $S$, the
conditions for the occurrence of a stationary point of the potential at $p\in S$,
can be conveniently written in the following form:

\eqll{\left\{\begin{array}{l}
f_\alpha(p) = 0,\qquad \alpha=1,\dots ,k,\\
 \dfrac {\partial}{\partial p_i} \left(\widehat
V^{(d)}_a(p)-\sum_{\alpha=1}^k\,\lambda_\alpha\,
f_\alpha(p)\right)=0, \qquad i= 1,\dots
,q,\end{array}\right.}{ext} where the $\lambda_\alpha$'s are real
Lagrange multipliers. A stationary point at $\overline{p}$ will be
a stable local minimum on the stratum if the Hessian matrix
$M^2_{\rm s}(\phi)$ of $V^{(d)}_a(\phi)$ is $\ge 0$ and has rank
$n$ minus the dimension $\nu$ of the orbit ($\nu$ equals the
number of Goldstone bosons), for any $\phi$ lying on the orbit of
equation $p(\phi)=\overline{p}$. These conditions can be
conveniently expressed in terms of the sums $K_i$ of the
(determinants of) the principal minors of $M^2_{\rm s}(\phi)$ in
the form $K_i>0$, $i=1,\dots ,n-\nu$. Being $M^2_{\rm s}(\phi)$ a
$G$-tensor of rank 2, the $K_i$'s are $G$-invariant polynomials in
the $\phi_i$'s and can, therefore be expressed as polynomials in
the elements of the MIB. As shown in \cite{AS3, AS2},  the squared
mass matrix of the scalars is reducible in the singular strata, so
the above conditions on its semi-positivity and rank are
equivalent to the following simpler conditions (for any $\phi$ on
the $G$-orbit of equation $p(\phi)=\overline{p}$:

\begin{enumerate}
\item the $n\times n$ matrix $\partial^2(V^{(d)}_a(\phi)-
\sum_{\alpha=1}^k\,\lambda_\alpha\,
f_\alpha(p(\phi)))/\partial \phi_i\partial \phi_j$ is $\ge 0$ and
its rank equals the dimensions $q-k$ of $S$;
\item  the $n\times n$ matrix $\partial^2(\sum_{\alpha=1}^k\,\lambda_\alpha\,
f_\alpha(p(\phi)))/\partial \phi_i\partial \phi_j$ is $\ge 0$ and its
rank equals the dimension $n-(q-k)-\nu$ of
the orthogonal space in $\real^n$ to the stratum at $\phi)$).
\end{enumerate}

Also these conditions can obviously  be expressed in terms of the
principal minors of the matrices, {\em i.e.\/} as polynomial
inequalities in the basic invariants. The requirement for
$V_a^{(d)}(\phi)$ being bounded from below is equivalent to the
condition that the constrained minimum of $V_a^{(d)}(\phi)$, in
the intersection of the orbit space with the hyperplane of
equation $p_1=$ const ($p_1$ is defined in (\ref{quadr})), is
positive\footnote{We impose the condition in a strong sense, {\em
i.e.\/} we require that $V_a^{(d)}(\phi)$ is a bounding
potential.}.

\section{Allowed and observable phases in some two-Higgs doublet extensions
of the Standard Model} The possibility of generating the observed
Baryon Asymmetry of our Universe (BAU) during the Electro-Weak
Phase Transition (EWPT) has been extensively studied since the
middle of the eighties by several authors (see for instance
\cite{Dine1, Dine2} and references therein). It is nowadays well
established that the Standard Model is not suited to account for
BAU, both because the amount of CP violation in the quark sector
is too tiny and because the experimental lower bounds on the Higgs
mass cause the phase transition not being enough strongly first
order to prevent the baryon excess generated at the EWPT from
being subsequently washed out by sphaleron effects.  In the
2HD models there is a natural additional source of
CP violation: the phase between the two VEV's of the Higgs fields.
Notwithstanding,  as was pointed out in \cite{Dine2}, since the
baryon production ceases at very small values of the Higgs fields,
models with only two Higgs doublets can hardly generate the right
amount of BAU, because at the EWPT they behave  as  models with
one light Higgs doublet, with the second heavier Higgs decoupling
and having small impact on the phase transition. More promising
has been the introduction of gauge singlet scalars which couple to
the Higgs. In particular, a model with a second Higgs doublet and
a complex gauge singlet has been analyzed in connection with
baryogenesis and the dark matter problem in \cite{MCDON}.

In this section, we shall characterize all the allowed and
tree-level observable phases and all possible phase transitions
between contiguous phases, for variants of a 2HD
extension of the SM. In particular, for each model we shall
determine a minimal set of basic polynomial invariants of $G$, the
geometrical features of the orbit space, {\em i.e.\/} its
stratification (including connectivity properties and bordering
relations of the strata) and the orbit-types of its strata. Since
our analysis will be stricly bounded to tree-level Higgs
potentials, {\em all our statements will be intended as tree-level
statements}, even when not explicitly claimed.

The conclusion will be that, if discrete
symmetries are added, the renormalizable versions of the models
are incomplete. In the following section we shall show that
renormalizability and completeness can be reconciled if these
models are extended by adding convenient scalar singlets.

\subsection{Model 0: The two-Higgs extension of the Standard Model}
In this subsection we analyze the basic two-Higgs extension of the
Standard Model. The symmetry group of the Lagrangian is SU$_2 \times \mathrm{U}_1$ and
there are two complex Higgs doublets $\Phi_1$ and $\Phi_2$ of
hypercharge $Y=1$:

\eq{\Phi_1=\left(\begin{array}{c} \phi_1 + i\,\phi_2\\
\phi_3 + i\,\phi_4\end{array}\right)\,,\qquad
\Phi_2=\left(\begin{array}{c} \phi_5 + i\,\phi_6\\
\phi_7 + i\,\phi_8\end{array}\right)\,,\qquad \phi_i\in\real.}

In this model, natural flavor conservation is violated by neutral
current effects in the phase (hereafter called ${\mathcal
F}^{(3)}$), that should correspond to the present phase of our
Universe. So the model is not realistic, but it provides a simple
example in which renormalizability does not exclude completeness.

The transformation induced by the element
$\hat{j} = (-\uno_2,\mathrm{e}^{\mathrm{i} \pi Y}) \in$
SU$_2\times \mathrm{U}_1$ leaves invariant the fields $\Phi_1$ and
$\Phi_2$. So, the linear group acting on the vector $\phi=
(\phi_1,\dots ,\phi_8)$ of the real Higgs fields of the model
is $G= \left((\mathrm{SU}_2\times \mathrm{U}(1))/\integer_2,\
{\underline 8}\right)$, where $\integer_2$ is the group
generated by $\hat{j}$.

A convenient choice for a MIB of real independent polynomial
$G$-invariants is the following:

\eqll{p_1=\Phi_1^\dagger\Phi_1 + \Phi_2^\dagger\Phi_2,\ \ p_2 =
\Phi_1^\dagger\Phi_1 - \Phi_2^\dagger\Phi_2,\ \ p_3 + i p_4 =
2\,\Phi_2^\dagger\Phi_1.}{IB1}

The relations defining $p(\real^4)$ and its strata, which are
listed in Table~\ref{tab1}, can be obtained from rank and
positivity conditions of the $\widehat P(p)$-matrix associated to
the MIB defined in Eq.~\eref{IB1}:

\eq{\widehat P(p)=\left(\begin{array}{cccc}
4\,p_1 & 4\,p_2 & 4\,p_3 & 4\,p_4 \\
4\, p_2     & 4\,p_1 & 0      & 0      \\
4\, p_3     & 0      & 4\,p_1 & 0      \\
4\, p_4      & 0      & 0      & 4\,p_1 \end{array}\right)\,.}

\TABLE{
\begin{tabular}{llccl}
\hline
Stratum   & Defining relations             & Symmetry   & Boundary &  Typical point $\phi$      \\
\hline
&&&&\\
$S^{(4)}$ & $p_1>\sqrt{p_2^2+p_3^2+p_4^2}$ & $\{\uno \}$            & ${\overline S}^{(3)}$ & $(\phi_1,0,\phi_3,\phi_4,0,0,\phi_7,0)$\\
&&&&\\
$S^{(3)}$ & $p_1=\sqrt{p_2^2+p_3^2+p_4^2}$ & U$_1^{\mathrm e.m.}$ & $S^{(0)}$ &  $(0,0,\phi_3,\phi_4,0,0,\phi_7,0)$\\
&&&&\\
$S^{(0)}$ & $p_1= p_2 = p_3 = p_4 = 0$     & $G$ & & $(0,0,0,0,0,0,0,0)$\\
&&&&\\
\hline
\end{tabular}

\caption{ Strata of the orbit space for the symmetry group $G=
(\mathrm{SU}_2\times \mathrm{U}_1)/\integer_2$ of Model 0. A
 bar denotes topological closure.
The group U$_1^{\mathrm e.m.}$ is formed by the elements $ \left
\{ \mathrm{e}^{ \mathrm{i} \theta (T_3+Y/2)} \right \}_{0 \leq
\theta < 2 \pi}$. For each stratum, a field configuration with the
same symmetry is supplied {\em (typical point)}. The $\phi_i$'s are generic non zero values. } \label{tab1} }

The orbit space is the half-cone bounded by the surface of
equation $p_1=\sqrt{\sum_{i=2}^4 p_i^2}$.
There are, evidently, only three primary strata of dimensions 0, 3 and 4.
They are connected sets and have, consequently, to be identified to the three
distinct symmetry strata:
the tip of the cone
corresponds to the stratum $S^{(0)}$, and the rest of the surface
to the stratum $S^{(3)}$, while the interior points form
$S^{(4)}$. There are no one-dimensional and two-dimensional
strata.

A general fourth-degree polynomial invariant Higgs
potential can be written in the following form:

\begin{equation}
\begin{array}{rcl}
\widehat V(p)&=& \sum_{i,j=1}^4 \,A_{ij}\,p_i\, p_j +
\sum_{i=1}^4\,\alpha_i\, p_i \\
&&\\
&=&  \sum_{i,j=1}^4 \,A_{ij}\,(p_i-\eta_i)(p_j-\eta_j)-
 \sum_{i,j=1}^4 \,A_{ij}\,\eta_i\,\eta_j,
\end{array}\label{V41}
\end{equation}
where, to make sure that $\widehat V(p)$ is bounded from below, we
assume that all the coefficients are real and the symmetric matrix
$A$ is positive definite\footnote{These are only  sufficient conditions.
Explicit necessary and sufficient conditions can easily be
determined, but they would be too cumbersome to write down and
would add nothing to our analysis. \label{notapi}}.
Moreover: $\eta_i= - \frac{1}{2} \sum_{j=1}^4
(A^{-1})_{ij}\,\alpha_j$.

In this simple case (convex orbit space), the constrained minima of $\widehat V(p)$ can be easily
determined from elementary geometrical considerations. To this
end, let us first choose $A\propto \uno$ and let us denote by
$C=C^+\cup C^-$  the closed double cone bounded by the
surfaces of equation $p_1=\pm\sqrt{\sum_{i=2}^4 p_i^2}$ in the $p$-space $\real^4$.
Then, since the potential is a constant plus the squared distance of
$p$ from $\eta$, for given values of the $\eta_i$'s, the potential
has a stable absolute minimum at the point $p$ of the orbit space
which is closest to $\eta$. One is left, therefore, with the
following possibilities:

\begin{itemize}
\item[i)] the minimum is stable  in $S^{(4)}$, at
$p=\eta$, for $\eta$ in the interior of $C^+$;
\item[ii)] the minimum is stable in $S^{(0)}$ ($p=0$) for $\eta$
in the interior of $C^-$;
\item[iii)]  the minimum is stable in $S^{(3)}$, at the nearest point
to $\eta$, for $\eta$ outside $C$;
\item[iii)]  the minimum is unstable in $S^{(3)}\cup S^{(0)}$,
at $p=\eta$, for $\eta$ on the surface of the double cone.
\end{itemize}

For a general fixed $A>0$, the results do not essentially change,
since one can revert to the case $A= \uno$ by means of a
convenient linear transformation of the $p_i$'s, which defines a
new (equivalent) MIB: as a result, $C^+$ and $C^-$ are simply
rotated and deformed by independent re-scalings along the
coordinate axes. So, in the space of the parameters
$(\alpha_1,\dots ,\alpha_4)$, that are independent linear
combinations of the $\eta_i$'s, there are three disjoint open
regions of stability of the three allowed phases associated to the
strata of the orbit space. These regions are separated by
inter-phase boundaries, formed by critical points where second
order phase transitions may start; moreover, first order phase
transitions cannot take place.

We can conclude that the model just discussed is both {\em
renormalizable} and (tree-level) {\em complete}.

\subsection{Model 1: A FCNC protecting version of Model 0}

The model we shall analyze in this subsection contains the same
set of fields as Model 1, but a discrete symmetry, generated by a
reflection $\hat \iota$ is added, to protect the theory from FCNC
processes (see, for instance \cite{1,2} and references therein).
So, the symmetry group of the Lagrangian is assumed to be
SU$_2\times\mathrm{U}_1\times \{\hat \iota\}$, where $\{\hat
\iota\}$ denotes the $\integer_2$ group generated by $\hat \iota$,
which is assumed to act on the Higgs fields in the following way:
$(\Phi_1,\Phi_2) \rightarrow (\Phi_1,-\Phi_2)$.

The phase transitions of Model 1 have been analyzed in \cite{Zar,
Zar1}. The possibility of two--stage phase transitions was
proposed in \cite{Zar} to reconcile the smallness of the
CP-breaking term explicitly introduced at tree-level and the
necessary amount of CP violation required to successfully account
for baryogenesis. The author asserts that ``Investigating the
whole parameter space would be too time consuming'', so he
performs only a numerical analysis of the nature of the phase
transition driven by the third degree finite temperature
corrections to the classical potential. In a more recent paper
\cite{Zar1}, the full contribution of the extra breaking terms
(not considering them as perturbations) is also examined. The
result is still a two--stage phase transition, but, in addition to
the CP violation, the phase transition to the charge conserving
vacuum generates some charge asymmetry in the presence of heavy
leptons, which is compared with the astrophysical bounds.

The orbit space approach enables us to study such problems analytically and in
full generality. Moreover, in an extended renormalizable and complete
version of Model 1, Model 1C, that we shall study in a subsequent subsection,
we shall check the possibility of spontaneous CP violation \cite{5}.

\FIGURE{\epsfig{file=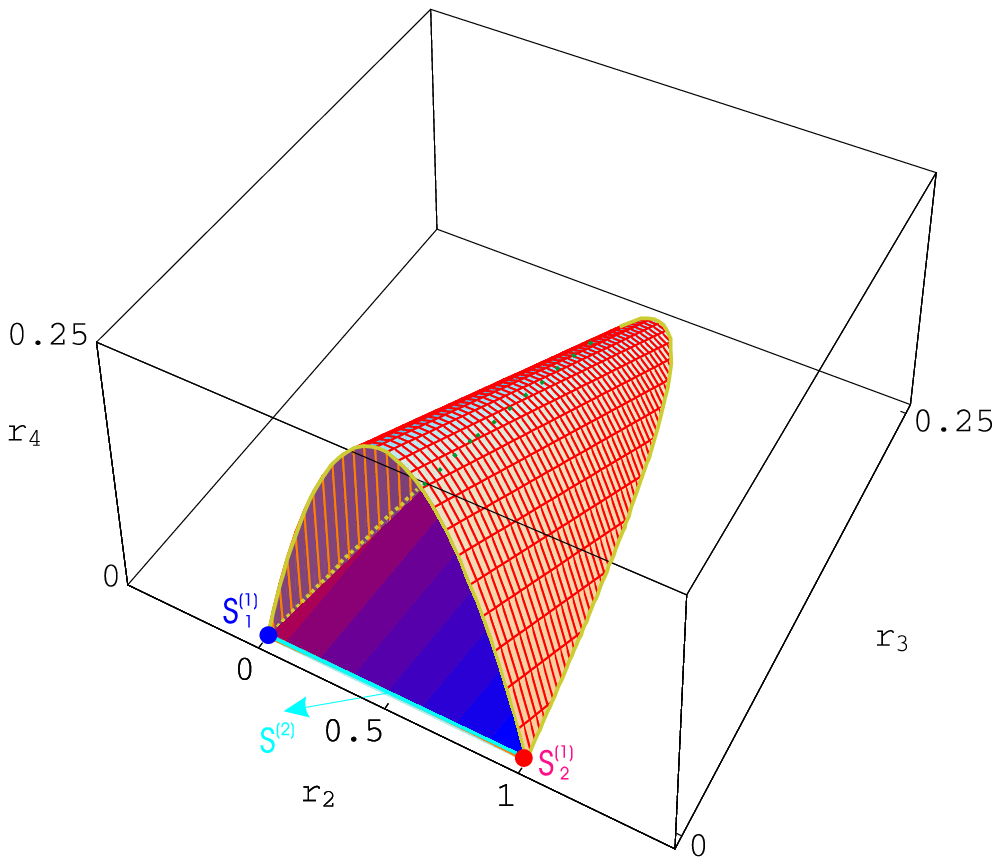,width=9cm}\caption{Representation
in $\real^3$ of the section $\Xi$ of one of the two layers
$p_5=\pm \sqrt{p_3 p_4}$ of the four dimensional orbit space of
Model 1 with the hyperplane of equation $p_1 + p_2=1$. In this
representation the primary stratification cannot be deduced
directly from the figure. The blue and red points represent the
one dimensional strata $S^{(1)}_1$ and $S^{(1)}_2$, respectively.
They are connected by a pale blue line representing the stratum
$S^{(2)}$. $\Xi$ looks like a cave: the union of the entrance, the
floor and the interior form the principal stratum $S^{(4)}$, while
the union of the ceiling and its green border represents the
stratum $S^{(3)}$. The relations defining the strata can be found
in Table~\ref{tab2}.
 }\label{F1DD}}


\TABLE{{\small

\begin{tabular}{llcll}
\hline
 ${S}$   & Defining relations   &  Symmetry & Boundary & {\em Typical point}  $\phi$ \\
 \hline
 &&&&\\
 $ {S}^{(4)}$ & $  {p}_5^2- {p}_3  {p}_4=0 $ & $\{ \uno \}$ &
 $\overline{ {S}^{(3)}}\,,\; \overline{ {S}^{(2)}}$ &
 $(\phi_1,0,\phi_3,\phi_4,0,0,\phi_7,0)$ \\
& $  {p}_1+ {p}_2 >0\,,\;$ &&&\\
& $ {p}_3+ {p}_4 >0\,,\;$ && & \\
& $ {p}_1  {p}_2
- {p}_3 - {p}_4>0$ && & \\
$ {S}^{(3)}$ &$  {p}_5^2- {p}_3  {p}_4=0 \,,$ &U$_1^{\rm e.m.}$&
$\overline{ {S}^{(1)}_1}\,,\; \overline{ {S}^{(1)}_2}$ &
 $(0,0,\phi_3,\phi_4,0,0,\phi_7,0)$ \\
 & $ {p}_1  {p}_2 - {p}_3 - {p}_4=0 \;$ &&&\\
& $  {p}_1\,, {p}_2 > 0\,, $ &&&\\
&$ {p}_3 ,\,
 {p}_4 \geq 0$ & && \\
$ {S}^{(2)}$ &$  {p}_j=0 \,,\;j\neq 1,2$ &$ \{ \alpha \hat{i}\}$&
 $\overline{ {S}^{(1)}_1}\,,\; \overline{ {S}^{(1)}_2}$ &
 $(\phi_1,0,0,0,0,0,\phi_7,0)$ \\
 &$ {p}_1 \,, {p}_2 >0 \;$&&&\\
$ {S}^{(1)}_1 $& $  {p}_j=0<  {p}_1 \,,\;j\neq 1$ & U$_1^{\rm
e.m.} \times \{\hat \iota\}$&
 $ {S}^{(0)}$ &
 $(0,0,\phi_3,0,0,0,0,0)$ \\
$ {S}^{(1)}_2 $& $ {p}_j=0< {p}_2\,,\;\;j\neq 2 \;$ & U$_1^{\rm
e.m.}
\times \{e^{i \pi Y}\,\hat \iota\}$ &  $ {S}^{(0)}$ &  $(0,0,0,0,0,0,\phi_7,0)$ \\
 $ {S}^{(0)}$   & $ {p}_i=0\,,$ $1 \leq i \leq 5$ &
 $G$ &    & $(0,0,0,0,0,0,0,0)$ \\
&&&&\\
\hline
\end{tabular}
} \caption{ Strata $S$ of the orbit space for the symmetry group
$G= (\mathrm{SU}_2\times$U$_1)/\integer_2 \times \{ \hat \iota \}$
of Model 1. A bar denotes topological closure. The group U$_1^{\rm
e.m.}$ is formed by the elements $ \left \{e^{i\theta (T_3+Y/2)}
\right \}_{0 \leq \theta < 2 \pi}$. For each stratum, a field
configuration with the same symmetry is supplied {\em (typical
point)}. The $\phi_i$'s are generic non zero values. Confining
strata are indicated, so that possible second order phase
transitions can  be easily identified. } \label{tab2}}


In Model 1, the linear group acting on the vector $\phi$ of real Higgs fields
is $G=\left((\mathrm{SU}_2\times \mathrm{U}(1))/\integer_2
\,\times \{\hat \iota\},\ \underline 8\right)$ and a MIB is the following:

\begin{equation}
\begin{array}{l}
{p}_1=\Phi_1^\dagger\Phi_1,\;\; {p}_2=\Phi_2^\dagger\Phi_2,\;\;
{p}_3=\left(\mathrm{Re}\left[\Phi_2^\dagger\Phi_1\right]\right)^2,\;\;\\
\\
{p}_4=\left(\mathrm{Im}\left[\Phi_2^\dagger\Phi_1\right]\right)^2,\;\;
{p}_5=\mathrm{Re}\left[\Phi_2^\dagger\Phi_1\right]\;\mathrm{Im}
\left[\Phi_2^\dagger\Phi_1\right]. \end{array} \label{IB1D}
\end{equation}

The elements of the MIB are related by a syzygy
$s=p_5^2-p_3\,p_4$, so the orbit space is the four dimensional
algebraic variety of equation $s=0$ in the 5--dimensional
$p$-space.
The defining relations of the strata of $ {p}(\real^8)$,
summarized in Table~\ref{tab2},
 are obtained from
positivity and rank conditions of the matrix $\widehat P( {p})$,
associated to the MIB of Eq.~\eref{IB1D}:

\eq{\widehat P(  p)= \left(\begin{array}{ccccc}
4  {p}_1 & 0             & 4  {p}_3                            & 4  {p}_4                            & 4  {p}_5                           \\
0             & 4  {p}_2 & 4  {p}_3                            & 4  {p}_4                            & 4  {p}_5                           \\
4  {p}_3 & 4  {p}_3 & 4  {p}_3\,( {p}_1+ {p}_2) & 0                                        & 2  {p}_5\,( {p}_1+ {p}_2)\\
4  {p}_4 & 4  {p}_4 & 0                                        & 4  {p}_4\,( {p}_1+ {p}_2) & 2 {p}_5\,( {p}_1+ {p}_2)  \\
4  {p}_5 & 4  {p}_5 & 2 {p}_5\,( {p}_1+ {p}_2) & 2 {p}_5\,( {p}_1+
{p}_2) &( {p}_1+ {p}_2)\,( {p}_3+ {p}_4)
\end{array}\right)}

The orbit space $  p(\real^8)\subset\real^5$ is formed by the
union of two layers of equations $p_5=\pm \sqrt{p_3p_4}$. The
intersections with the hyperplane of equation $ {p}_1+ {p}_2=1$
are isomorphic and can be represented, in a three dimensional
space, as the closed solid (semialgebraic set) $\Xi$ shown in
Figure~\ref{F1DD}. The full orbit space is the four dimensional
connected semi-algebraic set of $\real^5$ formed by the points $
{p}= ( {p}_1, {p}_2, {p}_3, {p}_4,  p_5)= \Pi^{-1} (r)$, $ r \in
\Xi$, where $\Pi^{-1}$ is the inverse projection defined as
follows:

\begin{equation} \label{proinvA}
\begin{array}{lccl}
\Pi^{-1}: & \real^3 \supset \Xi & \longrightarrow & \real^5\\
& (r_2,r_3,r_4) & \mapsto &
 \left( \lambda (1-r_2),\lambda r_2,\lambda^2 r_3,\lambda^2 r_4,\lambda^2 \sqrt{r_3r_4} \right)
 \cup \\
 &&&
 \hspace{2em}\left( \lambda (1-r_2),\lambda r_2,\lambda^2 r_3,\lambda^2 r_4,-\lambda^2 \sqrt{r_3r_4} \right)\,,
 \hspace{2em}\lambda \geq 0\, .
\end{array}
\end{equation}

For the ease of the reader, a scheme of the stratification  is
shown in Fig.~\ref{FF1D}, page~\pageref{FF1D}.

We shall consider two different dynamical versions of Model 1, an
incomplete re\-nor\-ma\-li\-za\-ble and a complete
non-re\-nor\-ma\-li\-za\-ble one, that we shall denominate Model
$1_1$ and $1_2$, respectively.

\subsubsection{Model $1_1$: An incomplete renormalizable version of Model 1}

A general four degree $G$-invariant polynomial can be written in the form

\begin{equation}
\begin{array}{l}
V^{(4)}(\phi) = \widehat V^{(4)}(p(\phi)),\\
\\
\widehat V^{(4)}(p) = \sum_{i,j=1}^2\,A_{i\,j}\,p_i\,p_j +
\sum_{i=1}^{5} \alpha_i p_i\,,
\end{array}
\end{equation}
where all the coefficients are real, $A_{1\,2}=A_{2\,1}$ and the
following set of conditions (necessary and sufficient) has to be
imposed to make sure that $V^{(4)}(\phi)$ diverges to $+\infty$
for $\|\phi\|\to \infty$:
\begin{eqnarray}
A_{11},A_{22},A_{12} + \sqrt{A_{11}\,A_{22}} >0, \quad
\alpha_3,\alpha_4 > -2\left(A_{12}+\sqrt{A_{11}\,A_{22}}\right),\nonumber\\
\label{ew9}\\
 \alpha_5^2 < 4\left[\alpha_3 +
2\left(A_{12}+\sqrt{A_{11}\,A_{22}}\right)\right]\times\left[\alpha_4
+ 2\left(A_{12}+\sqrt{A_{11}\,A_{22}}\right)\right]\,. \nonumber
\end{eqnarray}

The principal stratum $S^{(4)}$ is open in the orbit space, so
possible minima of $V(p)$ located in $S^{(4)}$ are necessarily
stationary points, which  are determined by the solutions of the
following set of equations, where $\lambda$ is a real Lagrange
multiplier:

\eq{\begin{array}{l}
\sum_{j=1}^2\,A_{i\,j}p_j + \alpha_i = 0,\qquad i=1,2\,,\\
\\
\alpha_3 +\lambda\,p_4 = 0\,,\\
\\
\alpha_4 +\lambda\,p_3 = 0\,,\\
\\
\alpha_5 -2\lambda\,p_5 = 0\,,\\
\\
p_5^2 - p_3\,p_4  = 0.
\end{array}}
Since there are solutions only for $\alpha_5^2 =
\alpha_3\,\alpha_4$, possible minima in the principal stratum
cannot be stable. So {\em the model is renormalizable but
incomplete}.

The results of a more complete analysis are summarized in
Table~\ref{tab3} and show that all the phases associated to the
other strata are, instead, observable.

%

\subsubsection{Model $1_2$: A complete non-renormalizable version of Model 1}
\FIGURE{\epsfig{file=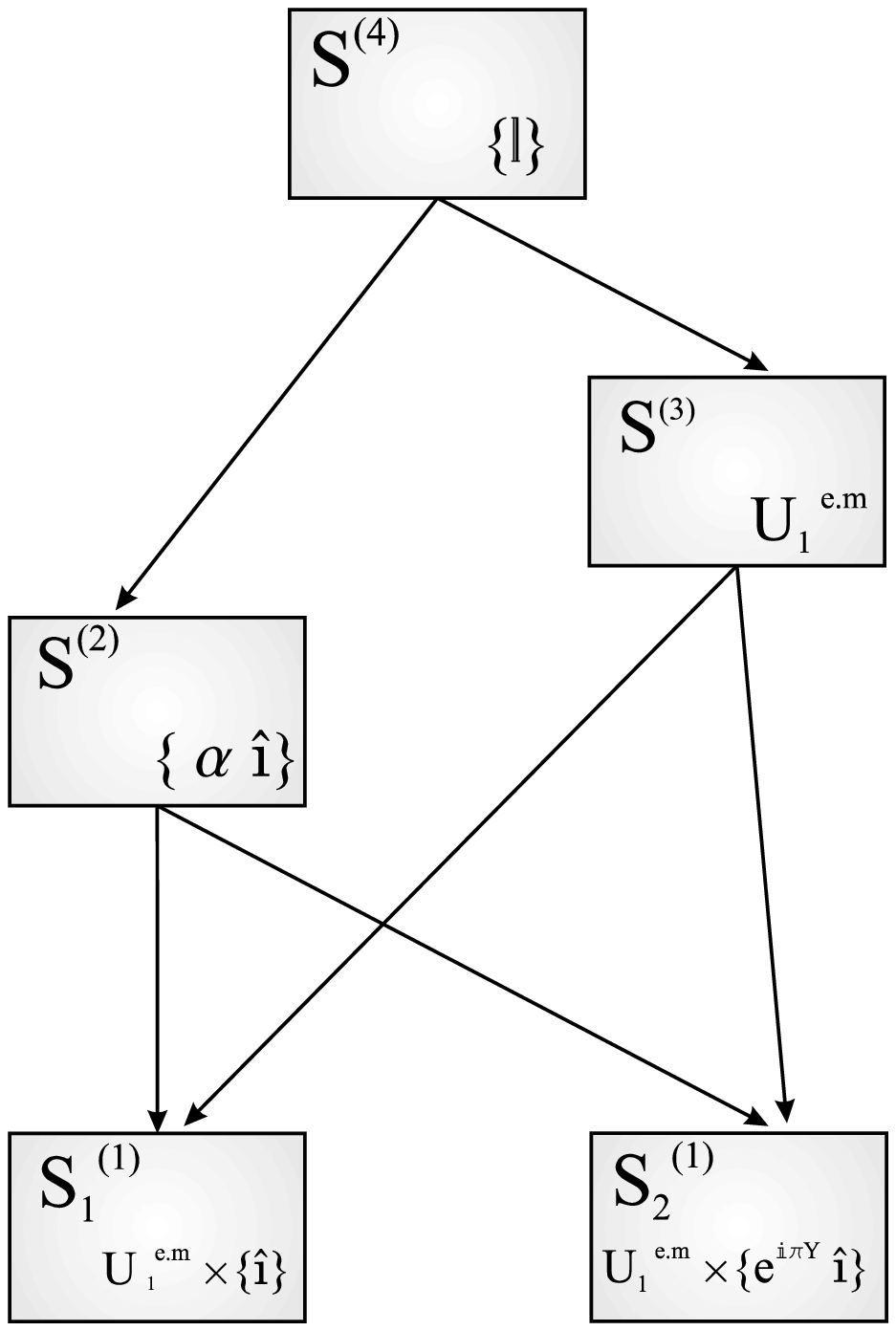,width=5cm}\caption{ Stratification
of the orbit space of model 1.
  Arrows connect higher dimensional (inner) strata
 to bordering lower dimensional ones. The stratum $S^{(0)}$ is not shown for simplicity. It would be connected
 by arrows issuing from $S^{(1)}_1$ and $S^{(1)}_2$  }\label{FF1D}}

Making use of geometrical arguments similar to those used in the
analysis of Model 0, we shall now prove that all the allowed
phases of the model become observable, if the requirement of
renormalizability is ignored and the dynamics of the Higgs sector
is determined by a $G$-invariant polynomial potential of degree
eight.

To this end, it will be sufficient to find a particular eight
degree $G$-invariant polynomial $\widehat V_0^{(8)}(p)$, which,
for convenient values of its coefficients, admits a stable
absolute minimum in each of the strata of the orbit space of $G$.
Stability is intended with respect to perturbations of $\widehat
V_0^{(8)}(p)$, induced by arbitrary $G$-invariant polynomials of
degree not exceeding eight.

The following simple potential is already sufficient,
to make observable all the allowed phases:

\begin{equation}
\begin{array}{rcl}
V^{(8)}(\phi)&=&\widehat V^{(8)}(p(\phi)),\\
&&\\
\widehat V^{(8)}(p) &=& \sum_{i=1}^5\,(p_i-\eta_i)^2 -
\sum_{i=1}^5\,\eta_i^2\,,\quad p\in p(\real^8).\end{array}
\label{V8}
\end{equation}
\TABLE{{\small

\begin{tabular}{ll}
\hline\\
Stratum & Structural stability conditions\\
\hline\\
&\\
$ {S}^{(3)}$ & $0<\lambda<2$, $\rho_1 < 0$,
$-\dfrac{2\lambda\,\rho_1^2}{(2-\lambda)^2} < \rho_2 <
\dfrac{\rho_1^2}{4}$, $\rho_3>-2\lambda$, \\
& $-\lambda(\lambda + \rho_3) < \rho_4 < \dfrac{\rho_1^2}{4}$,
$\lambda = \left(-\rho_3 + \sqrt{\rho_3^2 -
4\rho_4 + \alpha_5^2}\right)/2$ \\
&\\
$ {S}^{(2)}$ & $\rho_1<0$, $\rho_2>0$, $\rho_3>0$, $\alpha_5^2 < \mathrm{Min}\left(2\rho_4,4+2\rho_3+\rho_4\right)$ \\
&\\
$ {S}_{1}^{(1)}$ & $\alpha_{1}<0$, $\alpha_{2}>0$,
$\rho_3>-\dfrac{8\alpha_2}{\alpha_1^2}$,
$-\dfrac{16\alpha_2\left(4\alpha_2 + \alpha_1^2\,\rho_3\right)}{4\alpha_1^2} < \rho_4 < \dfrac{\rho_3^2}{4}$,\\
& $\alpha_{5}^2 <
4\dfrac{16\alpha_{2}^{2}+4\alpha_{1}^2\,\alpha_{2}\rho_{3}+\alpha_{1}^4\,\rho_{4}}
{\alpha_{1}^{4}}$\\
&\\
$ {S}_{2}^{(1)}$ & $\alpha_{1}>0,$ $\alpha_{2}<0$,
$\rho_3>-\dfrac{8\alpha_1}{\alpha_2^2}$,
$-\dfrac{16\alpha_1\left(4\alpha_1 + \alpha_2^2\,\rho_3\right)}{4\alpha_2^2} < \rho_4 < \dfrac{\rho_3^2}{4}$,\\
&$\alpha_5^2 <
4\dfrac{16\alpha_1^2+4\alpha_2^2\,\alpha_1\rho_3+\alpha_2^4\,\rho_4}
{\alpha_2^4}$\\
&\\
$ {S}^{(0)}$ & $\alpha_{1}>0,$ $\alpha_{2}>0$, $\rho_3>-4$,
$\rho_4 < \dfrac{\rho_3^2}{4}$, $\alpha_5^2<4+2\rho_3+\rho_4$\\
&\\
\hline
\end{tabular}
} \caption{ Necessary and sufficient conditions for structural
stability of the allowed phases in Model $1_1$, for $A=\uno$. The
phase corresponding to the principal stratum is dynamically
unattainable. In the table, the results are expressed also in
terms of the following functions of the control parameters:
$\rho_1=\alpha_1 + \alpha_2$,  $\rho_2=\alpha_1 \, \alpha_2$,
$\rho_3=\alpha_3 + \alpha_4$, $\rho_4=\alpha_3\, \alpha_4$.}
\label{tab3}}


In fact, it is easy to realize that, for each given choice of
$\eta$, thought of as a point in the $p$-space, the absolute
minimum of the potential $\widehat V^{(8)}(p)$ is located at the
points $\bar p$ of the orbit space which is nearest to $\eta$. The
minimum at the point $\bar p$, sitting on the stratum $\bar S$, is
non degenerate, if $\eta$ is close enough to $\bar S$  in the
intersection of the normal spaces at $\bar p$ to the higher
dimensional strata bordering $\bar S$. The geometrical feature of
the orbit space, that guarantees the existence and uniqueness of a
point of the orbit space at minimum distance from $\eta$, under
the conditions just specified, is the absence of intruding cusps
(see Figure~\ref{F1DD}).

The above statements have been checked analytically. Constraints
on the values of the control parameters $\eta_i$ which are
sufficient to guarantee the location and stability of the absolute
minimum in the different strata are listed in Table \ref{tab4}.

\TABLE{

\begin{tabular}{ll}
\hline\\
Stratum & Structural stability conditions\\
\hline\\
$S^{(4)}$ & $ \mu_1>0,\;\; \mu_2 >0,\;\;0<\mu_3<\mu_2$,
$\dfrac{2}{9} {\mu_3}^2 < \mu_4 < \dfrac{{\mu_3}^2}{4}$, $0 <
{\eta_5}^2< \mu_4$ \\
&\\
$S^{(3)}$ & $ \mu_1>0\,,\;\; \mu_2 >0\,,\;\;0<\mu_3<\dfrac{3}{2}
\mu_2$,
$- \mu_2 \, ( \mu_2 - \mu_3) < \mu_4 < \dfrac{{\mu_3}^2}{4}$ \\
&\\
$S^{(2)}$ & $ \eta_1>0\,,\;\; \eta_2 >0\,,\;\;\eta_3<0\,,\,\;\eta_4<0$ \\
&\\
${S}_{1}^{(1)}$ & $ \eta_1>0\,,\; \eta_2 <0$,  $\eta_3+\eta_4<-\dfrac{2 \eta_2}{\eta_1}$,
$\eta_3 \, \eta_4 < -\dfrac{4 \eta_2 \left[ \eta_2 + \eta_1 (\eta_3 + \eta_4) \right]}{3 {\eta_1}^2}$ \\
&\\
${S}_{1}^{(2)}$ & $ \eta_1<0$,  $ \eta_2 >0$,  $\eta_3+\eta_4<-\dfrac{2 \eta_1}{{\eta_2}^2}$,
$\eta_3 \, \eta_4 < -\dfrac{4 \eta_1 \left[ \eta_1 + {\eta_2}^2 (\eta_3 + \eta_4) \right]}{3 {\eta_2}^4}$ \\
&\\
$S^{(0)}$ &$ \eta_1\,,\; \eta_2 <0$\\
\hline
\end{tabular}

\caption{Sufficient conditions for structural stability of the
phases $\mathcal{F}^{(i)}_j$ associated to strata $S^{(i)}_j$ of
Model $1_2$. In the table, the results are expressed also in terms
of the following functions of the control parameters:
$\mu_1=\eta_1 + \eta_2$, $\mu_2=\eta_1 \, \eta_2$, $\mu_3=\eta_3 +
\eta_4$ and $\mu_4=\eta_3 \, \eta_4$.} \label{tab4}}


\subsection{Model 2: Implementing a CP--like discrete symmetry in Model 1}
In the model studied in \cite{Zar, Zar1} the role of the discrete
symmetry is fundamental in achieving the possibility of a two
stage phase transition. The main advantage advocated by the
authors is an {\em amplification} of the CP-violating effects.
Since the experimental information on the Higgs sector are at
present very weak and not enough to fully determine the discrete
symmetries in two-Higgs-doublet models, the addition of some
discrete symmetry could allow some subtler amplification pattern.
Moreover, from a technical point of way, adding some discrete
symmetry allows to construct symmetry group representations with a
lower level of non-coregularity\footnote{For the definition of
non-coregular linear group $G$, see page \pageref{defnoncor}.},
which implies easier computations
in the framework of the orbit space approach. Therefore in this
subsection we shall consider a model with the same set of fields
as Model 1, but with symmetry group SU$_2\times\mathrm{U}_1\times
\{\hat \iota\,,K\}$, where $\hat \iota$ is the generator of a
reflection group and $K$ is the generator of a $CP$-like
transformation\footnote{Since the most general $CP$ transformation
on a complex field $\chi$ contains a field--dependent phase
\cite{SWe}, i.e. $\chi \longrightarrow e^{i\theta} \chi^\ast$, the
$CP$ conservation is usually checked {\em a posteriori}. Note also
that the last cross in SU$_2\times$U$_1\times \{\hat i,\,K\}$ does
not denote a direct product, since $K$ does not commute with the
generators of SU$_2\times$U$_1$. }:
 $(\Phi_1,\Phi_2) \rightarrow (\Phi_1,-\Phi_2)$ and
 $(\Phi_1,\Phi_2) \rightarrow (\Phi_1^*,\Phi_2^*)$,
respectively. The addition of a new discrete symmetry will increase the number
of allowed phases.
\TABLE{

\begin{tabular}{lcl}
\hline
 Stratum   &  Symmetry         &   Typical point $\phi$ \\
 \hline
 $  {S}^{(4)} $  &  $\{\uno\}$ & $(\phi_1,0,\phi_3,\phi_4,0,0,\phi_7,0)$\\
 $ {S}^{(3)}_1$   & $\{\hat\iota\,K\}$  & $(\phi_1,0,\phi_3,0,0,0,0,\phi_8)$ \\
 $  {S}^{(3)}_2$ &  $\{K\}$            & $(\phi_1,0,\phi_3,0,0,0,\phi_7,0)$   \\
 $  {S}^{(3)}_3$ & U$_1^{\mathrm e.m.}$  & $(0,0,\phi_3,\phi_4,0,0,\phi_7,0)$  \\
 $  {S}^{(2)}_1$ &  $\{\alpha\,\hat\iota ,K\}$      & $(\phi_1,0,0,0,0,0,\phi_7,0)$  \\
$  {S}^{(2)}_2$  & U$_1^{\mathrm e.m.}\times\,\{\hat\iota\,K\}$ & $(0,0,\phi_3,0,0,0,0,\phi_8)$ \\
 $  {S}^{(2)}_3$ &  U$_1^{\mathrm e.m.}\times\{K\}$     & $(0,0,\phi_3,0,0,0,\phi_7,0)$   \\
$  {S}^{(1)}_1$ &  U$_1^{\mathrm e.m.}\,\times\{\hat\iota, K\}$      & $(0,0,\phi_3,0,0,0,0,0)$ \\
$  {S}^{(1)}_2$ &  U$_1^{\mathrm e.m.}\times \{e^{i\pi Y}\,\hat\iota, K\}$ & $(0,0,0,0,0,0,\phi_7,0)$ \\
$ {S}^{(0)}$   &  $(\mathrm{SU}_2\times \mathrm{U}_1)/\integer_2\,\times \{\hat \iota\,,\;K\}$       & $(0,0,0,0,0,0,0,0)$ \\
\hline
\end{tabular}

\caption{ Symmetries of the strata $ {S}$ of Model 2. The group
$\mathrm{U}_1^{\mathrm{e.m.}}$ is defined as in Table~\ref{tab1},
and $\alpha = \mathrm{e}^{\mathrm{i}\pi(T_3-Y/2)}$.
 Symmetries are specified by a {\em representative element} of the conjugacy class of isotropy
subgroups. Finite groups are defined through their generators
between brackets. For each stratum, a field configuration with the
same symmetry is supplied ({\em typical point\/}). The $\phi_i$'s
are generic non zero values.   }\label{tab5} }

\TABLE{\small{

\begin{tabular}{llll}
\hline Stratum    & Defining relations & Symmetry & Boundary\\
 \hline
&&&\\
$ {S}^{(4)}$   & $ {p}_1,  {p}_3,  {p}_4,  {p}_1  {p}_2 -  {p}_3 -  {p}_4 > 0$ & $\{\uno\}$                                                     & $\overline{  {S}_i^{(3)}}, i=1,2,3$\\
$ {S}^{(3)}_1$ & $  {p}_3 = 0 <  {p}_4 <  {p}_1  {p}_2\,,\;p_1+p_2>0$   & $\{\hat\iota \,K\}$   & $\overline{  {S}_i^{(2)}}, i=1,2$  \\
$ {S}^{(3)}_2$ & $  {p}_4 = 0 <  {p}_3 <  {p}_1  {p}_2\,,\;p_1+p_2>0$   &  $\{K\}$                      & $\overline{  {S}_i^{(2)}}, i=1,3$  \\
$ {S}^{(3)}_3$ & $ {p}_1   {p}_2 -  {p}_3 -  {p}_4 = 0 <  {p}_1+p_2,\,  {p}_3,\,  {p}_4$& U$_1^{\rm e.m.}$      & $\overline{  {S}_i^{(2)}}, i=2,3$  \\
$ {S}^{(2)}_1$ & $  {p}_3 =  {p}_4 = 0 <  {p}_1,\,  {p}_2$          & $\{\alpha\,\hat\iota ,K\}$                     & $\overline{  {S}_i^{(1)}}, i=1,2$  \\
$ {S}^{(2)}_2$ & $ {p}_1  {p}_2 -  {p}_4 =  {p}_3 = 0 <  {p}_1+{p}_2,\,p_4$     & U$_1^{\rm e.m.}\times\{\hat \iota \,K\}$                  & $\overline{  {S}_i^{(1)}}, i=1,2$  \\
$ {S}^{(2)}_3$ & $ {p}_1  {p}_2 -  {p}_3 =  {p}_4 = 0 <  {p}_1+{p}_2,\,p_3$     & U$_1^{\rm e.m.}\times\{K\}$                               & $\overline{  {S}_i^{(1)}}, i=1,2$  \\
$ {S}^{(1)}_1$ & $  {p}_2 =  {p}_3 =  {p}_4 = 0 <  {p}_1$              & U$_1^{\rm e.m.}\times \{\hat\iota,K\}$  & $ {S}^{(0)}$                        \\
$ {S}^{(1)}_2$ & $ {p}_1= {p}_3 =  {p}_4 = 0 <  {p}_2 $                   & U$_1^{\rm e.m.}\times\{ {\rm e}^{i\pi Y}\hat\iota ,K\}$                    & $ {S}^{(0)}$                        \\
$ {S}^{(0)}$   & $ {p}_1 =  {p}_2 =  {p}_3 =  {p}_4 = 0$                    & (SU$_2\times$U$_1)/\integer_2\,\times\{\hat\iota, K\}$       &                         \\
&&&\\
\\ \hline
\end{tabular}

\caption{ Symmetries and defining relations of the strata $ S$ of
Model 2. The symmetries are specified by means of a
``representative'' element of the conjugacy class of isotropy
subgroups.  Finite groups are specified through their generators
written between brackets and $\alpha = e^{i\pi(T_3-Y/2)}$.
Continuous phase transitions are possible only between bordering
strata. The group U$_1^{\rm e.m.}$ is the subgroup of
SU$_2\times$U$_1$ formed by the elements $({\rm
diag}\{e^{i\theta},e^{-i\theta}\},e^{i\theta})$ and U$_1^{\rm
e.m.}(\pi)$ denotes its element $({\rm diag}\{-1,1\},-1)$.
}\label{tab6} }}


\FIGURE{\epsfig{file=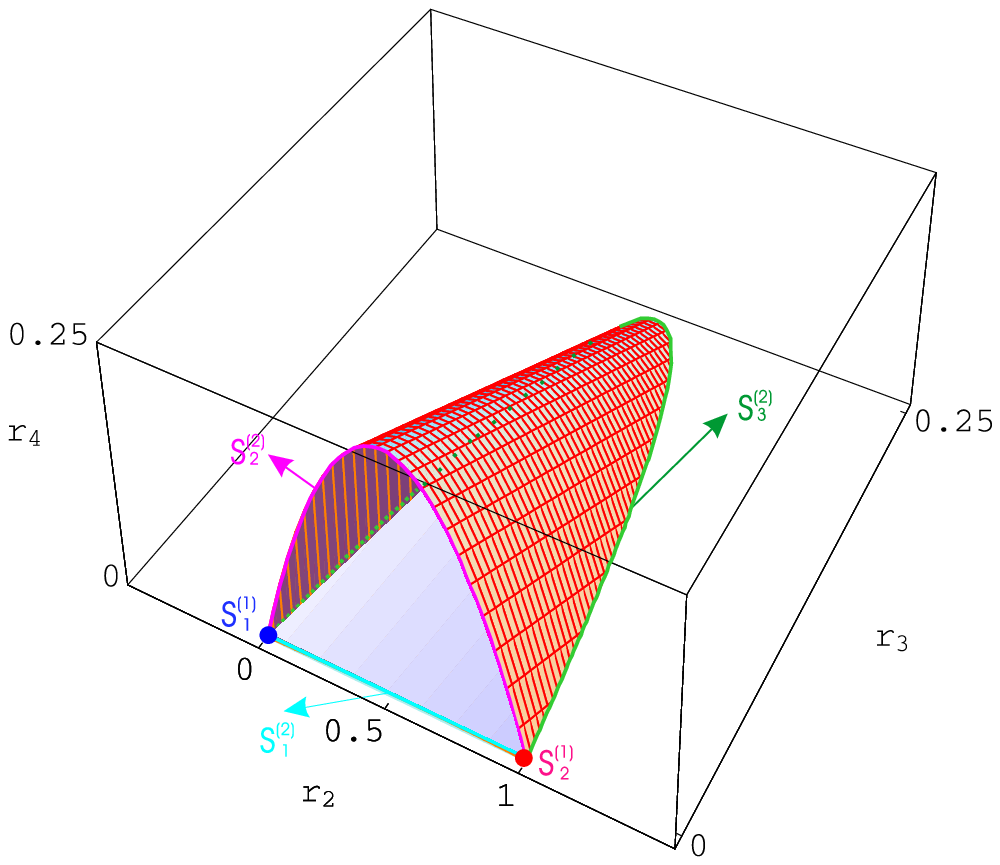,width=9cm}\caption{ Section $\Xi$
of the four dimensional orbit space of Model 2 with the hyperplane
of equation $p_1+p_2=1$. The one dimensional strata are
represented by the blue and red points. The pale blue, pink and
green lines represent the two dimensional strata $S^{(2)}_1$,
$S^{(2)}_2$ and $S^{(2)}_3$, respectively. $\Xi$ looks like a
cave, whose entrance, floor, ceiling and interior are formed by
the strata $S^{(3)}_1$, $S^{(3)}_2$, $S^{(3)}_3$ and $S^{(4)}$,
respectively. The relations defining the strata can be found in
Table~\ref{tab6}. }\label{F1}}

As in model 1, for our purposes it will
be equivalent, but simpler, to consider as a symmetry group of the
model $G=$ (SU$_2\times\mathrm{U}_1)/\integer_2\,\times \{\hat \iota\,,\, K\}$.
Under these assumptions, a MIB is the following:

\begin{equation}
\begin{array}{l}
 {p}_1=\Phi_1^\dagger\Phi_1,\;\;
 {p}_2=\Phi_2^\dagger\Phi_2,\;\;\\
\\
 {p}_3=\left(\mathrm{Re}\left[\Phi_2^\dagger\Phi_1\right]\right)^2,\;\;
 {p}_4=\left(\mathrm{Im}\left[\Phi_2^\dagger\Phi_1\right]\right)^2.\label{IB2}
\end{array}
\end{equation}
The defining relations of the strata of $ {p}(\real^8)$  can be
obtained from positivity and rank conditions of the symmetric
matrix $\widehat P( {p})$, associated to the MIB defined in
\eref{IB2}:

\eq{\widehat P(  p)= 4 \left( \begin{array}{cccc}
 {p}_1 &  0 &  {p}_3 &  {p}_4 \\
 0 &  {p}_2 &  {p}_3 &  {p}_4 \\
 {p}_3 &  {p}_3 & \,p_3( {p}_1+ {p}_2) & 0 \\
 {p}_4 &  {p}_4 & 0                           &  {p}_4\,( {p}_1+ {p}_2)
\end{array}\right).}
The results are shown in Tables~\ref{tab5} and \ref{tab6}.

The section $\Xi$ of $ {p}(\real^8) \subset \real^4$ with the
hyperplane of equation $ {p}_1+ {p}_2=1$ is the three dimensional
closed solid (semialgebraic set) shown in Figure~\ref{F1}. The
full orbit space is the four dimensional connected semi-algebraic
set formed by the points $ {p}= ( {p}_1, {p}_2, {p}_3, {p}_4)=
\Pi^{-1} (r)$, $ r \in \Xi$, where $\Pi^{-1}$ is the inverse
projection defined as follows:
\begin{equation} \label{proinvB}
\begin{array}{lccl}
\Pi^{-1}: & \real^3 \supset \Xi & \longrightarrow & \real^4\\
  & (r_2,r_3,r_4) & \mapsto &
 \left( \lambda (1-r_2),\lambda r_2,\lambda^2 r_3,\lambda^2 r_4 \right)\,,\hspace{2em}\lambda \geq 0\,.
\end{array}
\end{equation}

For the ease of the reader a scheme of the stratification of the
model is shown in Fig.~\ref{FFm2}.

We shall consider two different dynamical versions of Model 2, a
complete non-re\-nor\-ma\-li\-za\-ble and an incomplete
re\-nor\-ma\-li\-za\-ble one.
\FIGURE[ht]{\epsfig{file=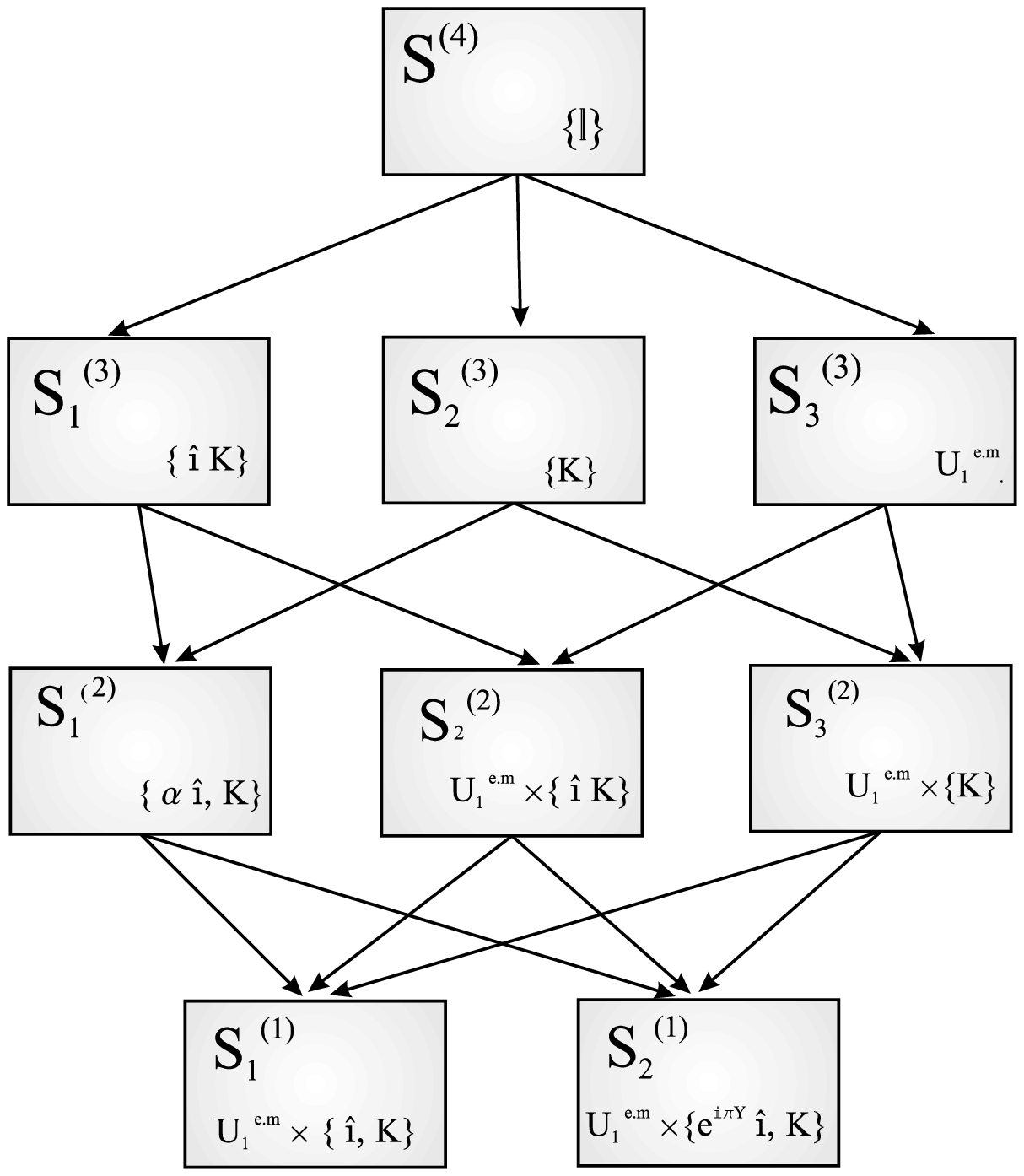,width=9cm}\caption{
Stratification of the orbit space of model 2.
  Arrows connect higher dimensional (inner) strata
 to bordering lower dimensional ones. The stratum $S^{(0)}$ is not shown for simplicity. It would be connected
 by arrows issuing from $S^{(1)}_1$ and $S^{(1)}_2$ }\label{FFm2}}

The $CP$-like transformation we have defined allows an easy
verification of $CP$ conservation. For example, with reference to
Tables~\ref{tab5} and \ref{tab6}, it is evident that $CP$
is broken in the stratum ${  S}^{(3)}_3$, while $CP$ is conserved
in ${ S}^{(2)}_2$: the transformations induced by $\hat{\iota}$
determine the right $CP$-phase $\theta$ for
each field of the theory.

\subsubsection{Model $2_1$: An incomplete renormalizable version of Model 2}
Let us now, in the frame of symmetries of Model 2, chose the Higgs
potential as the most general, bounded below invariant polynomial
of {\em degree four:}

\begin{equation}
\widehat V^{(4)}( {p})= \sum_{i,j=1}^2 \,A_{ij}\, {p}_i\, {p}_j +
\sum_{i=1}^4 \,\alpha_i\, {p}_i,\label{V3}
\end{equation}
where all the parameters are real, $A_{12}=A_{21}$ and the inequalities
in the first line
of Eq.~(\ref{ew9}) make sure that $V^{(4)}(
p(\phi))$ diverges to $+\infty$ for $\|\phi\|\to \infty$.

As stated in \cite{AS3, AS2}, since there are no relations among
the elements of the MIB and the potential is linear in the basic
invariants of degree four, its local minima can only sit on the
boundary of the orbit space, for general values of the
$\alpha_i$'s. The result of a detailed calculation is shown in
Table~\ref{tab7}: we have listed the values of the $\alpha_i$'s
that guarantee the location of a stable absolute minimum of
$\widehat V (p)$ in the different strata, for $A=\uno$. In fact,
there can be stationary points of the potential in the strata of
dimension $\ge 3$ only if the $\alpha_i$'s satisfy particular
conditions: $\alpha_3=\alpha_4=0$, $\alpha_4=0$, $\alpha_3=0$ and
$\alpha_3=\alpha_4$, respectively, for the strata $ {S}^{(4)}$, $
{S}^{(3)}_1$, $ {S}^{(3)}_2$ and $ {S}^{(3)}_3$.

\TABLE{{\small

\begin{tabular}{ll}
\hline\\
Stratum & Structural stability conditions\\
\hline\\
$ {S}^{(2)}_{1}$ & $\alpha_1<0,$ $\alpha_2<0,$  $\alpha_3>0,$  $\alpha_4>0$\\
&\\
$ {S}^{(2)}_{2}$ & $-2< \alpha_4 <0,$ $\alpha_3>\alpha_4,$
$\left(\alpha_1<0,\;\; \alpha_2<\dfrac{\alpha_1 \alpha_4}{2}\right)$ or   \\
& $\left(\alpha_1>0,\;\; \alpha_2>\dfrac{2 \alpha_1 }{\alpha_4}\right)$  \\
&\\
$ {S}^{(2)}_{3}$ & $-2< \alpha_3 <0,$ $\alpha_4>\alpha_3,$
$\left(\alpha_1<0,\;\; \alpha_2<\dfrac{\alpha_1 \alpha_3}{2}\right)$ or   \\
& $\left(\alpha_1>0,\;\; \alpha_2>\dfrac{2 \alpha_1 }{\alpha_3}\right)$  \\
&\\
$ {S}_{1}^{(1)}$ & $\alpha_{1}<0,$ $\alpha_{2}>0,$ $\alpha_{3}%
>\mathrm{Max}\left(2\dfrac{\alpha_{2}}{\alpha_{1}},-2 \right)$, $\alpha_{4}>\mathrm{Max}\left(2\dfrac{\alpha_{2}}{\alpha_{1}},-2\right)%
$ \\
&\\
$ {S}_{1}^{(2)}$ & $\alpha_{1}>0,$ $\alpha_{2}<0,$ $\alpha_{3}%
>\mathrm{Max}\left(2\dfrac{\alpha_{1}}{\alpha_{2}}, -2 \right)$,  $\alpha_{4}>\mathrm{Max}\left(2\dfrac{\alpha_{1}}{\alpha_{2}}, -2 \right)$
\\%
$ {S}^{(0)}$ & $\alpha_{1}>0,$ $\alpha_{2}>0,$ $\alpha_{3}>-2,$ $\alpha_{4}>-2$ \\
&\\ \hline
\end{tabular}

\caption{Necessary and sufficient conditions for structural
stability of the phases ${\mathcal F}^{(i)}_j$ associated to the
strata ${\mathcal S}^{(i)}_j$ of
 2$_1$, for $A=\uno$. The strata not appearing in
the table are physically unattainable. }\label{tab7}}}


These conditions reduce to zero the measures of the regions of
stability of the corresponding phases, in the space of the
parameters $\alpha_i$. So, there will not be stable phases
associated to the strata of dimension $\ge 3$. As a consequence,
it is impossible to generate spontaneous $CP$ violation in the
model. The general problem of spontaneous $CP$ breaking in
two-Higgs doublet models will be faced in a forthcoming paper
\cite{5}.

We can conclude that Model $2_1$ is renormalizable, but it is
{\em incomplete}.

\subsubsection{Model $2_2$: A complete non-renormalizable version of Model 2}

\TABLE{

\begin{tabular}{ll}
\hline\\
Stratum & Structural stability conditions\\
\hline\\
$S^{(4)}$ &  $\eta_1>0\,,\;\;\eta_2>0\,,\;\; 0 < \eta_3 < \eta_1 \eta_2\,,\;\;0< \eta_4 < \eta_1 \eta_2 - \eta_3$ \\
&\\
$S^{(3)}_1$ & $\eta_1>0\,,\;\;\eta_2>0\,,\;\;  \eta_3 < 0\,,\;\;0< \eta_4 < \eta_1 \eta_2 $\\
&\\
${S}^{(3)}_2$ & $\eta_1>0\,,\;\;\eta_2>0\,,\;\; 0 < \eta_3 < \eta_1 \eta_2\,,\;\; \eta_4 <0$\\
&\\
 $ {S}^{(3)}_3$ & $0 < \lambda <2\,,\;\; \mu_1>0\,,\;\; -\dfrac{2 \lambda}{(\lambda-2)^2} \mu_1^2 < \mu_2 < \dfrac{\mu_1^2}{4}$ \\
& $ \dfrac{\lambda}{2} < \eta_4 < \left( \dfrac{1}{2}+\dfrac{8\mu_1^2}{(\lambda^2-4)^2} \right) \lambda + \dfrac{4 \mu_2}{(\lambda+2)^2}$\\
& $ \eta_3= - \eta_4 + \dfrac{\lambda\left((\lambda^2-4)^2+8 \mu_1^2 \right) +4 \mu_2(\lambda-2)^2}{(\lambda^2-4)^2}$\\
 &\\
$ {S}^{(2)}_{1}$ & $\eta_1>0\,,\;\;\eta_2>0\,,\;\;  \eta_3 < 0\,,\;\; \eta_4 <0$\\
&\\
$ {S}^{(2)}_{2}$ & $0 < \lambda <2\,,\;\; \mu_1>0\,,\;\; -\dfrac{2 \lambda}{(\lambda-2)^2} \mu_1^2 < \mu_2 < \dfrac{\mu_1^2}{4}\,,\;\; \eta_3< \dfrac{\lambda}{2}$ \\
& $ \eta_4=  \dfrac{\lambda\left((\lambda^2-4)^2+16 \mu_1^2 \right) +8 \mu_2(\lambda-2)^2}{2(\lambda^2-4)^2}$\\
&\\
$ {S}^{(2)}_{3}$ &  $0 < \lambda <2\,,\;\; \mu_1>0\,,\;\; -\dfrac{2 \lambda}{(\lambda-2)^2} \mu_1^2 < \mu_2 < \dfrac{\mu_1^2}{4}\,,\;\; \eta_4< \dfrac{\lambda}{2}$ \\
& $ \eta_3=  \dfrac{\lambda\left((\lambda^2-4)^2+16 \mu_1^2 \right) +8 \mu_2(\lambda-2)^2}{2(\lambda^2-4)^2}$\\
  &\\
$ {S}_{1}^{(1)}$ & $\eta_1>0\,,\;\;\eta_2<0\,,\;\;  \eta_3 < -\dfrac{\eta_2}{\eta_1}\,,\;\; \eta_4 <-\dfrac{\eta_2}{\eta_1}$ \\
&\\
$ {S}_{1}^{(2)}$ & $\eta_1<0\,,\;\;\eta_2>0\,,\;\;  \eta_3 < -\dfrac{\eta_1}{\eta_2}\,,\;\; \eta_4 <-\dfrac{\eta_1}{\eta_2}$  \\
&\\
 ${S}^{(0)}$ & $\eta_1<0\,,\;\;\eta_2<0$ \\
&\\ \hline
\end{tabular}

\caption{Sufficient conditions for structural stability of the
phases ${\mathcal F}^{(i)}_j$ associated to the strata ${\mathcal
S}^{(i)}_j$ of
 2$_2$. In the table the results are
expressed also in terms of the following functions of the control
parameters: $\mu_1=\eta_1 + \eta_2$ and $\mu_2=\eta_1 \, \eta_2$. } \label{tab7b}}


If renormalizability conditions are dropped, the simple potential
defined in \eref{V41}, with the ${  p}_i$'s specified in
\eref{IB2}, is already sufficient  to make observable all the
allowed phases. It admits, in fact, a stable minimum in each of
the strata listed in Tables~\ref{tab5} and \ref{tab6}, for
suitable values of the $\eta_i$'s, as can be easily realized, with
the help of Figure~\ref{F1}, by conveniently modifying the
geometrical arguments exploited to determine the observable phases
of Model 1. The transformation in Eq.~\eref{proinvB} leads to a
four dimensional semialgebraic set which, contrary to $\Xi$, is
not convex, but, fortunately, like $\Xi$, has no intruding cusps.
In particular, for $A\propto \uno$ let us think of $\eta$ as a
point in the $p$-space.
 Then, if $\eta$ is within or near enough
to the orbit space, there is only one local minimum of the
potential (the absolute minimum) at the point $ {p}$ of the orbit
space which is nearest to $\eta$.

The above statements have been checked analytically: sufficient
conditions for structural stability are expressed in
Table~\ref{tab7b}, page~\pageref{tab7b}.

\section{Complete renormalizable two-Higgs-doublet $+$ singlet extensions of the SM}
In the previous sections we have shown that it is possible to
obtain complete models provided that renormalizability is given up
in the Higgs potential. This could be rightly assessed to be too
high a sacrifice.

In this section we shall propose a cheaper achievement of
completeness in the incomplete 2HD models studied in
Section 4. It is obtained by extending the models by adding one or
two scalar singlets with convenient transformation properties
under the discrete symmetries. The addition of scalar fields
obviously modifies the {\em linear} symmetry group $G$ of the
Higgs sector, {\em i.e.\/} the symmetry of the model, and extends
the set of allowed phases. The important point is that both the
allowed phases of the original models and the newly originated
ones turn out to be observable in the extended versions of the
models. Whether or not the increase in the number of phases is
welcome can only be decided on the basis of an analysis of the
phenomenological consequences of the models.

\subsection{Model 1C: A complete renormalizable extension of Model 1}

In this subsection, we shall show that the addition to Model 1 of
a real SU$_2\times \mathrm{U}_1$-singlet, $\phi_9$, with transformation rule
$\hat\iota\,:\;\phi_9 \rightarrow -\phi_9$ under the reflection
$\hat\iota$, is sufficient to make observable all the phases allowed
by the symmetry of the extended model, that we shall call Model 1C.

A MIB for the linear group $G= \left((\mathrm{SU}_2\times
\mathrm{U}(1))/\integer_2 \times \{ \widehat i \} ,\ {\underline
9}\right)$, acting on the nine independent scalar fields of the
model, is made up of the following eight invariants:

\eqll{\begin{array}{lcl} p_1=\Phi_1^\dagger\Phi_1, \;\;\;
p_2=\Phi_2^\dagger\Phi_2, &\hspace{1em} &p_3=\phi_9^2, \;\;\;
p_4=\mathrm{Re}\left[\Phi_2^\dagger\Phi_1\right]\phi_9,\\
&&\\
p_5=\mathrm{Im}\left[\Phi_2^\dagger\Phi_1\right]\,\phi_9,&\hspace{1em}&
p_6=\mathrm{Re}\left[\Phi_2^\dagger\Phi_1\right]^{2},\; \\
&&\\
p_7 = \mathrm{Im}\left[\Phi_2^\dagger\Phi_1\right]^{2}, & &p_8 =
\mathrm{Re}\left[\Phi_2^\dagger\Phi_1\right]\,
\mathrm{Im}\left[\Phi_2^\dagger\Phi_1\right]\;.
\end{array}}{IB3D}
\FIGURE[ht]{\epsfig{file=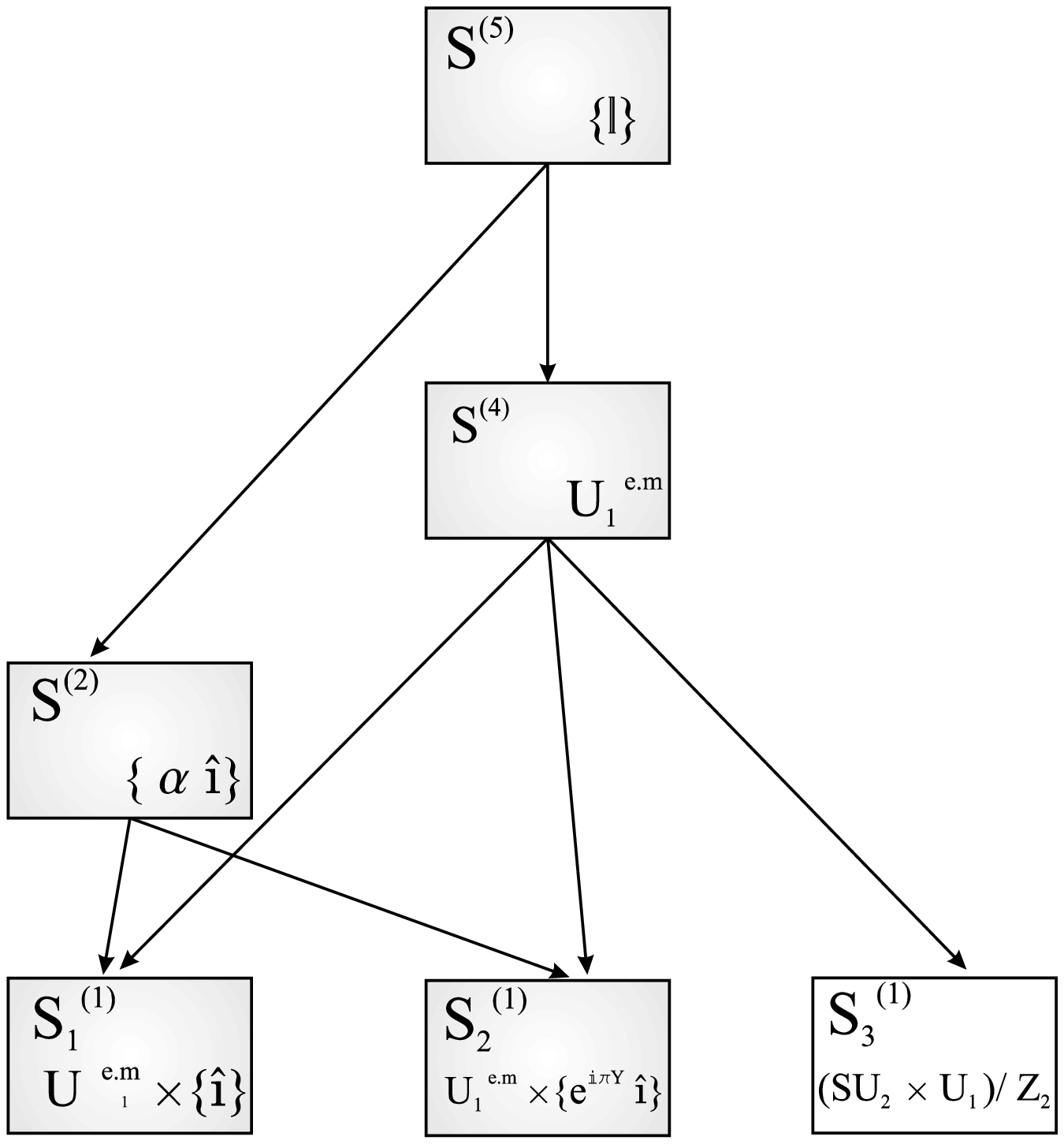,width=9cm}\caption{
Stratification of model 1C. A unique new phase ${\mathcal
F}_3^{(1)}$ associated to the stratum $S_3^{(1)}$  is added to the
set of allowed phases of Model 1 (grey boxes). The stratum
$S^{(0)}$ is not shown for simplicity. It would be connected by
arrows issuing from $S^{(1)}_1$, $S^{(1)}_2$ and
$S^{(1)}_3$.}\label{FF3D}}

The elements of the MIB have degrees $\{2,2,2,3,3,4,4,4\}$ and are
related by six syzygies $s_i=0$:
\eqll{\begin{array}{rcl} s_1 &=& p_4^2 - p_3\, p_6,\ s_2 = p_5^2 -
p_3\, p_7, \
s_3 = p_{8}^2 - p_6\, p_7,\\
s_4 &=&  p_4\, p_8 - p_5\, p_{6},\ s_5 = p_3\, p_{8} - p_4\, p_5,\
s_6 = p_5\, p_{8}- p_4\, p_7.
\end{array}}{SyR}
Only three of the syzygies are independent. Therefore, the orbit
space is a semialgebraic subset of the five dimensional algebraic
variety of the $p$-space $\real^8$, defined by the set of
equations $s_i=0$, $i=1,\dots ,6$. As usual, the relations
defining the orbit space and its stratification can be determined
from rank and positivity conditions of the $\widehat P(p)$-matrix
associated to the MIB defined in (\ref{IB3D}). The results are
reported in Tables~\ref{tab8} and \ref{tab9}.

The only non-vanishing upper triangular elements of the
$\widehat{P}$-matrix turn out to be the following:
\begin{eqnarray*}
\widehat{P}_{i\,i} &=& 4 p_i \;\; \mbox{\rm for} \;\; i=1,2,3\,,\\
\widehat{P}_{4\,4} &=&(p_1+p_2)p_3+p_6\,,\\
\widehat{P}_{5\,5} &=&(p_1+p_2)p_3+p_7\,,\\
\widehat{P}_{j\,j} &=& 4 \left(p_1+p_2\right)\,p_j \;\; \mbox{\rm for} \;\; j=6,7\,,\\
\widehat{P}_{8\,8} &=&(p_1+p_2)(p_6+p_7)\,,\\
  \widehat{P}_{i\,j} &=& 2 p_j \;\; \mbox{\rm for} \;\; i=1,2,3\,\;\mbox{\rm and} \, j=4,5\,,\\
  \widehat{P}_{i\,j} &=& 4 p_j \;\; \mbox{\rm for} \;\; i=1,2\,\;\mbox{\rm and} \, j=6,7\,,\\
    \widehat{P}_{i\,8} &=& 4 p_8 \;\; \mbox{\rm for} \;\; i=1,2\,, \\
  \widehat{P}_{4\,5} &=&  p_8\,,\\
   \widehat{P}_{4\,6} &=& 2 \widehat{P}_{5\,8} =  2(p_1+p_2) p_4\,,\\
   \widehat{P}_{5\,7} &=&2\widehat{P}_{4\,8} = 2(p_1+p_2) p_5\,,\\
    \widehat{P}_{i\,8} &=&  2(p_1+p_2) p_8\,,\mbox{\rm for} \;\; i=6,7\,.\\
\end{eqnarray*}

\TABLE{

\begin{tabular}{lcl}
\hline
 Stratum   &  Symmetry         &   Typical point $\phi$ \\
 \hline
 &&\\
 $S^{(5)}  $  &  $\{\uno\}$ & $(\phi_1,0,\phi_3,\phi_4,0,0,\phi_7,0,\phi_9)$\\
$S^{(4)}$   & U$_1^{\mbox{\rm e.m.}}$  & $(0,0,\phi_3,\phi_4,0,0,\phi_7,0,\phi_9)$ \\
$S^{(2)} $ &  $ \{ \alpha \hat{i}\}$      & $(\phi_1,0,0,0,0,0,\phi_7,0,0)$  \\
$S^{(1)}_1$ &  U$_1^{\mbox{\rm e.m.}} \times \{\hat i\}$      & $(0,0,\phi_3,0,0,0,0,0,0)$ \\
$S^{(1)}_2 $ &  U$_1^{\mbox{\rm e.m.}}\times \{e^{i \pi Y}\,\hat i\}$ & $(0,0,0,0,0,0,\phi_7,0,0)$ \\
$S^{(1)}_3$ & $(\mathrm{SU}_2\times \mathrm{U}_1)/\integer_2$  & $(0,0,0,0,0,0,0,0,\phi_{9})$ \\
$S^{(0)}$   &  $(\mathrm{SU}_2\times \mathrm{U}_1)/\integer_2\,\times \{\hat \iota\}$   & $(0,0,0,0,0,0,0,0,0)$ \\
&&\\
 \hline
\end{tabular}

\caption{  Symmetries of the strata of Model 1C. The group
$\mathrm{U}_1^{\mathrm{e.m.}}$ is defined as in Table~\ref{tab1},
and $\alpha = \mathrm{e}^{\mathrm{i}\pi(T_3-Y/2)}$.
 Symmetries are specified by a {\em
representative element} of the conjugacy class of isotropy
subgroups. Finite groups are defined through their generators
between brackets. For each stratum, a field configuration with the
same symmetry is supplied ({\em typical point}). The $\phi_i$'s
are generic non zero values. } \label{tab8}}



\TABLE{

\begin{tabular}{lll}
\hline
 Stratum   &  Defining relations       &   Boundary \\
\hline
&&\\
 $S^{(5)}  $  & $s_i=0\,,1 \leq i \leq 6$  & $\overline{S^{(4)}}\,,\overline{S^{(2)}}$\\
&$  p_1 p_2 -p_6 - p_7 > 0\,,\;p_1+p_2>0$&\\
&$  p_j \geq 0\,,\;\;j=1,2,3,6,7.$&\\
$S^{(4)}$   & $s_i=0\,,1 \leq i \leq 6$  & $\overline{S^{(1)}_{j}}\,,\;1 \leq j \leq 3$ \\
& $ p_6 + p_7 - p_1 p_2 =0\,,$ & \\
& $p_i+p_j>0\,,\; (i,j)=(1,2)\; \mbox{\rm{and}}\;(1,3)\; \mbox{\rm{and}}\; (2,3)$&\\
& $ p_k \geq 0 \,,\; 1 \leq k \leq 3$ &\\
$S^{(2)} $ & $p_j=0< p_1\,,p_2\,,\;\;\;j \neq 1,2$      & $\overline{S^{(1)}_{j}}\,,\;1 \leq j \leq 2$  \\
$S^{(1)}_1$ & $p_i=0< p_1\,,\;\;\;i \neq 1$        & $S^{(0)}$ \\
$S^{(1)}_2 $ & $p_i=0< p_2\,,\;\;\;i \neq 2$  & $S^{(0)}$ \\
$S^{(1)}_3$ & $p_i=0< p_3\,,\;\;\;i \neq 3$  & $S^{(0)}$\\
$S^{(0)}$   & $p_i=0,\quad 1 \leq i \leq 8$  &  \\
&&\\
\hline
\end{tabular}

\caption{ Stratification of the orbit space of Model 1C. The
syzygies are $s_1= p_4^2-p_3 p_6$, $s_2= p_5^2-p_3 p_7$, $s_3=
p_8^2-p_6 p_7$, $s_4=p_4 p_8 - p_5 p_6$, $s_5=p_3 p_8 - p_4 p_5$,
$s_6=p_5 p_8 - p_4 p_7$. Neighboring strata are indicated, so that
possible second order phase transitions can  be easily identified.
} \label{tab9}}

As expected, a new phase, $S^{(1)}_3$, is now allowed.

The most general invariant polynomial of  degree four in the
scalar fields of the model can be written in the following form:

\eqll{\widehat V^{(4)}(p)=\sum_{i=1}^8\,\alpha_i\,p_i +
\sum_{i,j=1}^3\,A_{ij}\,p_i\,p_j\,,}{WWW} where, to guarantee that
the potential is bounded from below, we assume that all the
coefficients are real, the symmetric matrix $A$ is positive
definite and\footnote{See footnote n.~\ref{notapi} on page~\pageref{notapi}.}
\begin{eqnarray}
A_{11},A_{22} >0, \quad
\alpha_6,\alpha_7 > -2\left(A_{12}+\sqrt{A_{11}\,A_{22}}\right),\nonumber\\
\label{ew9A}\\
 \alpha_8^2 < 4\left[\alpha_6 +
2\left(A_{12}+\sqrt{A_{11}\,A_{22}}\right)\right]\times\left[\alpha_7
+ 2\left(A_{12}+\sqrt{A_{11}\,A_{22}}\right)\right]\,. \nonumber
\end{eqnarray}

 The conditions for the occurrence of
a stationary point of $\widehat V^{(4)}(p)$ in a given stratum
are obtained from equation \eref{ext} and the explicit form of the
relations defining the strata can be read from Table~\ref{tab9}.
%


\TABLE[ht]{ \tiny{

\begin{tabular}{ll}
\hline
&\\
Stratum & Structural stability conditions\\
&\\
\hline
&\\
 $S^{(5)}  $  & $\alpha_2<0\,,\; \alpha_6>0\,,\; \alpha_7>0\,,\;4 \alpha_6 \alpha_7 -\alpha_8^2>0\,,\;\alpha_3<\dfrac{\alpha_5^2
 \alpha_6+\alpha_4^2 \alpha_7-\alpha_4\alpha_5\alpha_8}{4\alpha_6\alpha_7-\alpha_8^2}\,,\;$\\
&\\
 &$ \alpha_8 \left(-2\alpha_5\alpha_6+\alpha_4\alpha_8\right)
 \left(-2\alpha_4\alpha_7+\alpha_5\alpha_8\right)>0\,,\;$\\
 &\\
 &$
 \alpha_1<
 \dfrac{2\left[ {{{\alpha }_5}}^2\,{{\alpha }_6} +
      {{{\alpha }_4}}^2\,{{\alpha }_7} -
      {{\alpha }_4}\,{{\alpha }_5}\,{{\alpha }_8} +
      {{\alpha }_3}\,\left( -4\,{{\alpha }_6}\,{{\alpha }_7} +
         {{{\alpha }_8}}^2 \right)  \right]}{\left( 4\,\alpha_6\,\alpha_7 - \alpha_8^2 \right)^3\,} \times$  \\
&\\
        &$\hspace{5em}\times \dfrac{
    \left[ -4\,{{\alpha }_4}\,{{\alpha }_5}\,
       \left( {{\alpha }_6} + {{\alpha }_7} \right) \,{{\alpha }_8} +
      {{{\alpha }_5}}^2\,\left( 4\,{{{\alpha }_6}}^2 +
         {{{\alpha }_8}}^2 \right)  +
      {{{\alpha }_4}}^2\,\left( 4\,{{{\alpha }_7}}^2 +
         {{{\alpha }_8}}^2 \right)  \right]}{\alpha_2\,\alpha_9} <0$\\
&\\
$S^{(4)}$   & $ 0 < \lambda < 2 \,,\; \alpha_6 >-\lambda\,,\; \alpha_7 >-\lambda\,,\;
\rho_1<0 \,,\;
- \dfrac{2 \lambda}{(\lambda-2)^2} \rho_1^2 < \rho_2 <\dfrac{1}{4} \rho_1^2\,,\;$   \\
&\\
&$ \alpha_8^2 < 4
\left(\alpha_6+\lambda\right)\left(\alpha_7+\lambda\right)\,,\;
\alpha_8 \left[ \alpha_4 \alpha_8
-2\alpha_5\left(\alpha_6+\lambda\right)\right]
 \left[ \alpha_5 \alpha_8 -2\alpha_4\left(\alpha_7+\lambda\right)\right]>0 \,,\;$\\
 &\\
 &$\!\!\!\!\!\!\!\!\rho_1< \dfrac{\left( -2 + \lambda  \right) \,
    {\left( -4\,{{\alpha }_4}\,{{\alpha }_5}\,
         \left( 2\,\lambda  + {{\alpha }_6} +
           {{\alpha }_7} \right) \,{{\alpha }_8} +
        {{{\alpha }_5}}^2\,
         \left( 4\,{\left( \lambda  + {{\alpha }_6}
                \right) }^2 + {{{\alpha }_8}}^2 \right)  +
         {{{\alpha }_4}}^2\,
         \left( 4\,{\left( \lambda  + {{\alpha }_7}
                \right) }^2 + {{{\alpha }_8}}^2 \right)
        \right) }^2}{{\left( -2\,{{\alpha }_5}\,
         \left( \lambda  + {{\alpha }_6} \right)  +
        {{\alpha }_4}\,{{\alpha }_8} \right) }^2\,
    {\left( -4\,\left( \lambda  + {{\alpha }_6} \right) \,
         \left( \lambda  + {{\alpha }_7} \right)  +
        {{{\alpha }_8}}^2 \right) }^2}<0 $ \\
&\\
& $ \!\!\!\!\!\!\!\!{{\alpha }_3} = \dfrac{\lambda \,{{{\alpha
}_4}}^2 +
      \lambda \,{{{\alpha }_5}}^2 +
      {{{\alpha }_5}}^2\,{{\alpha }_6} +
      {{{\alpha }_4}}^2\,{{\alpha }_7} -
      {{\alpha }_4}\,{{\alpha }_5}\,{{\alpha }_8}}{4\,
       {\lambda }^2 + 4\,\lambda \,{{\alpha }_6} +
      4\,\lambda \,{{\alpha }_7} +
      4\,{{\alpha }_6}\,{{\alpha }_7} - {{{\alpha }_8}}^2}$\\
      &\\
$S^{(2)} $ & $\rho_1<0\,,\; 0<\rho_2<\dfrac{\rho_1^2}{4}\,,\;\alpha_6>-2\,,\;\alpha_7>-2\,,$\\
& $\alpha_3>\dfrac{\rho_1}{2} (\alpha_6+\alpha_7)\,,\quad |\alpha_8|<2\sqrt{(\alpha_6+2)(\alpha_7+2)}$\\
& $ 2 \left(\alpha_4^2+\alpha_5^2\right)-8\alpha_3
\left(\alpha_6+\alpha_7\right)+
\left( \alpha_1+\alpha_2 \right) \left(4\alpha_6 \alpha_7 -\alpha_8^2\right)<0\,,\;$\\
&\\
&$\alpha_5^2 \alpha_6 + \alpha_4^2 \alpha_7 -\alpha_4 \alpha_5
\alpha_8 -\alpha_3\left(
 4\alpha_6 \alpha_7 -\alpha_8^2\right)<0$ \\
&\\
$S^{(1)}_1$ & $\alpha_1<0\,,\;\alpha_2>0\,,\;\alpha_3>0\,,\;
\alpha_6>\mathrm{Max}\left(\dfrac{2
\alpha_2}{\alpha_1}\,,-2\right)\,,\;$
$\alpha_7>\mathrm{Max}\left(\dfrac{2 \alpha_2}{\alpha_1}\,,-2\right)\,,\;$\\
& $|\alpha_4|<2 \sqrt{\alpha_3 \left( -\dfrac{2\alpha_2}{\alpha_1}
+\alpha_6 \right)}\,,\quad $ $|\alpha_5|<2 \sqrt{\alpha_3
\left( -\dfrac{2\alpha_2}{\alpha_1} +\alpha_7 \right)}\,,\;$\\
& $\!\!\!\!\dfrac{1}{2\alpha_1 \alpha_3} \left( \alpha_1 \alpha_4
\alpha_5 + \sqrt{
 \left[ 8 \alpha_2 \alpha_3 +\alpha_1\left(\alpha_4^2 - 4 \alpha_3 \alpha_6 \right) \right]
 \left[ 8 \alpha_2 \alpha_3 +\alpha_1\left(\alpha_5^2 - 4 \alpha_3 \alpha_7 \right) \right]}
 \right)<\alpha_8$\\
& $\hspace{3em} \alpha_8< \dfrac{1}{2\alpha_1 \alpha_3} \left(
\alpha_1 \alpha_4 \alpha_5 - \sqrt{
 \left[ 8 \alpha_2 \alpha_3 +\alpha_1\left(\alpha_4^2 - 4 \alpha_3 \alpha_6 \right) \right]
 \left[ 8 \alpha_2 \alpha_3 +\alpha_1\left(\alpha_5^2 - 4 \alpha_3 \alpha_7 \right) \right]}
 \right)\,,$\\
 & $\alpha_8^2<4(\alpha_6+2)(\alpha_7+2)$\\
&\\
$S^{(1)}_2 $ & $\alpha_1>0\,,\;\alpha_2<0\,,\;\alpha_3>0\,,\;
\alpha_6>\mathrm{Max}\left(\dfrac{2\alpha_1}{\alpha_2}\,,-2\right)\,,\;$
$\alpha_7>\mathrm{Max}\left(\dfrac{2 \alpha_1}{\alpha_2}\,,-2\right)\,,\;$\\
&  $|\alpha_4| < 2 \sqrt{\alpha_3 \,\left(- \dfrac{2
\alpha_1}{\alpha_2}+\alpha_6 \right)}\,,\quad
 |\alpha_5| < 2 \sqrt{\alpha_3 \,\left(- \dfrac{2 \alpha_1}{\alpha_2}+\alpha_7 \right)}\,,\;$ \\
 & $\!\!\!\!\dfrac{1}{2\alpha_2 \alpha_3} \left( \alpha_2 \alpha_4 \alpha_5 + \sqrt{
 \left[ 8 \alpha_1 \alpha_3 +\alpha_2\left(\alpha_4^2 - 4 \alpha_3 \alpha_6 \right) \right]
 \left[ 8 \alpha_1 \alpha_3 +\alpha_2\left(\alpha_5^2 - 4 \alpha_3 \alpha_7 \right) \right]}
 \right)<\alpha_8$\\
 & $\hspace{3em} \alpha_8< \dfrac{1}{2\alpha_2 \alpha_3} \left( \alpha_2 \alpha_4 \alpha_5 - \sqrt{
 \left[ 8 \alpha_1 \alpha_3 +\alpha_2\left(\alpha_4^2 - 4 \alpha_3 \alpha_6 \right) \right]
 \left[ 8 \alpha_1 \alpha_3 +\alpha_2\left(\alpha_5^2 - 4 \alpha_3 \alpha_7 \right) \right]}
 \right)\,,$\\
& $\alpha_8^2<4(\alpha_6+2)(\alpha_7+2)$\\
 &\\
$S^{(1)}_3$ & $
\alpha_1>0\,,\;\alpha_2>0\,,\;\alpha_3<0\,,\,\;\alpha_6>-2\,,\;\alpha_7>-2\,,\;\alpha_8^2<4(\alpha_6+2)(\alpha_7+2)\;
$\\
&\\
 $S^{(0)}$ & $
\alpha_1>0\,,\;\alpha_2>0\,,\;\alpha_3>0\,,\;\alpha_6>-2\,,\;\alpha_7>-2\,,\;\alpha_8^2<4(\alpha_6+2)(\alpha_7+2)
$\\
&\\
\hline\\
\end{tabular}}

\caption{ Necessary and sufficient conditions for structural
stability of the phases ${\mathcal F}^{(i)}_j$ associated to the
strata ${\mathcal S}^{(i)}_j$ of model 1C, for $A=\uno$.  Some
solutions are given in terms of $\rho_1=\alpha_1+\alpha_2$ and
$\rho_2=\alpha_1 \alpha_2$.} \label{tab10}}


The high dimensionality of the orbit space prevents, in this case,
a simple geometric determination of conditions guaranteing the
existence of a stable local minimum on a given stratum. For this
model, a complete analytic solution of these conditions is
possible, even if high degree polynomial equations are involved. A
way to overcome this difficulty is  to express the structural
stability conditions in parametric form, at least for some higher
dimensional strata. Moreover, it is sometimes convenient to
symmetrize the solution, {\em i.e.\/} to express it in terms of
the functions $\alpha_i+\alpha_j$ and $\alpha_i \alpha_j$ of
couples of control parameters $\alpha_i$ and $\alpha_j$ appearing
in the Higgs potential.
More generally, the main mathematical problem one has to face is
the solution of systems of polynomial equalities and inequalities
in the phenomenological parameters $\alpha$. At the very end one
hopes to get a Cylindrical Algebraic Decomposition (CAD)\footnote{
For a precise definition  of CAD see \cite{Coste}, page 32.
Loosely speaking, the CAD form of the solution of a system of
inequalities $F_i(x_1,\ldots,x_{n_2})>0$ for $i=1,\ldots,n_1$ is
represented by a set of logical options $\mathrm{Op^{(1)}} ||
\mathrm{Op^{(2)}}||\ldots||\mathrm{Op^{(m)}}||\ldots$, where the
$m$-th option is written in the form:
\begin{eqnarray*}
\mathrm{Op}^{(m)} &=& L^{(m)}_{n_2}(x_1,\ldots,x_{n_2-1}) <x_{n_2}< U^{(m)}_{n_2}(x_1,\ldots,x_{n_2-1}) \&\& \\
&& \cdots \cdots \&\& L^{(m)}_{j}(x_1,\ldots,x_{j-1}) < x_{j} <U^{(m)}_{j}(x_1,\ldots,x_{j-1})\;\cdots \\
&& \cdots \cdots \&\& L^{(m)}_{1} < x_{1} <U^{(m)}_{1}
\end{eqnarray*}
and $L^{(m)}_{1}$ and $U^{(m)}_{1}$ are numbers (the symbols $||$  and $\& \&$ stand for the
boolean `` Or'' and ``And'', respectively).
Every different ordering of the set of variables $x_j$ leads to a different CAD form for the solution; also the
number of options $m$ generally varies.}
 which is sufficiently compact to be fitted in a
table. Since that is very often an impossible task, due to the
large amount of logical options involved, in this work we
contented ourselves with exhibiting sufficient conditions for
structural stability of Models $1_2$, $2_2$ (see above) and 2C
(see below).
In the case of model 1C, in Table~\ref{tab10} we  reported the
complete solution (necessary and sufficient conditions) but, in
the aim of keeping the table within reasonable dimensions, we
renounced to the standard CAD form, which can be derived and
written out with a reasonable effort.

Model 1C could be relevant in
the study of electro-weak baryogenesis: $CP$ violation is achieved
in phase $S^{(4)}_1$, so it is interesting to examine the
possibility of first order phase transitions to more symmetrical
phases \cite{5}.

\subsection{Model 2C: A complete renormalizable extension of Model 2}

Like Model 1, Model 2 can be completed, without giving up
renormalizability, through the addition of scalar singlets with
convenient transformation properties under the discrete symmetry
group. We shall call 2C, the the model obtained from Model 2 by
adding a couple of real SU$_2\times \mathrm{U}_1$-singlets,
denoted by $\phi_9$ and $\phi_{10}$, with transformation rules
$(\phi_9,\phi_{10})\rightarrow (-\phi_9,-\phi_{10})$ and,
respectively, $(\phi_9,\phi_{10})\rightarrow (-\phi_9,\phi_{10})$
under transformations induced by $\hat\iota$ and $K$.

The following set of invariants yields a MIB in the present case:
\eqll{\begin{array}{l} p_1=\Phi^\dagger_1\,\Phi_1\,, \hspace{1em}
\ p_2=\Phi^\dagger_2\,\Phi_2\,, \hspace{1em} p_3={\phi_9}^2\,,
\hspace{1em} p_4={\phi_{10}}^2,\ \
 \\
p_5=\mathrm{Im}\left[\Phi_2^\dagger\Phi_1\right]\phi_9\,,
\hspace{1em}
p_6=\mathrm{Re}\left[\Phi_2^\dagger\Phi_1\right]\phi_{10},\;
 \\
p_7 =
\left(\mathrm{Re}\left[\Phi_2^\dagger\Phi_1\right]\right)^2\,,
\hspace{1em} p_8 =
\left(\mathrm{Im}\left[\Phi_2^\dagger\Phi_1\right]\right)^2\; .
\end{array}}{IB3A}

The elements of the MIB have degrees $\{2,2,2,2,3,3,4,4\}$ and are
related by the  independent syzygies $s_1=0$ and $s_2=0$, where
\begin{equation}
\begin{array}{lcr}
s_1&=&p_6^2-p_4 p_7\,, \\
s_2&=&p_5^2-p_3 p_8\,.
\end{array}
\end{equation}

Therefore, the orbit space is a semialgebraic subset of the six
dimensional algebraic variety defined in the $p$-space $\real^8$
by the set of equations $s_1=s_2=0$. The relations defining the
orbit space and its stratification, reported in Tables~\ref{tab11}
and \ref{tab12}, can be determined from rank and positivity
conditions of the $\widehat P(p)$-matrix associated to the MIB
defined in (\ref{IB3A}), whose non-vanishing upper triangular
elements are listed below:

\begin{eqnarray*}
\widehat{P}_{i\,i} &=& 4 p_i, \qquad i=1,2,3,4\,,\\
\widehat{P}_{5\,5} &=&(p_1+p_2)p_3+p_8\,,\\
\widehat{P}_{6\,6} &=&\left( {p_1} + {p_2} \right) \,{p_4} + {p_7}\,,\\
\widehat{P}_{j\,j} &=& 4 \left(p_1+p_2\right)\,p_j ,\qquad j=7,8\,,\\
  \widehat{P}_{i\,5} &=& 2 p_5 \,,\qquad i=1,2,3\,,\\
  \widehat{P}_{i\,6} &=& 2 p_6 \,,\qquad i=1,2,4\,,\\
  \widehat{P}_{i\,7} &=& 4 p_7 \,,\qquad i=1,2\,, \\
  \widehat{P}_{i\,8} &=& 4 p_8 \,,\qquad i=1,2\,, \\
    \widehat{P}_{5\,8} &=& 2 \left( p_1 + p_2\right) p_5\,,\\
  \widehat{P}_{6\,7} &=& 2 \left( p_1 + p_2\right ) p_6\,.\\
\end{eqnarray*}


\TABLE{

\begin{tabular}{lcl}
\hline
 Stratum   &  Symmetry         &   Typical point $\phi$ \\
 \hline
 $S^{(6)}  $  &  $\{\uno\}$ & $(\phi_1,0,\phi_3,\phi_4,0,0,\phi_7,0,\phi_9,\phi_{10})$\\
 $S^{(5)}$   & U$_1^{\mathrm e.m.}$  &  $(0,0,\phi_3,\phi_4,0,0,\phi_7,0,\phi_9,\phi_{10})$\\
 $S^{(4)}_1$ &  $\{K\}$            & $(\phi_1,0,\phi_3,0,0,0,\phi_7,0,0,\phi_{10})$   \\
 $S^{(4)}_2$ & $\{\hat\iota\,K\}$  &$(\phi_1,0,0,\phi_4,0,0,\phi_7,0,\phi_9,0)$   \\
$S^{(3)}_1 $ &  U$_1^{\mathrm e.m.}\times\{K\}$     & $(0,0,\phi_3,0,0,0,\phi_7,0,0,\phi_{10})$   \\
$S^{(3)}_2 $  & U$_1^{\mathrm e.m.}\times\,\{\hat\iota\,K\}$ & $(0,0,0,\phi_4,0,0,\phi_7,0,\phi_9,0)$ \\
$S^{(2)}_1 $ &  $\{\alpha\,\hat\iota ,K\}$      & $(\phi_1,0,0,0,0,0,\phi_7,0,0,0)$  \\
$S^{(2)}_2$ &  $(\mathrm{SU}_2\times \mathrm{U}_1)/\integer_2$  & $(0,0,0,0,0,0,0,0,\phi_9,\phi_{10})$   \\
$S^{(1)}_1$ &  U$_1^{\mathrm e.m.}\,\times\{\hat\iota, K\}$      & $(0,0,\phi_3,0,0,0,0,0,0,0)$ \\
$S^{(1)}_2 $ &  U$_1^{\mathrm e.m.}\times \{e^{i\pi Y}\,\hat\iota, K\}$ & $(0,0,0,0,0,0,\phi_7,0,0,0)$ \\
$S^{(1)}_3$ & $(\mathrm{SU}_2\times \mathrm{U}_1)/\integer_2\times\{K\}$    & $(0,0,0,0,0,0,0,0,\phi_9,0)$ \\
$S^{(1)}_4$ & $(\mathrm{SU}_2\times \mathrm{U}_1)/\integer_2\times\{\hat\iota \,K\}$  & $(0,0,0,0,0,0,0,0,0,\phi_{10})$ \\
$S^{(0)}$   &  $(\mathrm{SU}_2\times \mathrm{U}_1)/\integer_2\,\times \{\hat \iota\,,\;K\}$       & $(0,0,0,0,0,0,0,0,0,0)$ \\
\hline
\end{tabular}

\caption{ Symmetries of the strata $S$ of Model 2C.
The group
$\mathrm{U}_1^{\mathrm{e.m.}}$ is defined as in Table~\ref{tab1},
and $\alpha = \mathrm{e}^{\mathrm{i}\pi(T_3-Y/2)}$.
 Symmetries are specified by a {\em
representative element} of the conjugacy class of isotropy
subgroups. Finite groups are defined through their generators
between brackets. For each stratum, a field configuration with the
same symmetry is supplied {\em (typical point)}. The $\phi_i$'s
are generic non zero values. } \label{tab11}}


\TABLE{

\begin{tabular}{lll}
\hline
 Stratum   &  Defining relations       &   Boundary \\
 \hline
 &&\\
 $S^{(6)}  $  & $s_1=s_2=0<q\,,p_1,p_2;  $  & $\overline{S^{(5)}}\,,\;\overline{S^{(4)}_{1}}\,,\;\overline{S^{(4)}_{2}}$\\
                                  & $p_4(p_1+p_2)+p_7\,,p_3(p_1+p_2)+p_8>0\,;$ &\\
                                  &  $p_i \geq 0$ for $i=3,4$                               &\\
  $S^{(5)}$ & $s_1=s_2=q=0\,;  $    &   $\overline{S^{(2)}_{2}}\,,\;\overline{S^{(3)}_{1}}\,\;\overline{S^{(3)}_{2}}$\\
                                   & $p_4 (p_1+p_2)+p_7\,,p_3(p_1+p_2)+ p_8\,,p_7+ p_8>0\,;$  & \\
                                    & $p_i \geq 0$ for $i=1,2,3,4$  &\\
 $S^{(4)}_1$ & $s_1=p_3=p_5=p_8=0<q\,, p_4(p_1+p_2)+ p_7 $        & $\overline{S^{(2)}_{1}}\,,\;\overline{S^{(3)}_{1}}$   \\
                                    &  $p_i \geq 0$ for $i=1,2,4$      &\\
$S^{(4)}_2$   & $s_2=p_4=p_6=p_7=0<q\,, p_3(p_1+p_2)+ p_8 $     &            $\overline{S^{(2)}_{1}}\,,\;\overline{S^{(3)}_{2}}$ \\
                                     &  $p_i \geq 0$ for $i=1,2,3$                   &\\
$S^{(3)}_1 $ & $s_1=p_3=p_5=p_8=q=0<p_1+p_2 \,,p_1+p_4 \,,p_2+p_4  $      &  $\overline{S^{(1)}_{1}}\,,\;\overline{S^{(1)}_{2}}\,,\;\overline{S^{(1)}_{4}}$  \\
                                     &  $p_i \geq 0$ for $i=1,2$                               &\\
$S^{(3)}_2 $  &$s_2=p_4=p_6=p_7=q=0<p_1+ p_2 \,,p_1+ p_3 \,,p_2+ p_3 $ & $\overline{S^{(1)}_{1}}\,,\;\overline{S^{(1)}_{2}}\,,\;\overline{S^{(1)}_{3}}$  \\
                                      &   $p_i \geq 0$ for $i=1,2$     &\\
$S^{(2)}_1 $ & $p_i=0\,,\; i \neq 1,2\,; \; 0<p_1\,,p_2$       &  $\overline{S^{(1)}_{1}}\,,\;\overline{S^{(1)}_{2}}$ \\
 $S^{(2)}_2$ & $p_i=0\,,\; i \neq 3,4\,; \; 0<p_3\,,p_4$    & $\overline{S^{(1)}_{3}}\,,\;\overline{S^{(1)}_{4}}$   \\
$S^{(1)}_1$ &  $p_i=0<p_1\,,\;i\neq 1$     & $S^{(0)}$ \\
$S^{(1)}_2 $ &  $p_i=0<p_2\,,\;i\neq 2$ &  $S^{(0)}$\\
$S^{(1)}_3$ &   $p_i=0<p_3\,,\;i\neq 3$  & $S^{(0)}$ \\
$S^{(1)}_4$ & $p_i=0<p_4\,,\;i\neq 4$  & $S^{(0)}$ \\
$S^{(0)}$   & $p_i=0$ for $ 1 \leq i \leq 8$      &  \\
&&\\ \hline
\end{tabular}

\caption{ Orbit space characterization of strata $S$ of Model 2C.
The syzygies are $s_1= p_6^2-p_4 p_7$, $s_2= p_5^2-p_3 p_8$, and $
q=p_1 p_2 - p_7 - p_8$.
Neighbouring strata are given, so that
possible second order phase transitions can  be easily identified.
} \label{tab12}}




\TABLE[ht]{

\tiny{
\begin{tabular}{ll}%
\hline
Stratum & Structural stability conditions\\
\hline
&\\
$S^{(6)}$ & $\alpha_{7}>0,$ $\alpha_{8}>0,$ $\rho_{1}<0,$
$\alpha_{5}^{2} -4\ \alpha_{3}\alpha_{8}>0,$  $\alpha_{6}^{2}-4\
\alpha_{4}\alpha_{7}
>0,$\\
 &$\rho_{1}^{2}>\dfrac{1}{2}\left(  \dfrac{\alpha_{6}^{2}\left(  \alpha
_{6}^{2}-4\ \alpha_{4}\alpha_{7}\right)
}{\alpha_{7}^{3}}+\dfrac{\alpha _{5}^{2}\left(  \alpha_{5}^{2}-4\
\alpha_{3}\alpha_{8}\right)  }{\alpha
_{8}^{3}}\right)  ,$\\
& $\ \ \dfrac{1}{8}\left(  \dfrac{\alpha_{6}^{2}\left(
\alpha_{6}^{2}- 4\ \alpha_{4}\alpha_{7}\right)
}{\alpha_{7}^{3}}+\dfrac{\alpha_{5}^{2} \left(  \alpha_{5}^{2}-4\
\alpha_{3}\alpha_{8}\right)  }{\alpha_{8}^{3}}\right)
 <\rho_{2}<\dfrac{\rho_{1}^{2}}{4}$\\
&\\
$S^{(5)}$ & $-2<\lambda<0,$ $\alpha_{7}>\lambda,$
$\alpha_{8}>\lambda,$ $\rho_{1}<0,$ \  $\dfrac{\alpha_{5}^{2}}{4\
\left(  \alpha_{8}-\lambda\right) }-\dfrac{2\left(
\alpha_{8}-\lambda\right)  ^{2}}{\alpha_{5}^{2}\left(
\lambda+2\right)  ^{2}}\
\rho_{1}^{2}<\alpha_{3}<\dfrac{\alpha_{5}^{2}}{4\
 \left(  \alpha_{8}-\lambda\right)  },$\\
& $\dfrac{\alpha_{5}^{2}\left(  \lambda-2\right)  ^{2}\left[
\alpha_{5}^{2}- 4\alpha_{3}\left(  \alpha_{8}-\lambda\right)
\right]  }{32\left( \alpha_{8}-\lambda\right)  ^{3}}+\dfrac{2\
\lambda}{\left(  2+\lambda\right)
^{2}}\ \rho_{1}^{2}<\rho_{2}<\dfrac{\rho_{1}^{2}}{4}$\\
& $\!\!\!\! {{\alpha}_4} = \dfrac{-\left({\left(-4 + {\lambda}^2
\right)}^2\, \ {{{\alpha}_6}}^4 \right) +
          64\, \lambda\, {{{\alpha}_1}}^2\, {\left( \lambda - {{\alpha}_7} \right)}^3 -
          32\, \left(4 + {\lambda}^2 \right)\, {{\alpha}_1}\, {{\alpha}_2}\, \
{\left( \lambda - {{\alpha}_7} \right)}^3 +
          64\, \lambda\, {{{\alpha}_2}}^2\, {\left( \lambda - {{\alpha}_7} \right)}^3}{4\,
          {\left(-2 + \lambda \right)}^2\, {\left(2 + \lambda \right)}^2\,
           {{{\alpha}_6}}^2\, \left( \lambda - {{\alpha}_7} \right)} +$\\
& \hspace{3em}
    $+\dfrac{{{{\alpha}_5}}^4\, {\left( \lambda - {{\alpha}_7} \right)}^2}
{4\, {{{\alpha}_6}}^2\, {\left(-\lambda + {{\alpha}_8} \right)}^3}
-
  \dfrac{{{\alpha}_3}\, {{{\alpha}_5}}^2\, {\left( \lambda - {{\alpha}_7}
  \right)}^2}{{{{\alpha}_6}}^2\, {\left(-\lambda + {{\alpha}_8} \right)}^2}$ \\
&\\
$S_{1}^{(4)}$ & $\alpha_{7}>0,$ $\ \rho_{1}<0,$  $\alpha_{8}>0,$
$\ \alpha _{5}^{2}\ -4\ \alpha_{3}\alpha_{8}<0,$ $\alpha_{6}^{2}\
-4\ \alpha_{4} \alpha_{7}>0,$
$\rho_{1}^{2}>\dfrac{\alpha_{6}^{2}\left(  \alpha_{6}^{2}\
 -4\ \alpha_{4}\alpha_{7}\right)  }{2\ \alpha_{7}^{3}},$\\
& $\dfrac{\alpha_{6}^{2}\left(  \alpha_{6}^{2}\ -4\
\alpha_{4}\alpha
_{7}\right)  }{8\ \alpha_{7}^{3}} <\rho_{2}<\dfrac{\rho_{1}^{2}}{4}$\\
&\\
$S_{2}^{(4)}$ & $\alpha_{7}>0,$ $\alpha_{8}>0,$ $\rho_{1}<0,$
$\alpha_{5}^{2}- 4\ \alpha_{3}\alpha_{8}>0,$ $\alpha_{6}^{2}-4\
\alpha_{4}\alpha_{7}<0,$ $\rho_{1}^{2}>\dfrac{\alpha_{5}^{2}\left(
\alpha_{5}^{2}\ -4\ \alpha
_{3}\alpha_{8}\right)  }{2\ \alpha_{8}^{3}}$\\
& $\dfrac{\alpha_{5}^{2}\left(  \alpha_{5}^{2}\ -4\
\alpha_{3}\alpha
_{8}\right)  }{8\ \alpha_{8}^{3}} <\rho_{2}<\dfrac{\rho_{1}^{2}}{4}$\\
&\\
$S_{1}^{(3)}$ & $-2<\lambda<0,$ $\rho_{1}<0,$ $\dfrac{2\
\lambda}{\left( 2+\lambda\right)  ^{2}}\
\rho_{1}^{2}<\rho_{2}<\dfrac{\rho_{1}^{2}}{4},$ $\ \
\alpha_{3}>0,$ $\alpha_{7}>\lambda,$ $\alpha_{8}>\lambda,$
 $\alpha_{5}^{2}-4\ \alpha_{3}\left(  \alpha_{8}-\lambda\right)  <0$\\
&${{{{\alpha}_4}} = {\dfrac{{\left(-4 + {\lambda}^2 \right)}^2\, \
{{{\alpha}_6}}^4 +
            32\, \left( \lambda\, {{\alpha}_1} - 2\, {{\alpha}_2} \right)\,
          \left(-2\, {{\alpha}_1} + \lambda\, {{\alpha}_2} \right)\, {\left(
          \lambda - {{\alpha}_7} \right)}^3}{4\, {\left(-4 + {\lambda}^2 \right)}^2\, \
{{{\alpha}_6}}^2\,
          \left(-\lambda + {{\alpha}_7} \right)}}}$\\
&\\
$S_{2}^{(3)}$ & $0<\lambda<2,$ $\rho_{1}<0,$ $\dfrac{2\
\lambda}{\left( 2+\lambda\right)  ^{2}}\
\rho_{1}^{2}<\rho_{2}<\dfrac{\rho_{1}^{2}}{4},$ $\ \
\alpha_{4}>0,$ $\alpha_{7}>\lambda,$ $\alpha_{8}>\lambda,$
 $\alpha_{6}^{2}-4\ \alpha_{4}\left(  \alpha_{7}-\lambda\right)  <0$\\
&${{{\alpha}_3}} = \dfrac{{\left(-4 + {\lambda}^2 \right)}^2\,
{{{\alpha}_5}}^4 \ + 32\, \left( \lambda\, {{\alpha}_1} - 2\,
{{\alpha}_2} \right)\,
      \left(-2\, {{\alpha}_1} + \lambda\, {{\alpha}_2} \right)\, {\left( \
\lambda - {{\alpha}_8} \right)}^3}{4\, {\left(-4 + {\lambda}^2
\right)}^2\, \
{{{\alpha}_5}}^2\, \left(-\lambda + {{\alpha}_8} \right)}$\\
&\\
$S_{1}^{(2)}$ & $\alpha_{1}<0,$ $\alpha_{2}<0,$ $\alpha_{3}>0,$
$\alpha_{4}>0,$  $\alpha_{7}$ $>\dfrac{\alpha_{6}^{2}}{4
\alpha_{4}},$
$\ \alpha_{8}$ $>\dfrac{\alpha_{5}^{2}}{4\;\alpha_{3}}$\\
$S_{2}^{(2)}$ & $\alpha_{1}>0,$ $\alpha_{2}>0,$ $\alpha_{3}<0,$
$\alpha _{4}<0,$ $|\alpha_{5}|$ $<$
$2\sqrt{2}\sqrt{-\dfrac{\alpha_{1\ }\alpha_{2}}{\alpha_{3}}},$
 $|\ \alpha_{6}|$ $<\sqrt{\dfrac{-8\ \alpha_{1}\alpha
_{2}-\alpha_{3}\alpha_{5}^{2}}{\alpha_{4}}}$, $\alpha_7>-2$, $\alpha_8>-2$ \\
$S_{1}^{(1)}$ & $\alpha_{1}<0,$ $0<\alpha_{2}<-\alpha_{1},$
$\alpha_{3}>0,$ $\alpha _{4}>0,$  $\alpha_{7}>\
\dfrac{8\;\alpha_{2}\alpha_{4}+\alpha_{1}\alpha_{6}^{2}}{4
\;\alpha_{1}\alpha_{4}},$ $\alpha_{8}>\ \dfrac
{8\ \alpha_{2}\alpha_{3}+\alpha_{1}\alpha_{5}^{2}}{4\ \alpha_{1}\alpha_{3}}$\\
$S_{2}^{(1)}$ & $\alpha_{1}>0,$ $-\alpha_{1}<\alpha_{2}<0,$
$\alpha_{3}>0,$ $\alpha _{4}>0,$ $\
\alpha_{7}>\dfrac{8\;\alpha_{1}\alpha_{4}+\alpha_{2}\alpha_{6}^{2}}{4\;\alpha_{2}\;\alpha_{4}},$
$\alpha_{8}
>\ \dfrac{8\ \alpha_{1}\alpha_{3}+\alpha_{2}\alpha_{5}^{2}}{4\ \alpha
_{2}\;\alpha_{3}}$\\
$S_{3}^{(1)}$ & $\ \alpha_{1}>0,$ $\alpha_{2}>0,$ $\alpha_{3}<0,$
$\alpha
_{4}>0,$ $|\alpha_{5}|$ $<$ $2\sqrt{2}\sqrt{-\dfrac{\alpha_{1\ }\alpha_{2}}{\alpha_{3}}}$,
 $\alpha_7>-2$, $\alpha_8>-2$\\
$S_{4}^{(1)}$ & $\ \alpha_{1}>0,$ $\alpha_{2}>0,$ $\alpha_{3}>0,$
$\alpha
_{4}<0,$ $|\alpha_{6}|$ $<$ $2\sqrt{2}\sqrt{-\dfrac{\alpha_{1\ }\alpha_{2}}{\alpha_{4}}}$,  $\alpha_7>-2$, $\alpha_8>-2$\\
&\\
$S^{(0)}$ & $\ \alpha_{1}>0,$ $\alpha_{2}>0,$ $\alpha_{3}>0$,
$\alpha
_{4}>0$,   $\alpha_7>-2$, $\alpha_8>-2$\\
&\\
\hline\\
\end{tabular}

}\caption{  Sufficient conditions for structural stability of
strata of model 2C, for $A=\uno$.  Some solutions are given in
terms of $\rho_1=\alpha_1+\alpha_2$ and $\rho_2=\alpha_1
\alpha_2$.} \label{tab13}}


As expected, three new phases, $S^{(2)}_2$,  $S^{(1)}_3$ and
$S^{(1)}_4$, are now allowed.

The most general invariant polynomial of  degree four in the
scalar fields of the model can be written in terms of the
following polynomial $\widehat V(p)$ in the $p_i$'s with degree
$\le 4$:

\eqll{\widehat V(p)=\sum_{i=1}^8\,\alpha_i\,p_i +
\sum_{i,j=1}^4\,A_{ij}\,p_i\,p_j\,,}{V3X'} where all the
coefficients are real and, to guarantee that the potential is
bounded from below, the symmetric matrix $A$ is positive
definite\footnote{See footnote n.~\ref{notapi} on page~\pageref{notapi}.}.

The conditions for the occurrence of a stationary point of $\widehat V(p)$ in a
given stratum  are obtained from equation \eref{ext} and the
explicit form of the relations defining the strata can be read
from Table~\ref{tab12}.
\FIGURE[ht]{\epsfig{file=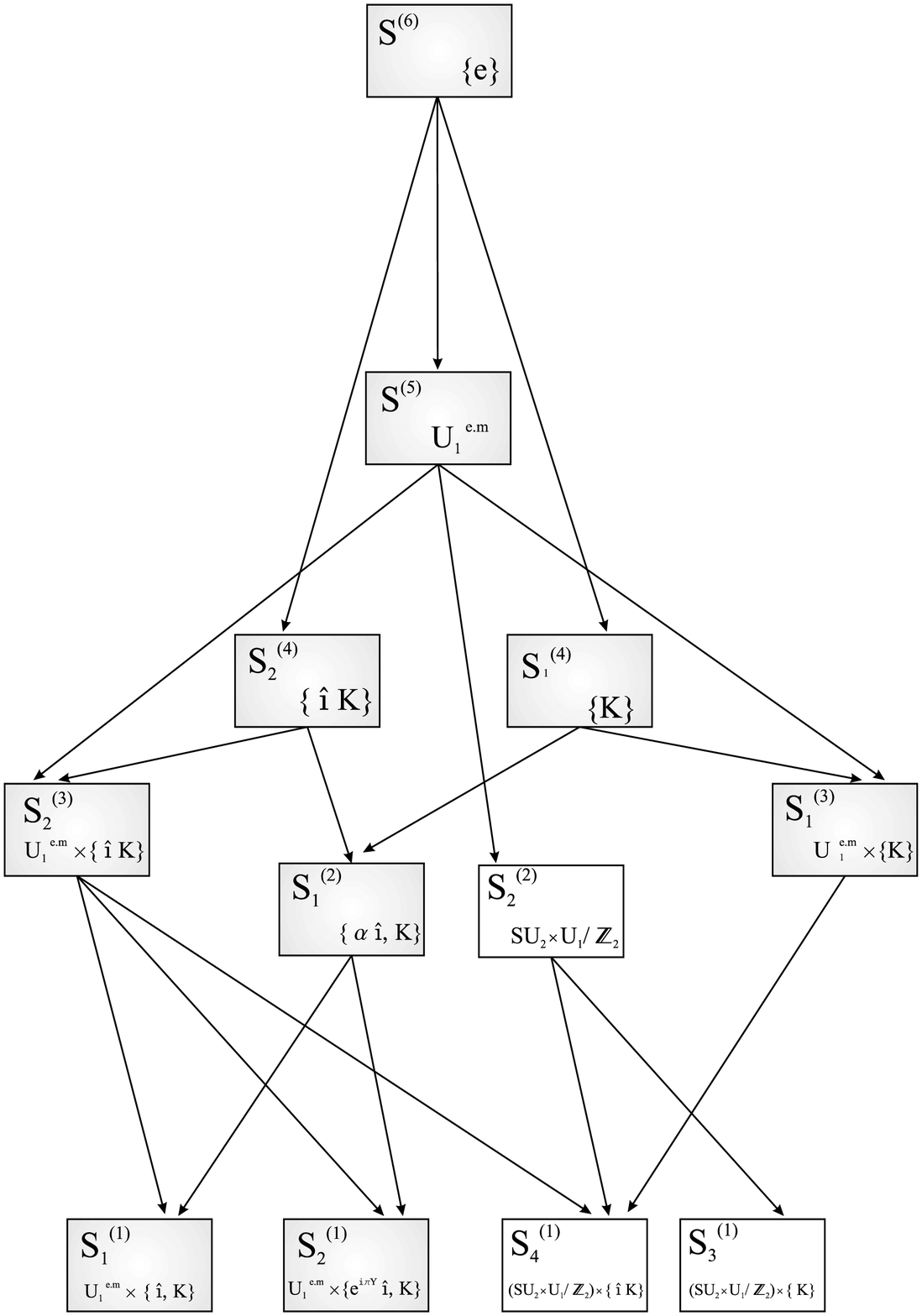,height=16cm}\caption{
Stratification of model 2C. Three new phases ${\mathcal
F}^{(2)}_2$, ${\mathcal F}^{(1)}_3$ and ${\mathcal F}^{(1)}_4$
associated to the strata $S^{(2)}_2$, $S^{(1)}_3$ and $S^{(1)}_4$
are added to the allowed phases of Model 2 (grey boxes). The
stratum $S^{(0)}$ is not shown for simplicity. It would be
connected
 by arrows issuing from each of the one dimensional strata.
}\label{FF3A}}

In this case too, the high dimensionality of the orbit space prevents
a simple geometric determination of the conditions guaranteing the
existence of a stable local minimum on a given stratum
and a complete analytic solution of these conditions
is impossible, since exceedingly high degree polynomial equations are
involved. Despite this, using convenient majorizations, we have
been able to prove
that all the phases allowed by the symmetry of Model 2C are
observable. In particular, for each allowed phase of symmetry
$[H]$, we have analytically determined an eight dimensional open
semialgebraic set ${R}_H$ in the space of the coefficients
$\alpha=(\alpha_1,\dots ,\alpha_8)$, such that, for all $\alpha\in
{R}_H$ and $A$ in a convenient neighborhood of the $4\times 4$
unit matrix, the potential $\widehat V(p(\phi))$, defined through
(\ref{V3X'}), has a stable absolute minimum in the stratum with
symmetry $[H]$.

In Table~\ref{tab13} we have listed the values of the $\alpha_i$'s
(in CAD form) that guarantee the location of the absolute minimum of $\widehat V
(p)$ in the different strata, for $A=\uno$.

Model 2C could be relevant in
the study of electro-weak baryogenesis: $CP$ violation is achieved
in the phase ${\mathcal F}^{(5)}$, so it is interesting to examine the
possibility of first order phase transitions to more symmetrical
phases \cite{5}.

\section{Comments and conclusions}

We have shown that in some renormalizable Quantum Field Theory
models with spontaneously broken gauge invariance, the request
that the Higgs potential is an invariant polynomial of degree not
exceeding four has the intriguing consequence of preventing the
observability, at tree-level, of some phases that would be,
otherwise, allowed by the symmetries of the models. Since
radiative corrections to the Higgs potential are invariant
polynomials of increasing degree at growing perturbative orders,
one could think that the problem can be solved by dynamics. We
have shown that this is not obvious at all. We have checked, in
fact, that the phenomenon can persist also if one-loop radiative
corrections are taken into account. This raises the doubt that
radiative corrections cannot be a general solution to the problem
of unobservability of some phases. In view also of the practical
difficulties which would be met to prove the completeness of the
perturbative solution of a model, we have proposed that tree-level
completeness should be accepted as a rule in building the Higgs
sector of any viable gauge model of electro-weak interactions.

We have proved that some popular 2HD extensions of
the SM, with discrete symmetries preventing NCFC
effects,  do not satisfy this requirement,
but the models can be made complete if the Higgs potentials are
allowed to be a sufficiently high degree polynomial in the scalar
fields. This choice might appear to be not very appealing, because it
implies giving up renormalizability. Thus, looking for a way to reconcile
completeness and renormalizability we found that a simple solution actually exists:
it is sufficient to extend the Higgs sector of these models through the addition
of scalar singlets with convenient transformation
properties under the discrete symmetries.

The advantages of matching symmetry and renormalizability are
quite obvious:

\begin{itemize}
\item[i):] It is possible to employ standard renormalizable
quantum field theory techniques also to deal with (possibly) new
physics phenomena.
\item[ii):] The analysis of standard 2HD models can
give important hints in the extensions of the SM Higgs sector.
\item[iii):] It is possible to conceive an Higgs sector extension
of the SM in which CP violation is spontaneously realized.
\end{itemize}
The phenomenological consequences of the last point are under
examination (\cite{5}).

The results we have obtained are relevant even if finite
temperature corrections to the effective potential are taken into
account. In fact, let us consider one loop thermal contribution to
the tree level Higgs potential: a high temperature series
expansion leads to the inclusion of two opposite contributions.
One is positive, symmetry restoring, and proportional to $\sum_i
(M^2_A)_i \,T^2$, where $(M^2_A)_i$ are the eigenvalues of the
gauge boson mass matrix, depending on the VEV's of the real Higgs
fields. The other one, which is negative and proportional to
$\sum_i {(M^2_A)_i}^{3/2} \,T$, contributes to the barrier in the
potential that makes the transition first order (see for instance
\cite{Cline} and references therein). We just note that the
inclusion of the symmetry restoring term is equivalent to the
increase (with temperature) in the values of the $\alpha$'s which
multiply second degree invariants (denoted by $\alpha^{(II)}$). It
is easy to realize, from the tables exhibiting structural
stability conditions for the different models, that a stable
minimum falls on $S^{(0)}$ whenever all the $\alpha^{(II)}$'s are
positive. The term linear in the temperature can be written as an
algebraic function of the basic polynomial
invariants of the linear group $G$, defining the symmetry of the
model. So, also in this case, an orbit space approach makes simpler the
analysis of possible spontaneous CP violation.
 In this case it becomes fundamental, not
only for a preliminary zero temperature analysis, to get the
complete symmetry breaking scheme of the model.

Let us conclude with some speculations concerning some (possible)
interpretations of the new singlet scalar fields appearing in the
completion of 2HD models studied in this paper. As for the
transformation properties under the symmetry group, the scalar
singlets behave like composite fields of a couple of doublet
fields, which enter in the construction of the basic polynomial
invariants. So, in the phenomenological approach (\`a la
Landau-Ginzburg) to the study of phase transitions that we are
considering, their introduction could be justified by the
necessity of accounting for the possible formation of bound states
of the Higgs doublets.

Alternatively, one might think that the observable phases are the
visible effects of a symmetry in a {\em field ``superspace''}, in
the spirit of the superspace group approach to quasi-crystals (for
a review, see \cite{Janner}). We just recall that in the
superspace group approach to quasi-crystals, the visible
diffraction structure exhibits some regularities, which can be
interpreted as the result of a projection of a super-crystal from
some super world to the physical one. Paralleling this framework,
one could also think that the new phases appearing after the
renormalizable completion of the Models are actually not visible.
In order to get a {\em weak isomorphism\/} between the phases of
the original model and the ones appearing in the Model enriched
with the new singlet fields, one could suitably restrict the
control parameter space, appealing some unknown dynamical reason,
in such a way that all the new phases are not stable (thus
unobservable), while the original ones are all attainable. For
instance, it would be sufficient to require that $\alpha_3>0$ for
Model 1C, and $\alpha_3,\, \alpha_4>0$ for Model 2C. It has not to
be forgotten, however, that the new singlet fields have some
indirect impact also in the scalar sector, since the number of the
eigenvalues of the mass matrix and their numerical values
generally depend also on the VEV of this new fields.
\acknowledgments
The authors would like to thank Antonio Masiero,
for helpful suggestions and discussions
on phenomenological aspects of our proposals. One of us (G.V.)
thanks Ferruccio Feruglio, Antonio Riotto,  and Anna Rossi
for helpful discussions and comments.

\end{document}